\documentclass[aps,prd,twocolumn,preprintnumbers,nofootinbib,amsmath,amssymb]{revtex4-2}

\usepackage{slashbox,latexsym,amssymb,amsfonts,amsmath,slashed,amsthm,bm,bbm}
\usepackage{graphicx}
\usepackage{mathrsfs}

\usepackage{xr}
\usepackage[pdftex, pdftitle={Dirac spectrum in the chirally symmetric phase of a gauge
  theory. II}]{hyperref}

\usepackage{xcolor}
\usepackage[normalem]{ulem}

\newcommand{\exref}[1]{{{\DSI}-\protect\NoHyper\ref{#1}\protect\endNoHyper}}

\makeatletter
\DeclareRobustCommand{\exeqref}[1]{\textup{\tagform@{\DSI-\protect\NoHyper\ref{#1}\protect\endNoHyper}}}
\newcount\dspbrk@lvl
\interdisplaylinepenalty\@M
\makeatother

\newcommand{\DSI}{DS1} 

\newcommand{\f}[2]{\frac{#1}{#2}}
\newcommand{\tf}[2]{{\textstyle\f{#1}{#2}}}

\newcommand{\de}{\partial}

\newcommand{\la}{\langle}
\newcommand{\ra}{\rangle}

\newcommand{\susc}{\mathcal{A}}

\newcommand{\ff}{f}

\newcommand{\ffh}{\hat{f}}

\newcommand{\Iu}{I^{(1)}}

\newcommand{\Id}{I^{(2)}}

\newcommand{\II}{I}

\newcommand{\chit}{\chi_t}
\newcommand{\lvol}{\mathrm{V}_4}
\newcommand{\svol}{\mathrm{V}_3}

\renewcommand{\Re}{{\rm Re}\,}
\renewcommand{\Im}{{\rm Im}\,}

\newcommand{\dcl}{\Delta}

\newcommand{\ah}{\epsilon}

\newcommand{\nn}{n} 
\newcommand{\nnd}{\nn_c^{(2)}}
\newcommand{\nndmax}{\nn_{c\,\mathrm{max}}^{(2)}}

\newcommand{\npeak}{n_{\mathrm{peak}}}
\newcommand{\nfree}{n_{\mathrm{free}}}

\makeatletter
\def\ignorecitefornumbering#1{%
     \begingroup
         \@fileswfalse
         #1
    \endgroup
}
\makeatother

\begin{document}

\title{Dirac spectrum in the chirally symmetric phase of a gauge
  theory. II}

\author{Matteo Giordano} \email{giordano@bodri.elte.hu}
\affiliation{Institute of Physics and Astronomy, ELTE E\"otv\"os
  Lor\'and University, P\'azm\'any P\'eter s\'et\'any 1/A, H-1117,
  Budapest, Hungary}

\date{\today}

\begin{abstract}
  I discuss the consequences of the constraints imposed on the Dirac
  spectrum by the restoration of chiral symmetry in the chiral limit
  of gauge theories with two light fermion flavors, with particular
  attention to the fate of the anomalous $\mathrm{U}(1)_A$ symmetry.
  Under general, physically motivated assumptions on the spectral
  density and on the two-point eigenvalue correlation function, I show
  that effective $\mathrm{U}(1)_A$ breaking in the symmetric phase
  requires specific spectral features, including a spectral density
  effectively behaving as $m^2\delta(\lambda)$ in the chiral limit, a
  two-point function singular at zero, and delocalized near-zero
  modes, besides an instanton gas-like behavior of the topological
  charge distribution.  I then discuss a $\mathrm{U}(1)_A$-breaking
  scenario characterized by a power-law divergent spectral peak
  tending to $O(m^4)/|\lambda|$ in the chiral limit and by a near-zero
  mobility edge, and argue that the mixing of the approximate zero
  modes associated with a dilute gas of topological objects provides a
  concrete physical mechanism producing the required spectral
  features, and so a viable mechanism for effective $\mathrm{U}(1)_A$
  breaking in the symmetric phase of a gauge theory.
\end{abstract}

\maketitle

\section{Introduction}
\label{sec:intro2}

A considerable amount of work has been done in recent years to
elucidate the nature of the finite-temperature transition in
QCD~\cite{Aoki:2006we,Borsanyi:2010bp,Bazavov:2011nk,
  Bhattacharya:2014ara,Bazavov:2016uvm}, the mechanisms behind it, and
the properties of the high-temperature phase. The most interesting
developments are related, to different extents, either to the
approximate symmetries of this phase, or to the spectrum and
eigenvectors of the Dirac operator, or to both. The existence of an
intermediate confining phase with an approximately restored, enlarged
chiral symmetry, between the low-temperature confined and chirally
broken phase and the high-temperature deconfined and chirally restored
phase, has been proposed and investigated
numerically~\cite{Glozman:2014mka,Glozman:2015qva,Rohrhofer:2019qwq,
  Glozman:2019fku,Glozman:2022lda,Glozman:2022zpy,Philipsen:2022wjj,
  Cohen:2023hbq,Chiu:2024jyz,Chiu:2024bqx,Aoki:2025mue} (see also
Ref.~\cite{Fujimoto:2025sxx}). A delayed deconfinement transition,
taking place at a much higher temperature than the usual QCD
crossover, has been suggested also in Ref.~\cite{Cardinali:2021mfh},
based on the observed persistence of monopole condensation, and in
Ref.~\cite{Mickley:2024vkm}, based on the observed persistence of
center-vortex percolation. An intermediate phase with a gas of
instanton-dyons was proposed in Ref.~\cite{Shuryak:2017fkh}.
References~\cite{Alexandru:2015fxa,Alexandru:2019gdm,Alexandru:2021pap,
  Alexandru:2021xoi,Alexandru:2023xho,Meng:2023nxf} proposed that a
true phase transition to an ``IR phase'', characterized by
scale-invariant features manifesting in the low-lying Dirac spectrum,
takes place at some temperature above the crossover. The fate of the
anomalous $\mathrm{U}(1)_A$ symmetry at and above the crossover
temperature has received much attention, with studies both at physical
and lower-than-physical quark masses~\cite{Chandrasekharan:1998yx,
  HotQCD:2012vvd,Cossu:2013uua,Buchoff:2013nra,Bhattacharya:2014ara,
  Dick:2015twa,Brandt:2016daq,Tomiya:2016jwr,HotQCD:2019xnw,
  Aoki:2020noz,Ding:2020xlj,Kaczmarek:2021ser,Kaczmarek:2023bxb,
  JLQCD:2024xey,Gavai:2024mcj,Fodor:2025yuj}, often involving a
detailed study of the low Dirac modes. These modes, and in particular
their localization properties, play an important role in attempts at
understanding quark deconfinement and its relation to chiral symmetry
restoration~\cite{Gockeler:2001hr,
  Gattringer:2001ia,GarciaGarcia:2005vj,GarciaGarcia:2006gr,
  Gavai:2008xe,Kovacs:2009zj,Kovacs:2010wx,Bruckmann:2011cc,Kovacs:2012zq,
  Giordano:2013taa,Nishigaki:2013uya,Ujfalusi:2015nha,
  Giordano:2015vla,Giordano:2016cjs,Giordano:2016vhx,
  Cossu:2016scb,Giordano:2016nuu,Kovacs:2017uiz,Holicki:2018sms,
  Giordano:2019pvc,Vig:2020pgq,Bonati:2020lal,Baranka:2021san,
  Kovacs:2021fwq,Cardinali:2021fpu,Baranka:2022dib,Kehr:2023wrs,
  Baranka:2023ani,Bonanno:2023mzj,Baranka:2024cuf,Bonanno:2025xuo}
(see Ref.~\cite{Giordano:2021qav} for a review).

Characterizing the Dirac spectrum of QCD in the chiral limit of
massless quarks in the chirally symmetric phase provides useful
insight into the fate of $\mathrm{U}(1)_A$ in this
phase~\cite{Cohen:1997hz,Aoki:2012yj,Kanazawa:2015xna,
  Azcoiti:2023xvu,Kanazawa:2014cua,Carabba:2021xmc}, and could be of
help in understanding the other issues pointed out above. Following
this strategy, Ref.~\cite{Aoki:2012yj} concluded that in the
two-flavor chiral limit, chiral symmetry restoration necessarily
leads to effective $\mathrm{U}(1)_A$ restoration in the correlation
functions of scalar and pseudoscalar flavor-singlet and
flavor-triplet fermion bilinears, in the sense that symmetry-breaking
effects are invisible in these correlation functions. With a simpler
analysis, Ref.~\cite{Kanazawa:2015xna} reached the same conclusion for
the simplest $\mathrm{U}(1)_A$ order parameter.  This partially
supported previous claims made in
Refs.~\cite{Cohen:1996ng,Cohen:1997hz} of $\mathrm{U}(1)_A$
restoration being necessary in the
$\mathrm{SU}(2)_L\times \mathrm{SU}(2)_R$ symmetric phase. This,
however, disagrees with the conclusion of
Refs.~\cite{Evans:1996wf,Lee:1996zy,Carabba:2021xmc} that
$\mathrm{U}(1)_A$ remains instead effectively broken in the symmetric
phase, i.e., $\mathrm{U}(1)_A$-breaking effects remain visible. The
main difference between the two approaches is that
Refs.~\cite{Cohen:1997hz,Aoki:2012yj,Kanazawa:2015xna} assume certain
analyticity properties of mass-independent observables of the theory
that depend only on gauge fields as functions of the squared
light-fermion mass, $m^2$, on top of certain technical assumptions on
the spectral density, $\rho(\lambda;m)$, as a function of the Dirac
eigenvalue, $\lambda$; while Refs.~\cite{Evans:1996wf,Lee:1996zy,
  Carabba:2021xmc} assume commutativity of the thermodynamic and
chiral limits. Both assumptions are well motivated in the chirally
symmetric phase, and their leading to opposite conclusions is rather
puzzling.  Moreover, making both the $m^2$-analyticity and the
commutativity assumption at the same time, Ref.~\cite{Azcoiti:2023xvu}
concluded that effective $\mathrm{U}(1)_A$ breaking in the symmetric
phase requires that the spectral density develops a term proportional
to $m^2 \delta(\lambda)$ at nonzero fermion mass.  As this is a
physically rather unlikely scenario, this result makes the situation
even more puzzling.

To shed some light on the issue of which assumptions must be kept and
which ones can be dropped, in the first paper of this
series~\cite{GM_DS1} I have revisited the foundations of the approach
to the problem of chiral symmetry restoration and the fate of
$\mathrm{U}(1)_A$ based on the Dirac
spectrum~\cite{Cohen:1997hz,Aoki:2012yj,Kanazawa:2015xna,
  Azcoiti:2023xvu,Kanazawa:2014cua,Carabba:2021xmc}, extending the
study of Refs.~\cite{Giordano:2024jnc,Giordano:2024awb}.  (In the
following this paper is referred to as \DSI, and all cross-references
to parts of that paper are preceeded by \mbox{\DSI-.}) The conclusion
is that chiral symmetry is restored in the scalar and pseudoscalar
sector, i.e., at the level of the susceptibilities of scalar and
pseudoscalar fermion bilinears, if and only if all these
susceptibilities are finite in the chiral limit (meaning
``non-divergent'' in this context, here and in the rest of this
paper). As a consequence, scalar and pseudoscalar susceptibilities are
$m^2$-differentiable (i.e., infinitely differentiable in $m^2$ at
zero), or $m$ times an $m^2$-differentiable function, depending on
whether they contain an even or odd number of isosinglet
bilinears. Under extended assumptions on chiral symmetry restoration,
requiring its realization also in susceptibilities involving nonlocal
gauge operators or external (partially quenched) fermion fields (see
Sec.~\exref{I-sec:chisymrest} and Appendix~\exref{I-sec:rhochirest}),
$m^2$-differentiability turns out to be a necessary property also of
the spectral density and other spectral quantities. Although obviously
different from the mathematical point of view, in the present context
$m^2$-differentiability is practically equivalent to $m^2$-analyticity
for most purposes, and so it essentially justifies the
$m^2$-analyticity assumptions of Refs.~\cite{Cohen:1997hz,Aoki:2012yj,
  Kanazawa:2015xna, Azcoiti:2023xvu,Kanazawa:2014cua}. On the other
hand, commutativity of the two relevant limits, as far as I know,
remains an unproven assumption.

Since scalar and pseudoscalar susceptibilities can be expressed solely
in terms of Dirac eigenvalues, the necessary and sufficient conditions
for chiral symmetry restoration discussed above imply a set of
constraints for the Dirac spectrum, involving the spectral density and
other spectral quantities. However, these constraints are in integral
form, and in order to exploit them to obtain detailed information on
the spectrum, one needs to make further technical assumptions on it.
As these assumptions are typically motivated by the available, and
usually limited, numerical results rather than justified from first
principles, they should be kept as general as possible, while still
allowing one to make progress.

The purpose of this paper is to extract information on the spectrum
under a more general set of assumptions than those employed in
Refs.~\cite{Cohen:1997hz,Aoki:2012yj,Kanazawa:2015xna}, extending also
those of Refs.~\cite{Giordano:2024jnc,Giordano:2024awb}, with
particular focus on the consequences for $\mathrm{U}(1)_A$ symmetry
after the constraints from chiral symmetry restoration are
imposed. Note that the approach of {\DSI} allows one to study not only
the spectral density, as in Refs.~\cite{Aoki:2012yj,Kanazawa:2015xna},
but eigenvalue correlations as well. Moreover, a careful scrutiny of
the conclusions of Ref.~\cite{Azcoiti:2023xvu} concerning the
implications of commutativity of the thermodynamic and chiral limits
is in order.

The main outcome of this study is that there is a relatively simple,
but at the same time highly constrained scenario in which
$\mathrm{U}(1)_A$ remains effectively broken in the symmetric phase,
even under the more restrictive, extended symmetry-restoration
assumptions leading to $m^2$-differentiability of spectral quantities,
already at the level of the simplest order parameter
$\Delta=\lim_{m\to 0}(\chi_\pi-\chi_\delta)/4$, where $\chi_\pi$ and
$\chi_\delta$ are the usual pion and delta susceptibilities. In this
scenario, a singular near-zero power-law peak forms in the spectral
density, with exponent tending to $-1$ in the chiral limit and a
prefactor suppressed at least like $m^4$. Moreover, the number of
modes within the peak per unit four-volume matches the topological
susceptibility, $\chit$, showing a close connection with the
topological features of gauge configurations. The singular peak
behaves then effectively as a term $\Delta m^2\delta(\lambda)$, with
$\chit = \Delta m^2 +O(m^4)$ by virtue of chiral symmetry
restoration. Since $\Delta\neq 0$ in the symmetric phase requires that
the topological charge be distributed in the chiral limit as in an
ideal instanton gas of density $\chit \propto m^2$ (see
Ref.~\cite{Kanazawa:2014cua} and Sec.~\exref{I-sec:top_iig}), the
density of peak modes matches that of (effective) topological
objects. Finally, the two-point eigenvalue correlation function is
singular at the origin, due to the appearance of a mobility edge close
to $\lambda=0$ that separates near-zero delocalized modes from the
well-known localized modes higher up in the low-lying
spectrum~\cite{Giordano:2021qav}. In such a scenario, all the
constraints imposed by chiral symmetry restoration on the spectral
density and the two-point function are satisfied, while
$\mathrm{U}(1)_A$ is effectively broken by $\Delta\neq 0$.

This scenario for effective $\mathrm{U}(1)_A$ breaking is not in
contradiction with the results of
Refs.~\cite{Aoki:2012yj,Kanazawa:2015xna}, as a singular peak was
simply not considered there, since it does not satisfy their more
restrictive assumptions on the spectrum.  Surprisingly,
$\mathrm{U}(1)_A$ breaking by a singular peak of the type described
above is also compatible with commutativity of the thermodynamic and
chiral limits, contrary to what one would expect from the results of
Ref.~\cite{Azcoiti:2023xvu} (and contrary to what was previously
claimed in Refs.~\cite{Giordano:2024jnc,Giordano:2024awb}). The
apparent contradiction, however, is resolved by noting that the
conditions on the spectral density derived in
Ref.~\cite{Azcoiti:2023xvu} from the requirements of
$m^2$-differentiability of the free energy density and commutativity
of limits do not, in fact, single out a delta-like behavior, and
allow for other solutions. Note that while compatible with
commutativity of limits, the singular peak by no means requires it in
order to comply with chiral symmetry restoration.

The scenario outlined above is motivated by, and in agreement with
numerical results supporting the existence of a near-zero peak in the
spectral density at nonzero fermion mass~\cite{Edwards:1999zm,
  HotQCD:2012vvd,Cossu:2013uua,Buchoff:2013nra,Dick:2015twa,
  Alexandru:2015fxa,Tomiya:2016jwr,Kovacs:2017uiz,Aoki:2020noz,
  Alexandru:2019gdm,Ding:2020xlj,Kaczmarek:2021ser,Vig:2021oyt,
  Kovacs:2021fwq,Meng:2023nxf,Kaczmarek:2023bxb,Alexandru:2024tel,
  JLQCD:2024xey,Fodor:2025yuj}, likely of topological
origin~\cite{Edwards:1999zm,HotQCD:2012vvd,Buchoff:2013nra,
  Dick:2015twa,Kovacs:2017uiz,Ding:2020xlj,Vig:2021oyt,
  Kaczmarek:2021ser,Kaczmarek:2023bxb,Kovacs:2023vzi}, although a
complete characterization of the peak is still lacking, and its
behavior in the chiral limit is still unclear. Moreover, numerical
results indicate delocalization of near-zero modes and support the
existence of a near-zero mobility edge at nonzero fermion
mass~\cite{Meng:2023nxf}, although no information is currently
available concerning its dependence on the mass. Finally, the simple
model of Ref.~\cite{Kovacs:2023vzi} for the near-zero Dirac spectrum,
described as originating from the zero modes associated with the
constituents of a dilute instanton gas, provides a physical mechanism
realizing the $\mathrm{U}(1)_A$ breaking scenario discussed above, and
supports its physical viability.

From the mathematical point of view, under specific assumptions on the
functional form of the spectral density, the singular peak discussed
above is the unique possibility leading to $\mathrm{U}(1)_A$ breaking
by $\Delta\neq 0$. However, although quite general and physically
motivated, these assumptions certainly do not cover all the physically
acceptable possibilities, and there is no reason to expect
uniqueness. Indeed, it is easy to include corrections to a pure
power-law behavior of the peak (that do not essentially change its
$1/|\lambda|$ behavior in the chiral limit) while preserving all the
properties discussed above. Nonetheless, if chiral symmetry is
restored in the extended sense, and more generally as long as the
spectral density is $m^2$-differentiable, then some sort of singular
behavior of the near-zero modes in the chiral limit is needed to
effectively break $\mathrm{U}(1)_A$ through $\Delta\neq 0$, even under
more general assumptions on the dependence of the spectral density on
$\lambda$.  This singular behavior need not be a divergent near-zero
peak at $m\neq 0$, but could be, e.g., a finite near-zero peak whose
height diverges in the chiral limit. This leaves open a larger set of
possibilities concerning the connection between near-zero modes and
topology, and the commutativity of the thermodynamic and chiral
limits. On the other hand, the scenario outlined above (including its
generalization to a non pure power-law behavior) is singled out and
remains favored on physical grounds, based on currently available
numerical results and especially on the existence of a simple,
concrete physical mechanism that could realize it.

The plan of this paper is the following. In Sec.~\ref{sec:constr_cons}
I briefly discuss the setup and summarize the relevant results of
\DSI.  In Sec.~\ref{sec:detailed_spdens} I study the consequences of
the constraints on the spectral density resulting from chiral symmetry
restoration, using first a definite but quite general
parameterization, both with and without imposing
$m^2$-differentiability, and later studying the $m^2$-differentiable
case under broader conditions.  In Sec.~\ref{sec:tpfunc} I study the
consequences of the constraints on the two-point correlation function
of non-zero eigenvalues, first assuming that it remains finite near
the origin, and later exploring the consequences of a localized
near-zero spectrum. Using the results of these investigations, in
Sec.~\ref{sec:u1breakingscenario} I discuss a physically viable
scenario for effective $\mathrm{U}(1)_A$ breaking compatible with
chiral symmetry restoration. In Sec.~\ref{sec:concl} I draw my
conclusions and show some prospects for future studies. Technical
details are discussed in Appendices \ref{sec:app_spdens_det},
\ref{sec:app_commlim}, and \ref{sec:tpf_app}.

\section{Finite-temperature gauge theories and Dirac spectrum of
  Ginsparg--Wilson fermions}
\label{sec:constr_cons}

In this section I summarize the setup and the main pieces of notation
defined in {\DSI}, to which I refer the reader for further details.  I
consider gauge theories at finite temperature on a hypercubic 3+1
dimensional lattice of temporal extension $1/\mathrm{T}$ and spatial
volume $\svol$, and total four-volume $\lvol=\svol/\mathrm{T}$.
Lattice units are used everywhere in this paper (unless explicitly
stated otherwise).  The gauge links, taking values in a generic
compact gauge group, are denoted collectively with $U$. Two flavors
of degenerate light fermions of mass $m$, that is eventually sent to
zero, and a number of massive fermions whose masses remain nonzero in
this limit, all transforming in (possibly different) irreducible
representations of the gauge group, are included. Periodic
(respectively antiperiodic) temporal boundary conditions are imposed
on gauge (respectively fermion) fields; periodic spatial boundary
conditions are imposed on all fields. Expectation values are denoted
with $\la \ldots\ra$.

The discretized gauge action and massive-fermions action need not be
specified, besides assuming that they respect the usual lattice
symmetries [translations, cubic rotations, reflections, and $CP$ --
see Eq.~\exeqref{I-eq:Trelf1}]. For the light fermions I use
Ginsparg--Wilson (GW) lattice discretizations $D(U)$ of the Dirac
operator~\cite{Ginsparg:1981bj,Kaplan:1992bt,Shamir:1993zy,
  Furman:1994ky,Borici:1999zw,Chiu:2002ir,Brower:2005qw,Brower:2012vk,
  Narayanan:1993sk,Narayanan:1993ss,Neuberger:1997fp,Neuberger:1998wv,
  Hasenfratz:1993sp,Bietenholz:1995cy,DeGrand:1995ji,Hasenfratz:1998ri,
  Hasenfratz:1998jp,Hasenfratz:2002rp,Chandrasekharan:1998wg,
  Niedermayer:1998bi,Luscher:1998pqa,Kikukawa:1998py,
  Horvath:1999bk,Hasenfratz:2002rp,Luscher:2004fu,Giusti:2004qd} that
are $\gamma_5$-Hermitean and obey the GW
relation~\cite{Ginsparg:1981bj,Hasenfratz:1998ri,Hasenfratz:1998jp,
  Luscher:1998pqa,Kikukawa:1998py,Neuberger:1998wv,Horvath:1999bk}
$\{D,\gamma_5\}=2DR\gamma_5D$ with $2R=\mathbf{1}$, and that
furthermore respect the usual lattice symmetries (e.g., domain
wall~\cite{Kaplan:1992bt,Shamir:1993zy,Furman:1994ky,
  Borici:1999zw,Chiu:2002ir,Brower:2005qw,Brower:2012vk} or overlap
fermions~\cite{Narayanan:1993sk,Narayanan:1993ss,
  Neuberger:1997fp,Neuberger:1998wv}). A doublet of massless GW
fermions possesses an exact $\mathrm{SU}(2)_L\times \mathrm{SU}(2)_R$
lattice chiral symmetry~\cite{Neuberger:1997fp,Neuberger:1998wv,
  Hasenfratz:1998ri,Luscher:1998pqa,Kikukawa:1998py,Horvath:1999bk},
possibly spontaneously broken, as well as an anomalous
$\mathrm{U}(1)_A$
symmetry~\cite{Ginsparg:1981bj,Hasenfratz:1998ri,Luscher:1998pqa}.  It
is assumed that the gauge group, fermion content, and temperature of
the system are such that the $\mathrm{SU}(2)_L\times \mathrm{SU}(2)_R$
symmetry is realized in the chiral limit $m\to 0$.

For a $\gamma_5$-Hermitean GW operator $D$ with $2R=\mathbf{1}$, the
corresponding eigenvalues, $\mu_n$, lie on the unit circle centered at
1, i.e., $2\Re\mu_n = |\mu_n |^2$. Here and in most of the following
their dependence on the gauge configuration $U$ is dropped for
notational simplicity. The eigenvalues of $D$ either form conjugate
pairs $\mu_n,\mu_n^*$ of complex eigenvalues ($\mu_n\neq\mu_n^*$), or
are real eigenvalues $\mu_n=0$ or $\mu_n=2$.  These zero and doubler
modes are chosen to be chiral, i.e., eigenvectors of $\gamma_5$. I
denote with $\lambda_n = |\mu_n|\,\mathrm{sgn}(\Im\mu_n)$ the signed
magnitude of the complex eigenvalues, $|\lambda_n|\in (0,2)$, and with
$N_\pm$ the number of exact chiral zero modes of chirality $\pm 1$ in
a gauge configuration. The total number of zero modes is denoted with
$N_0=N_++N_-$, and the topological charge~\cite{Hasenfratz:1998ri}
with $Q=N_+-N_-$. Finally, I assume that the gauge group and the
gauge-group representation under which the light fermions transform do
not lead to Kramers degeneracy of the spectrum (see
Ref.~\cite{Verbaarschot:2000dy}, and the discussion at the end of
Sec.~\exref{I-sec:comp_fdet}).

The main quantities of interest in this paper are the spectral
density,
\begin{equation}
  \label{eq:spqrecap1}
  \begin{aligned}
    \rho(\lambda;m) &= \lim_{\lvol\to\infty}   \f{\la \rho_U(\lambda) \ra}{\lvol}\,,\\
    \rho_U(\lambda)  &= \sum_n \delta \left(\lambda - \lambda_n(U)\right) \,,
  \end{aligned}
\end{equation}
where the sum runs over all the complex eigenvalues
$\mu_n\neq \mu_n^*$ of $D$, so $\rho_U(-\lambda)=\rho_U(\lambda)$; and
the connected two-point eigenvalue correlation function,
\begin{equation}
  \label{eq:spqrecap2}
    \begin{aligned}
      &  \rho_{c}^{(2)}(\lambda,\lambda';m) \\
      &  =\lim_{\lvol\to\infty}
        \f{\la \rho_U(\lambda)\rho_U(\lambda')\ra -
        \la\rho_U(\lambda)\ra \la\rho_U(\lambda')\ra}{\lvol}\\
      &\phantom{=}  -  \left[\delta(\lambda-\lambda')+
        \delta(\lambda+\lambda')\right]\rho(\lambda;m)\,,
  \end{aligned}
\end{equation}
see Eqs.~\exeqref{I-eq:rho_def3_th} and
\exeqref{I-eq:rho_def7}.\footnote{\label{foot:infty}Since only the
  infinite-volume case is considered in this paper, the subscript
  $\infty$ used in {\DSI} to denote the thermodynamic limit is dropped
  from the notation of spectral quantities and cumulants.} These
definitions are formal, and should be more precisely understood in the
distributional sense. For the spectral density, one starts from the
normalized mode number of a finite spectral interval, $\nn(\delta;m)$,
\begin{equation}
  \label{eq:rho_distr_def}
  \begin{aligned}
  \nn(\delta;m)  &\equiv \lim_{\lvol\to\infty}
\f{\left\la \nn_U(\delta)\right\ra}{\lvol}      
  \equiv \int_0^\delta d\lambda\, \rho(\lambda;m)\,,\\
    \nn_U(\delta) &\equiv \int_0^\delta  d\lambda\,\rho_U(\lambda)\,,
  \end{aligned}
\end{equation}
with $\delta\in [-2,2]$. The last passage on the first line defines
the spectral density as $\rho(\lambda;m)=\de_\lambda \nn(\lambda;m)$,
where $\de_x\equiv \de/\de_x$, and the derivative is understood in the
sense of distributions. The spectral density at $0$ and $\pm 2$ is
defined by continuity, i.e., $\rho(0;m)\equiv\rho(0^+;m)$ and
$\rho(\pm 2;m)\equiv \rho(\pm 2^\mp;m)$. By symmetry of the spectrum
one has $\nn(-\lambda;m)=-\nn(\lambda;m)$ and
$\rho(\lambda;m)=\rho(-\lambda;m)$. One proceeds similarly for the
two-point correlation function (see Sec.~\ref{sec:tpfunc}).

To obtain detailed constraints on the spectral density and the
two-point function, and possibly gain insight into the fate of
$\mathrm{U}(1)_A$ symmetry, the relevant constraints [see
Eqs.~\exeqref{I-eq:firstorder_again},
\exeqref{I-eq:firstorder_again1_2}, and \exeqref{I-eq:secondorder1_1}]
need to be supplemented with further, technical assumptions on the
properties of the spectrum. The assumptions made below are motivated
by the results of numerical studies of the spectral
density~\cite{Edwards:1999zm,HotQCD:2012vvd,Cossu:2013uua,Buchoff:2013nra,
  Dick:2015twa,Alexandru:2015fxa,Tomiya:2016jwr,Kovacs:2017uiz,
  Aoki:2020noz,Alexandru:2019gdm,Ding:2020xlj,Kaczmarek:2021ser,
  Vig:2021oyt,Kovacs:2021fwq,Meng:2023nxf,Kaczmarek:2023bxb,
  Alexandru:2024tel,JLQCD:2024xey} and of the localization properties
of low-lying Dirac modes~\cite{Gockeler:2001hr,Gattringer:2001ia,
  GarciaGarcia:2005vj,GarciaGarcia:2006gr,Gavai:2008xe,Kovacs:2009zj,
  Kovacs:2010wx,Kovacs:2012zq,Giordano:2013taa,Nishigaki:2013uya,
  Ujfalusi:2015nha,Cossu:2016scb,Giordano:2016nuu,Kovacs:2017uiz,
  Holicki:2018sms,Giordano:2019pvc,Vig:2020pgq,Bonati:2020lal,
  Baranka:2021san,Kovacs:2021fwq,Cardinali:2021fpu,Baranka:2022dib,
  Kehr:2023wrs,Baranka:2023ani,Bonanno:2023mzj,Baranka:2024cuf,
  Bonanno:2025xuo,Giordano:2021qav} in lattice QCD and other lattice gauge
theories. It goes without saying that they should nevertheless be
directly verified, and may have to be updated in the future.

Besides detailed assumptions on the dependence of spectral quantities
on the position in the spectrum and on $m$, in the following I will
always assume that the index theorem is realized in a minimal way, as
argued in Ref.~\cite{Blum:2001qg}, and so $N_\pm$ obey $N_+N_-=0$
almost everywhere in configuration space. This implies that the
density of zero modes vanishes in the thermodynamic limit,
$n_0 =\lim_{\lvol\to\infty}\f{\la N_0\ra}{\lvol}=0$. For the
second-order cumulant of $N_0$, $b_{N_0^2}$ (see footnote
\ref{foot:infty}), and the topological susceptibility, $\chit$,
\begin{equation}
  \label{eq:n0_cum2def_chit}
  b_{N_0^2} =\lim_{\lvol\to\infty}\f{\la N_0^2\ra -\la N_0\ra^2}{\lvol}\,,  \qquad
  \chit =\lim_{\lvol\to\infty}\f{\la Q^2\ra}{\lvol}\,,
\end{equation}
this implies
\begin{equation}
  \label{eq:n0_cum}
  b_{N_0^2}-\chit= -\lim_{\lvol\to\infty} \f{\la N_0\ra^2}{\lvol} \,.
\end{equation}
This simplifies the form of the constraints [see
Eqs.~\exeqref{I-eq:firstorder_again2_bis} and
\exeqref{I-eq:secondorder4}].

\section{Spectral density}
\label{sec:detailed_spdens}

In this section I discuss the consequences of chiral symmetry
restoration for the spectral density, Eqs.~\eqref{eq:spqrecap1} and
\eqref{eq:rho_distr_def}, after making additional technical
assumptions.

Generally, $\rho(\lambda;m)$ is expected to exist in the thermodynamic
limit, but does not need to be an ordinary function and may contain
Dirac deltas~\cite{van_Hemmen_1982}. However, there seems to be no
particular physical reason for any delta singularity to appear
anywhere, at least for $m\neq 0$, and for any divergence to be present
at $\lambda\neq 0$. On the other hand, $\lambda=0$ is special, as it
is singled out as a symmetry point in the spectrum by the exact chiral
symmetry of GW fermions.\footnote{\label{foot:singedge}Another
  exception may be the edge of the spectrum, $\lambda=\pm 2$, but
  singularities there, if present, would play no role in the chiral
  limit and could be ignored.} Indeed, a divergence in the spectral
density at $\lambda=0$ is known to appear in certain systems with
chiral symmetry (see, e.g., Refs.~\cite{Ziman_1982,S_N_Evangelou_2003}
and references therein). Nonetheless, the typical repulsion between
eigenvalues should prevent an accumulation of near-zero (but nonzero)
modes leading to a term $\delta(\lambda)$ in the thermodynamic limit,
even taking the chiral limit after that. Note that these
considerations apply independently of the fate of chiral symmetry as
$m\to 0$.  I will then assume that, for any $m\neq 0$,
$\rho(\lambda;m)$ is an ordinary function, finite at $\lambda\neq 0$;
and that its chiral limit, denoted for simplicity as
$\rho(\lambda;0)\equiv\lim_{m\to
  0}\rho(\lambda;m)$,\footnote{\label{foot:rhomz}This generally
  differs from the spectral density of the massless theory, obtained
  by setting $m=0$ before taking the thermodynamic limit. On the other
  hand, since $\rho(\lambda;m)=\rho(\lambda;-m)$ thanks to the
  non-anomalous $\mathbb{Z}_{2A}$ symmetry of the massless theory (see
  footnote~\exref{I-foot:z2a}), one has
  $\lim_{m\to 0^+}\rho(\lambda;m)=\lim_{m\to 0^-}\rho(\lambda;m)$,
  even in the chirally broken phase.} is also an ordinary function,
finite at $\lambda\neq 0$. Here and everywhere else, unless explicitly
stated, the chiral limit is taken after the thermodynamic limit.

If chiral symmetry is manifest also in susceptibilities involving
nonlocal functionals of gauge fields (``nonlocal restoration'') or in
the presence of external (partially quenched) fermion fields, it was
shown in \DSI\ (see discussion in Secs.~\exref{I-sec:chisymrest} and
\exref{I-sec:gauge_ext}, and Appendix~\exref{I-sec:rhochirest}) that
$\rho(\lambda;m)$ is $m^2$-differentiable, i.e., it is a $C^\infty$
function of $m^2$ at $m=0$. When making use of
$m^2$-differentiability, which will be always explicitly specified, I
will assume that the $m^2$-derivatives of $\rho$ at $m\neq 0$ are
ordinary functions of $\lambda$, while the presence of integrable
singularities, and in this case also Dirac deltas, at $\lambda=0$ in
the chiral limit will be allowed, in line with the special status of
this point.

The first requirement from chiral symmetry restoration in the scalar
and pseudoscalar sector that affects the spectral density directly is
the finiteness in the chiral limit of the pion susceptibility,
$\chi_\pi$ [Eq.~\exeqref{I-eq:coeff_rel1}]. This requirement leads to
the following constraint on the spectral density in the symmetric
phase [see Eq.~\exeqref{I-eq:firstorder_again}],
\begin{equation}
  \label{eq:firstorder_again_again}
  \lim_{m\to 0}\f{\chi_\pi}{4} =    \lim_{m\to 0}    \Iu[\ff] <\infty\,,
\end{equation}
where\footnote{\label{foot:limex}In the spectral representation of
  $\chi_\pi$ and of the other relevant observables, the thermodynamic
  limit and integration over $\lambda$ are expected to
  commute~\cite{van_Hemmen_1982}.}
\begin{equation}
  \label{eq:firstorder_again_again2}
  \begin{aligned}
    \Iu[g]&=  \int_0^2 d\lambda\, g(\lambda) \rho(\lambda;m)\,,
    &&& &\\
    \ff(\lambda;m) &= \f{h(\lambda)}{\lambda^2 + m^2h(\lambda)}\,,
    &&&   h(\lambda) &=1-\f{\lambda^2}{4}\,.
  \end{aligned}
\end{equation}
The requirement Eq.~\eqref{eq:firstorder_again_again} is equivalent to
\begin{equation}
  \label{eq:int_new20_bis}
     \lim_{m\to 0}  \II_0(\delta;m)<\infty\,,
\end{equation}
where
\begin{equation}
  \label{eq:int_new20_def}
  \II_n(\delta;m)\equiv  \int_0^\delta d\lambda\,
  \f{m^{2n}\rho(\lambda;m)}{(\lambda^2 + m^2)^{n+1}} \,,
\end{equation}
with $0<\delta\le 2$ an arbitrary $m$-independent cutoff. The
remaining contributions to $ \Iu[\ff]$ are automatically finite, since
the normalized mode number, Eq.~\eqref{eq:rho_distr_def}, is finite
for an arbitrary spectral interval (see
Appendix~\ref{sec:coeff_contr}).

The other direct constraint on the spectral density originates in the
finiteness of
$\chi_{\pi\delta}=\f{\chi_\pi-\chi_\delta}{m^2}-\f{4\chit}{m^4}$ [see
Eq.~\exeqref{I-eq:coeff_real1_quater_bis}], that requires [see
Eqs.~\exeqref{I-eq:firstorder_again1} and
\exeqref{I-eq:firstorder_again1_2}]
\begin{equation}
  \label{eq:firstorder_again1_2_again}
  \f{\chi_\pi -\chi_\delta}{4} = 2m^2\Iu[f^2] = \f{\chit}{m^2} +  O(m^2)\,,
\end{equation}
where $\chi_\delta$ is the usual delta susceptibility
[Eq.~\exeqref{I-eq:coeff_rel1}]. This implies for the order parameter
$\Delta$
\begin{equation}
  \label{eq:firstorder_again2_again}
  \dcl =  \lim_{m\to 0}  \f{\chi_\pi -\chi_\delta}{4} 
  =\lim_{m\to 0} 2m^2\Iu[f^2] = \lim_{m\to 0}\f{\chit}{m^2} \,.
\end{equation}
Since finiteness in the chiral limit of $\chi_\pi$ implies finiteness
of $\chi_\delta$ and so of $\Delta$ (see
Sec.~\exref{I-sec:constr_first}), this requires in particular that
$\chit=O(m^2)$. Using again finiteness of the normalized mode number
in any spectral interval and the required finiteness of $\Iu[\ff]$,
one shows that Eq.~\eqref{eq:firstorder_again2_again} reduces to (see
Appendix~\ref{sec:coeff_contr})
\begin{equation}
  \label{eq:int_new20_bis_delta}
  \dcl = 2\lim_{m\to 0} \II_1(\delta;m) \,,
\end{equation}
with $\II_1$ defined in Eq.~\eqref{eq:int_new20_def}, and all the
other contributions to $2m^2\Iu[\ff^2]$ vanishing in the chiral
limit. Since $\II_1\le \II_0$, finiteness of $\II_1$ in the chiral
limit is guaranteed by that of $\II_0$. Note that
Eq.~\eqref{eq:firstorder_again1_2_again} requires
$2\II_1(\delta;m) - \Delta = O(m^2)$, since $\f{\chit}{m^2}$ must be
$m^2$-differentiable (see Sec.~\exref{I-sec:constr_first}),
$\f{\chit}{m^2} = \Delta + O(m^2)$, and since
$m^2\Iu[\ff^2] -\II_1(\delta;m) = O(m^2)$ (see
Appendix~\ref{sec:coeff_contr}). Finally, since
Eq.~\eqref{eq:int_new20_bis_delta} holds for any $\delta$ one has also
\begin{equation}
  \label{eq:int_new20_bis_delta3}
  \dcl  =   2\lim_{\epsilon\to 0^+} \lim_{m\to 0}  \II_1(\epsilon;m)\,.
\end{equation}
To work out the consequences of the constraints
Eqs.~\eqref{eq:firstorder_again_again} and
\eqref{eq:firstorder_again1_2_again} I will initially use an explicit
parameterization of the spectral density motivated by numerical
results (Sec.~\ref{sec:spdens_expl}), that gives one full analytic
control of the quantities of interest. This allows one to explore in
detail the restrictions imposed by the constraints
(Sec.~\ref{sec:chipi}), and the stronger restrictions imposed by the
request of $m^2$-differentiability
(Secs.~\ref{sec:mdiffspd}--\ref{sec:furtherrho}), or by commutativity
of the thermodynamic and chiral limits (Sec.~\ref{sec:ctc}). A more
general setup, independent of a specific functional form of the
spectral density as a function of $\lambda$, but requiring
$m^2$-differentiability as well as making some physically motivated
assumptions on the $\lambda$- and $m$-dependence, is discussed in
Sec.~\ref{sec:mdiffspdgen}.

\subsection{Explicit parameterization}
\label{sec:spdens_expl}

Motivated by the results of Refs.~\cite{Edwards:1999zm,
  HotQCD:2012vvd,Cossu:2013uua,Buchoff:2013nra,Dick:2015twa,
  Alexandru:2015fxa,Tomiya:2016jwr,Kovacs:2017uiz,Aoki:2020noz,
  Alexandru:2019gdm,Ding:2020xlj,Kaczmarek:2021ser,Vig:2021oyt,
  Kovacs:2021fwq,Meng:2023nxf,Kaczmarek:2023bxb,Alexandru:2024tel,
  JLQCD:2024xey,Fodor:2025yuj}, I assume that the spectral density
near zero is dominated by power-like contributions, possibly divergent
as $\lambda\to 0$. Specifically, I assume that the spectral density is
of the form
\begin{equation}
  \label{eq:spdens_gen1}
  \rho(\lambda;m) = \sum_{i=1}^s C_i(m)|\lambda|^{\alpha_i(m)} + \bar{\rho}(\lambda;m) \,, 
\end{equation}
with possibly $m$-dependent, continuous exponents $\alpha_i(m)$,
$-1\le \alpha_1(0)<\alpha_2(0)<\ldots <\alpha_s(0)\le 1$, and a
remainder $\bar{\rho}$, continuous in $m$, that obeys the bound
\begin{equation}
  \label{eq:barrhobound}
  |\bar{\rho}(\lambda;m)|\le A |\lambda|^{1+\zeta}\,,
\end{equation}
for some mass-independent $A,\zeta>0$, for $|\lambda|\in[0,2]$, i.e.,
$\bar{\rho}=O(|\lambda|^{1+\zeta})$ $\forall m$, including in the
limit $m\to 0$. In general, $\alpha_i(m)>-1$ is required at $m\neq 0$
to ensure integrability. Since $\rho(\lambda,0)$ is assumed to be an
ordinary function, the coefficients $C_i$ cannot diverge in the chiral
limit: If they did, divergences could not cancel out among the various
terms due to their different dependence on $\lambda$ near zero.

The functional form Eq.~\eqref{eq:spdens_gen1} is quite general, and
includes as special cases those considered in
Refs.~\cite{Aoki:2012yj,Kanazawa:2015xna,Giordano:2024jnc}, namely a
spectral density admitting an expansion in integer non-negative powers
near $\lambda=0$~\cite{Aoki:2012yj,Kanazawa:2015xna,Giordano:2024jnc},
possibly up to one additional power-law term $|\lambda|^\alpha$ with
non-integer $\alpha>0$~\cite{Aoki:2012yj};\footnote{The assumptions of
  Ref.~\cite{Aoki:2012yj} actually concern the near-zero behavior of
  the infinite-volume limit of the spectral density on a fixed
  configuration, assumed to exist.}  and a spectral density dominated
by a single, possibly singular and mass-dependent power-law
term~\cite{Giordano:2024jnc}. No assumption is made initially on the
$m^2$-differentiability of $\rho$. The restrictions imposed by
$m^2$-differentiability are discussed in
Secs.~\ref{sec:mdiffspd}--\ref{sec:furtherrho}.

\subsubsection{Constraints from finiteness of \texorpdfstring{$\chi_\pi$}{chi\textunderscore pi}}
\label{sec:chipi}

For a spectral density of the form Eq.~\eqref{eq:spdens_gen1},
finiteness of $\chi_\pi$ in the chiral limit requires that [see
Eq.~\eqref{eq:int_new20_bis}]
\begin{equation}
  \label{eq:int_new20_bis_delta2_0}
  \II_0(\delta;m) = \sum_{i=1}^s C_i(m)\mathcal{I}_{\alpha_i(m)}(\delta;m)+
  \int_0^\delta d\lambda\, \f{\bar{\rho}(\lambda;m)}{\lambda^2+ m^2 }
\end{equation}
remains finite as $m\to 0$. Here
\begin{equation}
  \label{eq:inte1_again}
  \mathcal{I}_{\alpha(m)}(\delta;m)\equiv \int_0^\delta d\lambda\, \f{\lambda^{\alpha(m)}}{\lambda^2+ m^2 }\,,
\end{equation}
for $\alpha(m)$ any continuous function of $m$ with $\alpha(m)>-1$ for
$m\neq 0$ and $ \alpha(0)\ge -1$. In the chiral limit,
$ \mathcal{I}_{\alpha(m)}$ remains finite if $\alpha(0)> 1$,
$\mathcal{I}_{\alpha(m)}(\delta;m)\to
\f{\delta^{\alpha(0)-1}}{\alpha(0)-1}$, and diverges if
$|\alpha(0)|\le 1$, with a $\delta$-independent behavior (see
Appendix~\ref{sec:inte}),
\begin{equation}
  \label{eq:int_new15}
  \mathcal{I}_{\alpha(m)}(\delta;m) \mathop{\sim}_{m\to 0}
  \left\{
    \begin{aligned}
      &  \f{ |m|^{\alpha(m)-1}}{1+ \alpha(m)}\,,
      &&& \alpha(0)&=-1\,,\\
      &  \f{|m|^{\alpha(m)-1}}{\f{2}{\pi}\cos\f{\pi \alpha(0)}{2}}\,,
      &&& |\alpha(0)|&<1 \,,\\
      &  \f{|m|^{\alpha(m)-1}-1}{1-\alpha(m)}\,,
      &&& \alpha(0)&=1\,. 
    \end{aligned}
  \right.
\end{equation}
While in principle the contributions $C_i \mathcal{I}_{\alpha_i}$ of
the individual power-law terms could separately diverge but add up to
a finite quantity, one can show that this possibility is excluded due
to the positivity of the spectral density, and each contribution must
be separately finite (see Appendix~\ref{sec:finX0}). The contribution
of the remainder is finite due to the assumed bound on $\bar{\rho}$,
Eq.~\eqref{eq:barrhobound}. One has then
$C_i \mathcal{I}_{\alpha_i}=O(1)$, and so
\begin{equation}
  \label{eq:X0fin1_0}
  C_i = O(1/\mathcal{I}_{\alpha_i})=o(1)\,, \quad 1\le i\le s\,,
\end{equation}
and $\rho(\lambda;0)=\bar{\rho}(\lambda;0)=O(|\lambda|^{1+\zeta})$
with $\zeta>0$.

For the $\mathrm{U}(1)_A$ order parameter $\Delta$ one finds [see
Eqs.~\eqref{eq:int_new20_bis_delta} and 
\eqref{eq:u1abreak}]
\begin{equation}
  \label{eq:int_new20_bis_delta2bis_0}
  \Delta = \sum_{i=1}^s \left[1-\alpha_i(0)\right]
  \left[\lim_{m\to 0} C_i(m)\mathcal{I}_{\alpha_i(m)}(\delta;m)\right]\,.
\end{equation}
Since the term in brackets is finite for every $i$ separately and need
not vanish, under the current assumptions one can have $\Delta\neq 0$
without contradicting finiteness of $\chi_\pi$ in the chiral
limit. This requires that at least one of the coefficients obeys
$C_i\propto |m|^{1-\alpha_i(m)}$ if $|\alpha_i(0)|<1$, or
$C_1\propto |m|^{1-\alpha_1(m)}(1+\alpha_1(m))$ if $\alpha_1(0)= -1$,
to leading order in $m$. The result
Eq.~\eqref{eq:int_new20_bis_delta2bis_0} is independent of $\delta$,
as it should be [see Eq.~\eqref{eq:int_new20_bis_delta3}], since a
nonzero contribution to $\Delta$ is possible only if the leading-order
term in $C_i$ exactly compensates the $\delta$-independent divergence
of $\mathcal{I}_{\alpha_i(m)}(\delta;m)$. Notice that if
$\alpha_s(0)=1$ the corresponding term does not contribute to
$\Delta$. Finally, note that $\Delta=0$ requires that the term in
square brackets in Eq.~\eqref{eq:int_new20_bis_delta2bis_0} vanishes
in the chiral limit for every $i$, except possibly for $i=s$ if
$\alpha_s(0)=1$, again due to positivity of the spectral density (see
Appendix~\ref{sec:finX0}).

If $\rho(\lambda;m)$ admits a convergent expansion in powers of
$|\lambda|$ around $\lambda=0$, within a mass-independent finite
radius $\delta_\rho$ (as assumed in
Refs.~\cite{Aoki:2012yj,Kanazawa:2015xna}),
  \begin{equation}
  \label{eq:app_sppow1}
  \rho_{\mathrm{series}}(\lambda;m) \equiv \sum_{n=0}^\infty\rho_n(m)|\lambda|^n\,,
  \qquad 0\le |\lambda| < \delta_\rho\,,
\end{equation}
choosing $\delta<\delta_\rho$ one finds from Eqs.~\eqref{eq:int_new15}
and \eqref{eq:X0fin1_0} [see also Eq.~\eqref{eq:inteI_integer}] that
finiteness of $\chi_\pi$ requires
\begin{equation}
  \label{eq:app_sppow_cons}
  \rho_0(m)=O(|m|)\,, \qquad \rho_1(m)=O\left(1/\ln\tf{1}{|m|}\right)\,,
\end{equation}
while $\rho_{n}(m)=O(m^0)$ for $n\ge 2$, as they contribute to
$\bar{\rho}$, for which the bound Eq.~\eqref{eq:barrhobound} holds
with $\zeta=1$. Note that finiteness of $\rho_n$ in the chiral limit
is guaranteed by the same argument used for $C_i$, see under
Eq.~\eqref{eq:barrhobound}. For the $\mathrm{U}(1)_A$ order parameter
one finds
\begin{equation}
  \label{eq:Delta_series}
  \Delta = \f{\pi}{2}\lim_{m\to 0} \f{\rho_0(m)}{|m|}\,,
\end{equation}
so $\Delta\neq 0$ only if $\lim_{m\to 0}\rho_0(m)/|m| >0$ is strictly
nonzero. The conclusions above do not change if $\rho$ additionally
contains one or more terms $\sim |\lambda|^{\alpha_i(m)}$ with
noninteger $\alpha_i(0)\ge -1$, except of course for
Eq.~\eqref{eq:Delta_series} that could receive further nonvanishing
contributions.

Without restrictions on the functional dependence of the spectral
density on $m$ it is then not difficult to achieve effective
$\mathrm{U}(1)_A$ breaking by $\Delta\neq 0$, even if the spectral
density is finite (including if it vanishes) at
$\lambda=0$. Similarly, the requirement
$2m^2\Iu[\ff^2]-\f{\chit}{m^2} = O(m^2)$,
Eq.~\eqref{eq:firstorder_again1_2_again}, constraining the corrections
to the leading contribution, can be easily satisfied.

\subsubsection{\texorpdfstring{$m^2$}{m\textasciicircum 2}-differentiable spectral density:
  singular peak}
\label{sec:mdiffspd}

Under the additional assumption of nonlocal restoration (see
Sec.~\exref{I-sec:gauge_ext}), or of restoration in the presence of
external fields (see Appendix~\exref{I-sec:rhochirest}), the spectral
density $\rho$ must be $m^2$-differentiable. For a spectral density of
the form Eq.~\eqref{eq:spdens_gen1}, this requires that $C_i$ and
$\alpha_i$, $i=1,\ldots, s$, as well as $\bar{\rho}$, all be
$m^2$-differentiable (see Appendix~\ref{sec:mdiff}). In particular,
$\alpha_i(m)-\alpha_i(0)=O(m^2)$, so
\begin{equation}
  \label{eq:Cmdiff}
  \begin{aligned}
    C_i&=O\left(m^{1-\alpha_i(0)}\right)\,, &&& &\text{if }|\alpha_i(0)|<1\,,\\
    C_s&=O\left(1/\ln \tf{1}{|m|}\right)\,, &&& &\text{if }\alpha_s(0)=1\,,\\
    C_1&=O\left(m^2\left(1+\alpha_1(m)\right)\right)\,, &&& &\text{if }\alpha_1(0)=-1\,,
  \end{aligned}
\end{equation}
see Eqs.~\eqref{eq:int_new15} and \eqref{eq:X0fin1_0}. Then, if
$-1<\alpha_i(0)\le 1$ the requirement of $m^2$-differentiability
further constrains $C_i=O(m^2)$, so that
$C_i\mathcal{I}_{\alpha_i}=o(1)$ and these terms contribute neither to
$\II_0$ in the chiral limit nor to $\Delta$, and cannot lead to
effective $\mathrm{U}(1)_A$ breaking by $\Delta\neq 0$.  Instead, if
$\alpha_1(0)=-1$ the minimal requirement from finiteness of $\chi_\pi$
is that $C_1(m)= m^2\f{1+\alpha_1(m)}{2}\bar{C}(m)$ for some
$\bar{C}(m)$ with $\bar{C}(0)$ finite, and $m^2$-differentiability of
$C_1$ and $\alpha_1$ requires that of $\bar{C}$. A nonvanishing
$\bar{C}(0)$ is then allowed, and effective $\mathrm{U}(1)_A$ breaking
by $\Delta= \bar{C}(0)\neq 0$ is possible.

In summary, for an $m^2$-differentiable spectral density of the form
Eq.~\eqref{eq:spdens_gen1}, the only way to obtain effective
$\mathrm{U}(1)_A$ breaking by $\Delta\neq 0$, compatibly with
finiteness of $\chi_\pi$, is if $\rho$ has a singular near-zero peak
of the form
\begin{equation}
  \label{eq:rhom2diff_peak2}
  \rho_{\mathrm{peak}}(\lambda;m) \equiv
  \f{1}{2}m^2\gamma(m^2)\left[\Delta+B(m^2)\right]|\lambda|^{-1+\gamma(m^2)} \,,
\end{equation}
with $m^2$-differentiable $\gamma(m^2)\equiv 1+\alpha_1(m) = O(m^2)$
and $B(m^2)\equiv \bar{C}(m)-\Delta=O(m^2)$. The prefactor makes this
term at least $O(m^4)$ at fixed $\lambda$.  Further power-law terms
$\propto |\lambda|^{\alpha(m)}$ with $-1<\alpha(0)\le 1$ are allowed,
but do not contribute to $\mathrm{U}(1)_A$ breaking since they must be
suppressed by $O(m^2)$ coefficients.  In the chiral limit one still
finds that
$\rho(\lambda;0)=\bar{\rho}(\lambda;0)=O(|\lambda|^{1+\zeta})$ with
$\zeta>0$.

In the case of $\rho = \rho_{\mathrm{series}}$ admitting a
power-series expansion around zero, Eq.~\eqref{eq:app_sppow1},
$m^2$-differentiability of $\rho$ requires that of $\rho_n$ (see
Appendix~\ref{sec:mdiff}), and so it strengthens the constraints on
$\rho_0$ and $\rho_1$ coming from finiteness of $\chi_\pi$,
Eq.~\eqref{eq:app_sppow_cons}, to
\begin{equation}
  \label{eq:app_sppow_cons2}
  \rho_0=O(m^2)\,, \qquad \rho_1=O(m^2)\,,
\end{equation}
while $\rho_n=O(m^0)$ for $n\ge 2$ remains unchanged. In particular,
one finds that $\Delta=0$, and so $\mathrm{U}(1)_A$ must be restored
(at least at this level) in the symmetric phase, in agreement with the
findings of Refs.~\cite{Aoki:2012yj,Kanazawa:2015xna}.  This
conclusion does not change if power-law terms with non-integer
exponent are included, not only for positive $m$-independent
exponents, as already pointed out in Ref.~\cite{Aoki:2012yj}, but for
negative and for $m$-dependent exponents as well --- unless a term
$\rho_{\mathrm{peak}}$, Eq.~\eqref{eq:rhom2diff_peak2}, is present.

\subsubsection{Relation to topology}
\label{sec:spdenstop}

If present, the singular, $\mathrm{U}(1)_A$-breaking peak
$\rho_{\mathrm{peak}}$, Eq.~\eqref{eq:rhom2diff_peak2}, provides the
only contribution of the Dirac spectrum to $\Delta$, and so by
Eq.~\eqref{eq:firstorder_again2_again} it should be strongly related
to the topological properties of the gauge-field configurations. This
is shown explicitly by the following result for the normalized mode
number of the singular peak,
\begin{equation}
  \label{eq:rhom2diff_peakdens}
  \npeak(\delta;m) \equiv 2 \int_0^{\delta}d\lambda \,  \rho_{\mathrm{peak}}(\lambda;m)\,,
\end{equation}
where $\delta$ is an arbitrary cutoff $0< \delta\le 2$, and having
accounted also for the modes of $D$ with negative imaginary part. In
the chiral limit one finds
\begin{equation}
  \label{eq:nasp9bis}
  \begin{aligned}
    \lim_{m\to  0} \f{\npeak(\delta;m)}{m^2}
    & = \lim_{m\to  0} \left[\Delta+B(m^2)\right] \delta^{\gamma(m^2)} \\
    & = \Delta =\lim_{m\to 0} \f{\chit}{m^2}\,,
  \end{aligned}
\end{equation}
independently of the cutoff. In the last passage I have used the
symmetry-restoration condition
Eq.~\eqref{eq:firstorder_again2_again}.\footnote{One can also replace
  the fixed cutoff $\delta$ with a mass-dependent cutoff $c(m)$ and
  still obtain the same result as long as $\gamma(m^2)\ln c(m)\to 0$
  as $m\to 0$. To resolve the inner structure of the peak one needs an
  exponentially small cutoff, $c_e(m;z)= c_0 z^{1/\gamma(m^2)}$ with
  $0< z< 1$, leading to
  $ \lim_{m\to 0}m^{-2}\npeak(c_e(m;z);m) = z\Delta $.} In the chiral
limit the peak is then relevant only infinitesimally close to
$\lambda=0$: as it becomes more singular, tending to $1/|\lambda|$, it
also becomes more suppressed by the prefactor, with the normalized
mode number of the peak vanishing at least like $m^2$,
$\npeak \simeq \Delta m^2\simeq \chit$. This density matches precisely
(to leading order in $m$) that of the effective instanton gas
describing the topological properties of the theory in the chiral
limit if $\mathrm{U}(1)_A$ is effectively broken by $\Delta\neq 0$
(see Ref.~\cite{Kanazawa:2014cua} and Sec.~\exref{I-sec:top_iig}).

\subsubsection{\texorpdfstring{$m^2$}{m\textasciicircum
    2}-differentiability of
  \texorpdfstring{$\chi_\pi$}{chi\textunderscore pi} and
  \texorpdfstring{$\chi_\delta$}{chi\textunderscore delta}}
\label{sec:mdiffchipi}

While the requirements discussed above in Sec.~\ref{sec:chipi}
guarantee finiteness of $\chi_\pi$ and $\chi_\delta$ in the chiral
limit, chiral symmetry restoration requires that $\chi_\pi$ and
$\chi_\delta$ be $m^2$-differentiable (see
Sec.~\exref{I-sec:chisymrest_nsc}). Even imposing
$m^2$-differentiability of $\rho$, that leads to $m^2$-differentiable
coefficients $C_i$ and exponents $\alpha_i$, the contributions to
$\II_0(\delta;m)$ of the various power-like terms in
Eq.~\eqref{eq:spdens_gen1} are not manifestly
$m^2$-differentiable. Indeed, they are of the form
$C_i \mathcal{I}_{\alpha_i} = b_i(m) c_i(m^2) + d_i(m^2)$, with
$b_i(m)=m^{\alpha_i(m)+1}$ if $-1\le \alpha_i(0)< 1$ and
$b_s(m)=m^2\tf{m^{\alpha_s(m)-1}-1}{\alpha_s(m)-1}$ if
$\alpha_s(0)=1$, and with $c_i(m^2)$ and $d_i(m^2)$
$m^2$-differentiable functions. In general
$C_i \mathcal{I}_{\alpha_i}$ contain then terms logarithmic in $m$,
and suitable cancelations must take place in order to achieve
$m^2$-differentiability of $\chi_\pi$ and $\chi_\delta$.  While it may
seem difficult to do this while complying with positivity of the
spectral density, it is actually easy to come up with a simple example
that automatically yields $m^2$-differentiable $\chi_\pi$ and
$\chi_\delta$. For a spectral density of the form
\begin{equation}
  \label{eq:mdiffex1}
  \begin{aligned}
    &  \rho_{\mathrm{example}}(\lambda;m)\\
    &= [\lambda^2+m^2h(\lambda)]^2
      \left[\f{\Delta \gamma(m^2)}{2 m^2}|\lambda|^{-1+\gamma(m^2)}+ \hat{\rho}(\lambda;m)\right]\,,
  \end{aligned}
\end{equation}
with $m^2$-differentiable
$\hat{\rho}(\lambda;m) = O(|\lambda|^{-1+\hat{\zeta}})$ with
$\hat{\zeta}>0$, $\forall m$, and $m^2$-differentiable
$\gamma=O(m^2)$, one finds
\begin{equation}
  \label{eq:mdiffex2}
  \begin{aligned}
    \f{\chi_\pi}{2}
    &= \Delta\, \beta(m^2)\left(1+ \tf{\gamma(m^2)}{m^2}\right) \\
    &\phantom{=} + 2 \int_0^2d\lambda \, h(\lambda)\left[ \lambda^2+m^2h(\lambda)\right]
      \hat{\rho}(\lambda;m)\,,\\
    \beta(m^2)
    &\equiv  2^{\gamma(m^2)+3}\left[2+\gamma(m^2)\right]^{-1}\left[4+\gamma(m^2)\right]^{-1}\,,
  \end{aligned}
\end{equation}
and
\begin{equation}
  \label{eq:mdiffex3}
  \f{\chi_\pi-\chi_\delta}{4}
  =\Delta \, \beta(m^2)  + 2 m^2 \int_0^2d\lambda \,h(\lambda)^2\hat{\rho}(\lambda;m)\,,
\end{equation}
that are manifestly $m^2$-differentiable. The issue of
$m^2$-differentiability of $\chi_\pi$ and $\chi_\delta$ is then not
particularly concerning.\footnote{For the singular peak
  Eq.~\eqref{eq:rhom2diff_peak2} there is another way in which one can
  guarantee the $m^2$-differentiability of the relevant
  susceptibilities, without invoking cancelations: If $1+\alpha_1(m)$
  vanishes faster than any power of $m$, all the $m^2$-derivatives of
  the potentially problematic factor $m^{1+\alpha_1(m)}$ vanish at
  $m=0$, and the contributions of the peak to $\chi_\pi$ and
  $\chi_\delta$ are automatically $m^2$-differentiable.}

\subsubsection{Further constraints}
\label{sec:furtherrho}

Since $\f{\chit}{m^2}$ must be $m^2$-differentiable (see
Sec.~\exref{I-sec:constr_first}), the constraint
Eq.~\eqref{eq:firstorder_again1_2_again} together with
Eq.~\eqref{eq:firstorder_again2_again} amounts to requiring the
existence of the first $m^2$-derivative of
$\chi_\pi-\chi_\delta$. Having shown above that one can satisfy the
requirement of $m^2$-differentiability to all orders for an arbitrary
($m^2$-differentiable) $\gamma(m^2)$, this constraint cannot provide
further restrictions on the leading behavior of the singular peak, if
present, and so on the possibility of effectively breaking
$\mathrm{U}(1)_A$ by $\Delta\neq 0$.

On the other hand, for the purpose of comparison with
Ref.~\cite{Aoki:2012yj}, it is worth checking the consequences of this
constraint for an $m^2$-differentiable spectral density allowing a
power-series expansion around zero, Eq.~\eqref{eq:app_sppow1}. In this
case $\rho_n$ must be $m^2$-differentiable (see
Appendix~\ref{sec:mdiff}), and moreover $\Delta=0$ and
$\f{\chit}{m^2}=O(m^2)$. The constraint becomes [see
Eq.~\eqref{eq:inteJ_integer}]
\begin{equation}
  \label{eq:ps_extra1}
  \begin{aligned}
    \left.\f{\chi_\pi-\chi_\delta}{4}\right|_{\rho_{\mathrm{series}}} 
    &  = \rho_0(m)\left(\f{\pi}{2|m|} + O(m^0)\right)  \\
    &\phantom{=}+ \rho_1(m)\left(\f{1}{2}+O(m^2)\right)\\
    &\phantom{=} +  m^2  \rho_2(m) \left(\f{\pi}{2|m|} + O(m^0)\right)\\
    &\phantom{=} +  \rho_3(m)\left(2 m^2\ln\tf{1}{|m|}+ O(m^2)\right) \\
    &\phantom{=}+ O(m^2) = O(m^2)\,,
  \end{aligned}
\end{equation}
with $\rho_n = \rho_n^{(0)} + m^2\rho_n^{(1)}+\ldots$ (possibly up to
unimportant terms vanishing faster than any power of $m^2$, see
Sec.~\exref{I-sec:gauge_ext}). Since finiteness of $\chi_\pi$ already
requires $\rho_{0,1}=O(m^2)$, Eq.~\eqref{eq:app_sppow_cons2}, one
finds the constraints
\begin{equation}
  \label{eq:ps_extra2}
  \rho_0(m) + m^2 \rho_2(m) = O(m^4)\,, \qquad \rho_3(m)=O(m^2)\,.
\end{equation}
The first constraint was already found in Ref.~\cite{Aoki:2012yj}; the
second one is new. Since $\rho_0(m) = m^2\rho_{0}^{(1)} + O(m^4)$ and
$\rho_2(m) = \rho_{2}^{(0)} + O(m^2)$, the first constraint implies
$\rho_{0}^{(1)} + \rho_{2}^{(0)} = 0$. However, $\rho_0^{(1)}\neq 0$
is incompatible with positivity of the spectral density: since
$\rho_{0}^{(1)}\ge 0$, and so $\rho_2^{(0)}\le 0$, for
$\lambda=|m|\sqrt{1+\epsilon^2}$ one finds
\begin{equation}
  \label{eq:ps_extra4}
  \begin{aligned}
    \rho(|m|\sqrt{1+\epsilon^2};m)
    &= m^2\rho_0^{(1)} +  \rho_2^{(0)}m^2(1+\epsilon^2) + O(m^3)\\
    &=- m^2\rho_0^{(1)}\epsilon^2+ O(m^3)\,,
  \end{aligned}
\end{equation}
and so
\begin{equation}
  \label{eq:ps_extra5}
  0\le \lim_{m\to 0} \f{\rho(|m|\sqrt{1+\epsilon^2};m)}{m^2} = -\rho_0^{(1)}\epsilon^2\,, 
\end{equation}
requiring $\rho_0^{(1)}=\rho_2^{(0)}=0$. If
$\rho = \rho_{\mathrm{series}}$ then
\begin{equation}
  \label{eq:ps_extra6}
  \begin{aligned}
    \rho_0&= O(m^4)\,,&&& \rho_1&= O(m^2)\,,\\ \rho_2&= O(m^2)\,, &&& \rho_3&= O(m^2)\,.
  \end{aligned}
\end{equation}
The resulting constraint $\rho(\lambda;0) = O(\lambda^4)$ is stronger
than the one claimed in Ref.~\cite{Aoki:2012yj}, i.e.,
$\rho(\lambda;0)=O(|\lambda|^3)$, which is the behavior found for
free continuum fermions at $\mathrm{T}=0$. For this system chiral
symmetry is restored at the level of correlators but not at the level
of susceptibilities (see Sec.~\exref{I-sec:chisymrest}, in particular
footnote~\exref{I-foot:contfree}), so the corresponding behavior of
the spectral density, $\rho(\lambda;m)= K|\lambda|^3$ with
$m$-independent $K$, should indeed be excluded by the present
analysis.\footnote{To see this directly, imposing a UV cutoff
  $\Lambda$ on the Dirac spectrum for regularization purposes, one
  finds
  $$
  \f{\chi_\pi}{2K} = \int_0^\Lambda d\lambda\,
  \f{2\lambda^3}{\lambda^2+m^2} =
  \Lambda^2-m^2\ln\left[1+\left(\tf{\Lambda}{m}\right)^2\right]\,,
  $$
  which is not $m^2$-differentiable. Note that here $\lambda$ and $m$
  are in physical units.} Notice, however, that the restriction
$\rho(\lambda;0)=O(\lambda^4)$ is not valid in general, and depends on
the assumptions one makes on the spectral density (here the
possibility of expanding it in powers of $|\lambda|$): in particular,
$\rho(\lambda;0)=O(|\lambda|^3)$ is not completely excluded, as it
could be obtained, e.g., as the chiral limit of a term
$\propto |\lambda|^{3+\gamma(m^2)}$ [see Eq.~\eqref{eq:mdiffex1}].

On the other hand, Ref.~\cite{Aoki:2012yj} finds a stronger constraint
on $\rho_0(m)$, namely that it vanishes faster than any power of
$m$.\footnote{Since Ref.~\cite{Aoki:2012yj} assumes that $\rho$ is
  analytic in $m^2$ around zero, this would imply that $\rho(0;m)$
  vanishes identically for $|m|$ below some nonzero value.  This is an
  instance in which $m^2$-analyticity gives a stricter condition than
  $m^2$-differentiability.} However, the derivation of this result
contains a technical flaw.  Essentially, Ref.~\cite{Aoki:2012yj}
claims that if an expectation value vanishes in the chiral limit, then
it receives contributions only from a portion of configuration space
of vanishing measure. This is incorrect, as there are other possible
reasons why an expectation value has a vanishing chiral limit (e.g.,
the probability distribution of the observable is peaked on regions
where it takes vanishingly small values as $m\to 0$). This will be
discussed in detail elsewhere. The validity of this constraint on
$\rho_0(m)$, even under the stated assumptions on $\rho$, is then
dubious.

\subsubsection{Commutativity of thermodynamic and chiral limits}
\label{sec:ctc}

The correct order of limits (thermodynamic, $\lvol\to\infty$, followed
by chiral, $m\to 0$) is always used in the derivation of the general
symmetry-restoration conditions on the susceptibilities carried out in
{\DSI}, and no assumption is made on the commutativity of the two
limits. The same applies to the more detailed study of constraints
done above for a spectral density of the functional form
Eq.~\eqref{eq:spdens_gen1}.  However, commutativity of the
thermodynamic and chiral limits is a reasonable assumption to make in
the symmetric phase~\cite{Evans:1996wf}, so it is worth checking what
are its consequences, and what these entail for a spectral density of
the form Eq.~\eqref{eq:spdens_gen1}.

It was shown in Ref.~\cite{Azcoiti:2023xvu} that if the free energy
density is analytic in $m^2$ at $m=0$, and the thermodynamic and
chiral limits commute for the susceptibilities $\chi_\pi$ and
$\chi_\delta$ and for the spectral density, then the following
relation holds,\footnote{Reference~\cite{Azcoiti:2023xvu} actually
  uses only the fact that the free energy density is a $C^2$ function
  of $m$ with first derivatives vanishing at $m=0$, and so a $C^1$
  function of $m^2$ including at $m=0$. This is warranted by the
  $m^2$-differentiability in the symmetric phase of the free energy
  $-\mathcal{W}_{\!\scriptscriptstyle\infty}(0,0;m)=-\susc_{000}(m^2)$,
  proved in Sec.~\exref{I-sec:chisymrest_nsc}. Commutativity of the
  relevant limits for the spectral density is assumed implicitly in
  Ref.~\cite{Azcoiti:2023xvu}.}
\begin{equation}
  \label{eq:commlim6_0}
  \lim_{\epsilon\to 0^+}\lim_{m\to 0} \II_0(\epsilon;m)=
  \lim_{\epsilon\to 0^+}\lim_{m\to 0} \II_1(\epsilon;m)= \f{\Delta}{2}\,.
\end{equation}
This is the same as Eq.~(49) of Ref.~\cite{Azcoiti:2023xvu} in a
different notation, and having dropped terms that vanish in the chiral
limit. This relation is the key consequence of commutativity of the
thermodynamic and chiral limits in the present context. Through
further manipulations, not requiring any additional assumption, one
shows that Eq.~\eqref{eq:commlim6_0} implies
\begin{equation}
  \label{eq:commlim17_0}
  \lim_{m\to 0}  \f{\rho(|m| z;m)}{|m|} = \Delta\delta(z)\,,  \qquad z\in[-1,1]\,, 
\end{equation}
which is the same as Eq.~(62) of Ref.~\cite{Azcoiti:2023xvu} (up to
unimportant terms). One can further show that from
Eq.~\eqref{eq:commlim6_0} follows more generally
\begin{equation}
  \label{eq:commlim6_0_gen}
  \lim_{\epsilon\to 0^+}\lim_{m\to 0} \II_n(\epsilon;m)
  =\f{\Delta}{2}\,, \quad n\ge 0\,. 
\end{equation}
To keep the analysis self-contained, I rederive
Eqs.~\eqref{eq:commlim6_0} and \eqref{eq:commlim17_0} and obtain
Eq.~\eqref{eq:commlim6_0_gen} using the methods of this paper in
Appendix~\ref{sec:app_commlim}.

Reference~\cite{Azcoiti:2023xvu} concludes that the only acceptable
solutions of Eq.~\eqref{eq:commlim17_0} are spectral densities
containing a highly singular term
$\rho_{\mathrm{sing}}(\lambda;m)\equiv \Delta m^2 \delta(\lambda)$.
This would mean that a spectral density of the form
Eq.~\eqref{eq:spdens_gen1} that effectively breaks $\mathrm{U}(1)_A$
is incompatible with commutativity of the two limits, including if it
is $m^2$-differentiable [in which case
$\rho \sim \rho_{\mathrm{peak}}$, Eq.~\eqref{eq:rhom2diff_peak2}, near
$\lambda=0$]. However, while $\rho_{\mathrm{sing}}$ clearly solves
Eq.~\eqref{eq:commlim17_0}, so does any $\f{\rho(|m| z;m)}{|m|}$ that
is a regularization of the Dirac delta (a ``nascent delta function''),
and without further requirements there seems to be no reason to single
$\rho_{\mathrm{sing}}$ out as the only possible solution.

In fact, from Eqs.~\eqref{eq:int_new20_bis_delta2_0} and
\eqref{eq:int_new20_bis_delta2bis_0} one sees immediately that
$\rho=\rho_{\mathrm{peak}}$, Eq.~\eqref{eq:rhom2diff_peak2}, satisfies
Eq.~\eqref{eq:commlim6_0}, and so Eqs.~\eqref{eq:commlim17_0} and
\eqref{eq:commlim6_0_gen}, and is therefore compatible with
commutativity of the two limits (as well as with
$m^2$-differentiability). Indeed,
\begin{equation}
  \label{eq:commlim9_0}
  \begin{aligned}
    \f{\Delta}{2}
    &=  \lim_{\epsilon\to 0^+} \lim_{m\to 0}
      \int_0^\epsilon d\lambda\,\f{m^2\rho_{\mathrm{peak}}(\lambda;m)}{(\lambda^2+m^2)^2}\\
    &    =   \lim_{\epsilon\to 0^+} \lim_{m\to 0} C_1(m)\mathcal{I}_{\alpha_1(m)}(\epsilon;m)\\
    &=\lim_{\epsilon\to 0^+} \lim_{m\to 0}
      \int_0^\epsilon d\lambda\, \f{\rho_{\mathrm{peak}}(\lambda;m)}{\lambda^2+m^2} \,.
  \end{aligned}
\end{equation}
That $\rho_{\mathrm{peak}}$ satisfies Eq.~\eqref{eq:commlim17_0}, thus
providing a nascent delta function, can also be shown directly [see
Eqs.~\eqref{eq:commlim19} and \eqref{eq:commlim20} in
Appendix~\ref{sec:app_commlim_der2}]. This shows explicitly that
$\mathrm{U}(1)_A$ can be effectively broken by $\Delta\neq 0$ even if
the thermodynamic and chiral limits commute, without having to resort
to the clearly unphysical behavior $\rho_{\mathrm{sing}}$. This may
seem impossible at first, since a divergent peak
$\rho \sim C_\alpha|\lambda|^\alpha$ with $\alpha<0$ cannot form in a
finite volume. However, as shown above, chiral symmetry restoration
requires that if such a peak appears in the spectral density $\rho$
obtained in the infinite-volume limit, then the prefactor $C_\alpha$
vanishes as $m\to 0$, and so the peak will be absent in the chiral
limit for either choice of the order of limits.  Such a peak in $\rho$
is (possibly) effective only when computing observables, where one
takes the relevant limits, in the correct order, after integrating
over the whole spectrum. When taking limits in the ``wrong'' order,
$\mathrm{U}(1)_A$-breaking effects originate instead in the
contribution of the zero modes associated with topologically
nontrivial gauge configurations [see Eq.~\eqref{eq:commlim2_1}]. The
two mechanisms provide the same value of $\Delta$ if
Eq.~\eqref{eq:commlim17_0} is satisfied.

More generally, for a spectral density of the form
Eq.~\eqref{eq:spdens_gen1} the requirement of commutativity of limits
singles out
\begin{equation}
  \label{eq:rhom2diff_peak2_commut}
  \begin{aligned}
    \tilde{\rho}_{\mathrm{peak}}(\lambda;m)
    &\equiv\f{1}{2} m^2 \tilde{\gamma}(m)\tilde{C}(m)
      |\lambda|^{-1+\tilde{\gamma}(m)} \,, \\
    \tilde{C}(m)
    &= \tilde{c}(m)|m|^{-\tilde{\gamma}(m)} \,,
  \end{aligned}
\end{equation}
with $\tilde{\gamma}(0)=0$ and $\tilde{c}(0)=\Delta$, as the only
possible $\mathrm{U}(1)_A$-breaking contribution, independently of
whether one requires $m^2$-differentiability or not. In the latter
case one further needs $\tilde{\gamma}(m)=\gamma(m^2)$ to be
$m^2$-differentiable, and $ \tilde{C}(m) = \Delta + B(m^2)$ with
$m^2$-differentiable $B$, obtaining of course $\rho_{\mathrm{peak}}$,
Eq.~\eqref{eq:rhom2diff_peak2}.\footnote{If one requires
  $m^2$-differentiability, terms other than $\rho_{\mathrm{peak}}$ do
  not contribute to either side of Eq.~\eqref{eq:commlim6_0},
  satisfying it trivially and not causing $\mathrm{U}(1)_A$ breaking,
  see Sec.~\ref{sec:mdiffspd} and Appendix~\ref{sec:finX0}. In
  general, $\bar{\rho}(\lambda;m)$ does not contribute to $\Delta$
  [see Eq.~\eqref{eq:spdens_gen4}], and since
  $\bar{\rho}(\lambda;m)=O(|\lambda|^{1+\zeta})$, and so
  $\bar{\rho}(\lambda;0)/\lambda^2$ is integrable, it does not
  contribute to the first quantity in Eq.~\eqref{eq:commlim6_0}
  either.} This is shown in Appendix~\ref{sec:finX0}. 

To summarize, for a spectral density of the form
Eq.~\eqref{eq:spdens_gen1}, a term $\tilde{\rho}_{\mathrm{peak}}$ is
the only $\mathrm{U}(1)_A$-breaking contribution compatible with limit
commutativity; and $\rho_{\mathrm{peak}}$ is the only
$\mathrm{U}(1)_A$-breaking contribution compatible with both limit
commutativity and $m^2$-differentiability. (Of course, a behavior
$\rho \sim \tilde{\rho}_{\mathrm{peak}}$ or
$\rho \sim \rho_{\mathrm{peak}}$ does not necessarily imply
commutativity of the limits.)

\subsection{\texorpdfstring{$m^2$}{m\textasciicircum 2}-differentiable
  spectral density: general results}
\label{sec:mdiffspdgen}

For a spectral density of the form Eq.~\eqref{eq:spdens_gen1}, a
$1/|\lambda|$ behavior in the chiral limit (with a vanishingly small
prefactor) is singled out as the only one leading to $\mathrm{U}(1)_A$
breaking, if one requires either $m^2$-differentiability or
commutativity of the thermodynamic and chiral limits. Although quite
general, the functional form Eq.~\eqref{eq:spdens_gen1} is certainly
not the most general functional form of the spectral density, and one
would like to determine under what conditions $\mathrm{U}(1)_A$
breaking by $\Delta\neq 0$ is possible in a broader setting. As shown
in Ref.~\cite{Azcoiti:2023xvu} (see also
Appendix~\ref{sec:app_commlim}), commutativity of limits strongly
restricts this possibility, requiring the presence of a nascent delta
function, Eq.~\eqref{eq:commlim17_0}, in the spectral
density. Commutativity of limits, however, is not necessary for chiral
symmetry restoration, whether at the level of scalar and pseudoscalar
susceptibilities only, or in its extended sense discussed in
Sec.~\ref{sec:intro2}. The requirement of $m^2$-differentiability in
the symmetric phase, on the other hand, follows from extended chiral
symmetry restoration, and is therefore more fundamental. Under quite
general assumptions, also this requirement severely restricts the
possibility to effectively break $\mathrm{U}(1)_A$: as I now show,
$\Delta\neq 0$ requires that $\rho$ effectively develops a singular
term $a_0 m^2 \delta(\lambda)$ in the chiral limit, with $a_0$
providing an upper bound on $\Delta$.

To make statements precise, it is convenient to work with the
normalized mode number $\nn(\lambda;m)$, Eq.~\eqref{eq:rho_distr_def},
which is an ordinary function. Following the discussion at the
beginning of this section, the expected absence of integrable
divergences or Dirac deltas in the spectral density at $\lambda\neq 0$
translates into assuming that $\nn(\lambda;m)$ is differentiable in
$\lambda$ at $\lambda\neq 0$, $\forall m$ (see also
footnote~\ref{foot:singedge}).  The absence of a Dirac delta in the
spectral density at $\lambda=0$ translates into $\nn(0;m)=0$,
$\forall m\neq 0$ (see below for $m=0$). As with $\rho$, I denote
$\nn(\lambda;0)=\lim_{m\to 0}\nn(\lambda;m)$, that does not depend on
how $m=0$ is approached (see footnote~\ref{foot:rhomz}).

At $\lambda\neq 0$ and $m\neq 0$, $\nn(\lambda;m)$ should be
infinitely differentiable in $m$ and so in $m^2$, as its mass
derivatives are equal to (normalized) connected correlators of the
(signed) number of modes in a spectral interval, $\nn_U(\lambda)$ [see
Eq.~\eqref{eq:rho_distr_def}], and scalar isosinglet bilinears [see
Eq.~\exeqref{I-eq:mder1}, that can be easily generalized to multiple
derivatives]. Under the assumption of nonlocal chiral symmetry
restoration, or of restoration in the presence of external fermion
fields, $\nn(\lambda;m)$ is $m^2$-differentiable, i.e., its
$m^2$-derivatives exist also at $m=0$. The expected absence of
singularities in the $m^2$-derivatives of the spectral density at
$\lambda\neq 0$ translates again into assuming that
$\de_{m^2}^k\nn(\lambda;m)$ is differentiable in $\lambda$ at
$\lambda\neq 0$, $\forall m$.

Even though $\nn(0;m)=0$ identically for $m\neq 0$, which implies
$\de_{m^2}\nn(0;m)=0$, $\forall m\neq 0$, and so
$\lim_{m\to 0}\nn(0;m) =\lim_{m\to 0}\de_{m^2}\nn(0;m)=0$, in general
neither $\nn(0^+;0)=\lim_{\lambda\to 0^+}\lim_{m\to 0}\nn(\lambda;m)$
nor
$\de_{m^2}\nn(0^+;0)=\lim_{\lambda\to 0^+}\lim_{m\to
  0}\de_{m^2}\nn(\lambda;m)$ have to vanish. However,
$\nn(0^+;0)\neq 0$ would imply the presence of a term
$\delta(\lambda)$ with $O(m^0)$ coefficient in the spectral density,
which is not expected. In any case, such a term would lead to a
divergent $\chi_\pi$, so one must have $\nn(0^+;0)=0$ in the symmetric
phase. On the other hand, $\de_\lambda\nn(\lambda;0)$ may be divergent
as $\lambda\to 0$, as long as the divergence is integrable.

Concerning $\de_{m^2}\nn(\lambda;m)$, as $m\to 0$ modes are expected
to be more strongly repelled from the origin due to the increased
suppressing effect of the fermion determinant. This repulsion should
become weaker as one moves away from $\lambda=0$ and towards the bulk
of the spectrum, which should be the least sensitive to a change in
$m$. For small enough $\lambda$ and $m$ one then expects
$\de_\lambda\de_{m^2}\nn(\lambda;m)\ge 0$, and therefore
$\de_{m^2}\nn(\lambda;m)\ge 0$ for small positive $\lambda$.
Moreover, in the chirally symmetric phase the low-lying spectrum
should be depleted and show a pseudogap, due to the ordering of the
Polyakov loop~\cite{Bruckmann:2011cc,
  Giordano:2015vla,Giordano:2016cjs,Giordano:2016vhx,
  Giordano:2021qav,Baranka:2022dib}.\footnote{A confining theory is
  expected to spontaneously break chiral
  symmetry~\cite{Casher:1979vw}, so a fairly ordered Polyakov loop is
  expected in a chirally symmetric phase.} This ordering persists in
the chiral limit (at least in QCD)~\cite{Clarke:2020htu}, so repulsion
should be effective up to some finite distance from zero even in the
chiral limit.  Based on these considerations, one expects
$\de_{m^2}\nn(\lambda;m)\ge 0$ for small enough $|\lambda|<\lambda_0$
and $|m|<m_0$, i.e., up to a nonzero, $m$-independent distance
$\lambda_0$ from the origin of the spectrum, and for small enough $m$,
including in the limit $m\to 0$, and so
$\pm \de_\lambda\de_{m^2}\nn(\lambda;m)\ge 0$ for
$0\le \pm \lambda <\lambda_0$.

Since one does not expect the repulsion from the origin to change
singularly at $m=0$, $\de_{m^2}\nn(0^+;0)$ should be finite (and
possibly zero). Moreover, a divergent $\de_{m^2}\nn(\lambda;0)$ as
$\lambda\to 0$ would generally (although not necessarily) lead to a
divergent $\chi_\pi$, which is forbidden in the symmetric phase (see
Appendix~\ref{sec:genm2rho_1}).

The assumptions above are conveniently reformulated by writing
$\nn(\lambda;m)= \nn(\lambda;0) + m^2\nn_1(\lambda;m)$, with
$\nn(\lambda;0)$ and
$\nn_1(\lambda;m)\equiv[\nn (\lambda;m) - \nn (\lambda;0)]/ m^2$ both
odd functions of $\lambda$. Under the stated assumptions, the
normalized mode number in the chiral limit, $\nn(\lambda;0)$, vanishes
as $\lambda\to 0$, and $\de_\lambda\nn(\lambda;0)$ is an integrable
ordinary function.  For $\lambda\ge 0$, the quantity
$\nn_1(\lambda;m)$ measures the average change with $m^2$ of the
number of modes in the interval $[0,\lambda]$, so for
$0\le \lambda<\lambda_0$ and $|m|< m_0$ it is nonnegative, with
$\de_\lambda\nn_1(\lambda;m)\ge 0$.\footnote{Nonnegativity of
  $\de_\lambda\nn_1(\lambda;m)$ follows from that of
  $\de_\lambda\de_{m^2}\nn(\lambda;m)$ under mild technical
  assumptions: writing
  $ \nn_1(\lambda;m) = m^{-2}\int_0^{m^2} d\mu^2\, \de_{\mu^2}
  \nn(\lambda;\mu)$, it suffices that one can exchange differentiation
  with respect to $\lambda$ and integration over $\mu^2$.} For
$m\neq 0$ clearly $\nn_1(0^+;m)=0$, and $\de_\lambda\nn_1(\lambda;m)$
is an ordinary integrable function.  Finally, $\nn_1(\lambda;m)$ is
$m^2$-differentiable, with
$\nn_1(\lambda;0)= \de_{m^2}\nn(\lambda;m)|_{m=0}$ and
$0\le \nn_1(0^+;0)<\infty$, and $\de_\lambda\nn_1(\lambda;0)$ is an
ordinary integrable function plus possibly a Dirac delta at
$\lambda=0$.

For the spectral density $\rho(\lambda;m)=\de_\lambda\nn(\lambda;m)$
one has then
\begin{equation}
  \label{eq:ua1b_gen1}
  \rho(\lambda;m)= \rho(\lambda;0)+ m^2\rho_1(\lambda;m)\,,  
\end{equation}
where $\rho_1(\lambda;m)\equiv\de_\lambda\nn_1(\lambda;m)$. Notice
that since $\nn(2;m) = 2N_c$ is independent of $m$, $\nn_1(2;m) = 0$
and so $\rho_1(\lambda;m)$ must change sign at least once.  Under the
assumptions above, $\rho(\lambda;0)$ has at most an integrable
singularity at $\lambda=0$, while
$\rho_1(\lambda;0)= a_0\delta(\lambda) + b_0(\lambda)$ near
$\lambda=0$, with $b_0(\lambda)$ integrable and
\begin{equation}
  \label{eq:rho1_integral}
  \lim_{\epsilon\to 0^+}\int_{-\epsilon}^\epsilon d\lambda\,
  \rho_1(\lambda;0) = a_0 = 2\lim_{\epsilon\to 0^+} \nn_1(\epsilon;0)\,.
\end{equation}
Moreover, $\rho_1(\lambda;m)>0$ for $|\lambda|<\lambda_0$ and
$|m|<m_0$. The assumption of monotonicity in $\lambda$ of
$\de_{m^2}\nn(\lambda;m)$, leading to positivity of
$\rho_1(\lambda;m)$ at small $\lambda$, could be relaxed to that of
the existence of separately well-defined chiral limits for the
positive and negative components of $\rho_1$, leading to similar
results. This is discussed in Appendix~\ref{sec:genm2rho_2}.

The existence and finiteness of $\lim_{m\to 0}\chi_\pi$ and so of
$\lim_{m\to 0} \II_0(\delta;m)$ [see Appendix~\ref{sec:coeff_contr},
Eqs.~\eqref{eq:lim_mdiff1}--\eqref{eq:lim_mdiff4}] requires the
existence and finiteness of the chiral limit of
\begin{equation}
  \label{eq:ua1b_gen6}
  \II_0^{(0)}(\delta;m) \equiv \int_0^\delta d\lambda\, \f{\rho(\lambda;0)}{\lambda^2+m^2}\,,
\end{equation}
at least for $\delta<\lambda_0$. Indeed, this limit certainly exists
and is positive since $\II_0^{(0)}$ is a monotonically increasing
function of $m$. Under the present assumptions, one finds for
$0<\delta<\lambda_0$
\begin{equation}
  \label{eq:lim_mdiff_text}
  \begin{aligned}
    & \lim_{m\to 0}\II_0(\delta;m) \\
    &= \lim_{m\to 0}\left[
      \II_0^{(0)}(\delta;m) + \left(
      \II_0(\delta;m)-\II_0^{(0)}(\delta;m)\right)\right]\\
    &\ge \lim_{m\to 0} \II_0^{(0)}(\delta;m) 
      + \liminf_{m\to 0}\left(
      \II_0(\delta;m)-\II_0^{(0)}(\delta;m)\right)
    \\
    &\ge \lim_{m\to 0} \II_0^{(0)}(\delta;m) 
      - \lim_{m\to 0}\int_0^\delta d\lambda\,\rho_1(\lambda;m) \\
    &=\lim_{m\to 0} \II_0^{(0)}(\delta;m)   - \nn_1(\delta;0)
  \end{aligned}
\end{equation}
so $ \lim_{m\to 0} \II_0^{(0)}(\delta;m)$ is finite. Then
\begin{equation}
  \label{eq:ua1b_gen7}
  \begin{aligned}
    \infty &> \lim_{m\to 0}\II_0^{(0)}(\delta;m) 
             \ge \lim_{m\to 0}\int_m^\delta d\lambda\, \f{\lambda^2}{\lambda^2+m^2}
             \f{\rho(\lambda;0)}{\lambda^2} \\
           &\ge \f{1}{2}\lim_{m\to 0}\int_m^\delta d\lambda\, \f{\rho(\lambda;0)}{\lambda^2}\,,
  \end{aligned}
\end{equation}
implying that $\rho(\lambda;0)/\lambda^2$ is integrable near zero.

For the $\mathrm{U}(1)_A$ breaking parameter one finds from
Eq.~\eqref{eq:int_new20_bis_delta3}
\begin{equation}
  \label{eq:ua1b_gen10}
  \f{\Delta}{2} = \lim_{\epsilon\to 0^+}\lim_{m\to 0}
  \int_0^\epsilon d\lambda\, \f{m^4\rho_1(\lambda;m)}{(\lambda^2+m^2)^2}\,,
\end{equation}
having used the fact that
\begin{equation}
  \label{eq:ua1b_gen9}
  \lim_{\epsilon\to 0^+}\lim_{m\to 0}\int_0^\epsilon d\lambda\, \f{m^2\rho(\lambda;0)}{(\lambda^2+m^2)^2} 
  \le \lim_{\epsilon\to 0^+}\int_0^\epsilon d\lambda\, \f{\rho(\lambda;0)}{\lambda^2} = 0\,,
\end{equation}
since $\rho(\lambda;0)/\lambda^2$ is integrable.  Finiteness of
$\Delta$ in the symmetric phase requires the existence and finiteness
of the double limit on the right-hand side of
Eq.~\eqref{eq:ua1b_gen10}. Moreover, under the present assumptions
\begin{equation}
  \label{eq:ua1b_gen11}
  \f{\Delta}{2}  \le \lim_{\epsilon\to 0^+}\lim_{m\to 0}\int_0^\epsilon d\lambda\,  \rho_1(\lambda;m)
  = \lim_{\epsilon\to 0^+}n_1(\epsilon;0)  = \f{a_0}{2} \,.
\end{equation}
An integrable divergence at $\lambda=0$ in $\rho_1$ in the chiral
limit is then not sufficient to effectively break $\mathrm{U}(1)_A$,
and a Dirac delta at zero is required. In particular,
$\mathrm{U}(1)_A$ is effectively restored if
$\rho_1(\lambda;m)\le C|\lambda|^{-1+\zeta}$ for small $\lambda$ and
$m$, with $m$-independent $C,\zeta$, and $\zeta>0$. This agrees with
the analysis of Sec.~\ref{sec:mdiffspd}.

The presence of a Dirac delta at zero in $\rho_{1}(\lambda;0)$ is a
necessary condition for effective $\mathrm{U}(1)_A$ breaking, under
the stated assumptions. It is not, however, a sufficient condition,
nor is it equivalent to Eq.~\eqref{eq:commlim17_0}, and so it does not
necessarily imply compatibility of the result with the commutativity
of the thermodynamic and chiral limits. This is seen explicitly by
means of a few examples that show the variety of possible outcomes.
It suffices to consider $\rho_1$ of the form
$\rho_1(\lambda;m)=
\rho_{1,\,\mathrm{sing}}(\lambda;m)+\rho_{1,\,\mathrm{reg}}(\lambda;m)$,
with $\rho_{1,\,\mathrm{sing}}(\lambda;m)\ge 0$, and
$\rho_{1,\,\mathrm{reg}}(\lambda;m)\ge 0$ for
$|\lambda|\le \tilde{\lambda}_0$ for some $\tilde{\lambda}_0$, so that
$\rho_1(\lambda;m)\ge 0$ for $|\lambda|\le\lambda_0$ for some
$\lambda_0\ge \tilde{\lambda}_0$; with
$\rho_{1,\,\mathrm{reg}}(\lambda;0)$ an integrable function; and with
$\int_0^2d\lambda\,\rho_{1,\,\mathrm{reg}}(\lambda;m) =
-\int_0^2d\lambda\,\rho_{1,\,\mathrm{sing}}(\lambda;m)$. The function
$\rho_{1,\,\mathrm{reg}}(\lambda;m)$ need not be further specified, as
it gives at most a finite contribution to $\chi_\pi$ and none to
$\Delta$, since
\begin{equation}
  \label{eq:rho1_reg}
  \begin{aligned}
    &  \lim_{\epsilon\to 0}\lim_{m\to 0}\int_0^\epsilon d\lambda
      \left(\f{m^2}{\lambda^2+m^2}\right)^n \rho_{1,\,\mathrm{reg}}(\lambda;m) \\
    &\le\lim_{\epsilon\to 0}\int_0^\epsilon d\lambda\,\rho_{1,\,\mathrm{reg}}(\lambda;0)=0\,.
  \end{aligned}
\end{equation}
For the sake of example, a function with the required properties is
$\rho_{1,\,\mathrm{reg}}(\lambda;m) = K(m)\lambda\sin(\pi\lambda)$,
with
$K(m)= \f{\pi}{2}
\int_0^2d\lambda\,\rho_{1,\,\mathrm{sing}}(\lambda;m)$. Details of
the following calculations are found in Appendix~\ref{sec:genm2rho_3}.

The first example are singular peaks of the form
\begin{equation}
  \label{eq:genU1abr12}
  \rho_{1,\,\mathrm{sing}}(\lambda;m) =
  \f{\gamma(m^2)}{|\lambda|}\phi\left(\gamma(m^2)\ln\f{2}{|\lambda|}\right)\,,
\end{equation}
with $\phi(x)$ positive and $C^\infty$, integrable in $[0,\infty)$,
and with $m^2$-differentiable $\gamma(m^2)=O(m^2)$. For the singular
peak $\rho_{\mathrm{peak}}$, Eq.~\eqref{eq:rhom2diff_peak2},
$\rho_{1,\,\mathrm{peak}}=\rho_{\mathrm{peak}}/m^2$ is of this form
with $ \phi_{\mathrm{peak}}(x) = \Delta 2^{-1+\gamma(m^2)} e^{-x} $,
up to $O(m^2)$ corrections. In this case one finds
\begin{equation}
  \label{eq:genU1abr18}
  \lim_{m\to 0}\int_0^\epsilon d\lambda \,\left(\f{m^2}{\lambda^2+m^2}\right)^n
  \rho_{1,\,\mathrm{sing}}(\lambda;m) =  \int_{0}^\infty dw \,\phi(w)\,,
\end{equation}
$\forall n\ge 0$, which implies in particular that in this case
\begin{equation}
  \label{eq:genU1abr_final}
  \begin{aligned}
    \f{\Delta}{2} &= \lim_{\epsilon\to 0^+}\lim_{m\to 0}\II_0(\epsilon;m) =
                    \lim_{\epsilon\to 0^+}\lim_{m\to 0}\II_1(\epsilon;m)\\
                  & = \lim_{m\to 0} \nn_1 (\epsilon;m)\,.
  \end{aligned}
\end{equation}
One finds then effective $\mathrm{U}(1)_A$ breaking, compatibility
with the commutativity of chiral and thermodynamic limits [see
Eq.~\eqref{eq:commlim6_0}], and the same relation
$2m^2\nn_1(\epsilon;m)=\chit +O(m^2) $ between the topological
susceptibility and the contribution of the singular peak
$ m^2\rho_{1,\,\mathrm{sing}}$ to the normalized mode number as for
$\rho_{\mathrm{peak}}$ [see Eqs.~\eqref{eq:rhom2diff_peakdens} and
\eqref{eq:nasp9bis}; note that
$\nn_{\mathrm{peak}}=2m^2\nn_{1,\,\mathrm{peak}}$].

Another example is given by the functional form
\begin{equation}
  \label{eq:genU1abr19}
  \rho_{1,\,\mathrm{sing}}(\lambda;m) =
  \f{1}{|m|\varepsilon(m)} \phi\left(\f{|\lambda|- |m|\xi}{|m|\varepsilon(m)}\right)
\end{equation}
with positive and $m^2$-differentiable $|m|\varepsilon(m)=O(m^2)$, and
$\phi(x)$ positive, $C^\infty$, and integrable in $(-\infty,\infty)$,
and so bounded. If $\xi> 0$ this corresponds to having two smooth
peaks around $\lambda=\pm |m|\xi$ in the spectral density, getting
sharper and shifting toward zero as $m\to 0$; if $\xi= 0$, there is a
single smooth peak around $\lambda=0$, sharpening as $m\to 0$; if
$\xi<0$ there is a single cusped peak around $\lambda=0$. One finds
\begin{equation}
  \label{eq:genU1abr21}
  \begin{aligned}
    & \lim_{m\to 0} \int_0^\epsilon d\lambda\,\left(\f{m^2}{\lambda^2+m^2}\right)^n
      \rho_{1,\,\mathrm{sing}}(\lambda;m) \\
    &  =\left\{
      \begin{aligned}
        &       \left(\xi^2+1\right)^{-n}  \int_{-\infty}^{\infty} dz\,\phi(z)\,,        &&& \xi &>0\,,\\
        &       \int_{0}^{\infty} dz\,\phi(z)\,,        &&& \xi &=0\,,\\
        &       0\,, &&&   \xi &<0\,.
      \end{aligned}\right.
  \end{aligned}
\end{equation}
If $\xi>0$, then $\Delta\neq 0$ and $\mathrm{U}(1)_A$ is broken,
chiral and thermodynamic limits do not commute, and
$2m^2\nn_1(\epsilon;m)= (1+\xi^2)^2\Delta m^2 = (1+\xi^2)^2\chit$ to
leading order in $m$.  If $\xi=0$, $\mathrm{U}(1)_A$ is broken, limits
may commute, and $2m^2\nn_1 (\epsilon;m)=\chit$ to leading order in
$m$.  If $\xi<0$ no Dirac delta appears, $\Delta=0$ and so
$\mathrm{U}(1)_A$ may be restored, and limits may commute.

In all the cases above, when $\rho_1(\lambda;0)$ contains a term
$a_0\delta(\lambda)$ one gets effective $\mathrm{U}(1)_A$ breaking by
$\Delta\neq 0$, with $\Delta\le a_0$.  One can, however, modify
Eq.~\eqref{eq:genU1abr19} by considering a shift $|m|\xi(m)$, with an
$m$-dependent $\xi(m)>0$ that diverges in the chiral limit while
$|m|\xi(m)\to 0$. In this case
\begin{equation}
  \label{eq:genU1abr20_bis}
  \begin{aligned}
    &\lim_{m\to 0} \int_0^\epsilon d\lambda\,
      \left(\f{m^2}{\lambda^2+m^2}\right)^n \rho_{1,\,\mathrm{sing}}(\lambda;m)  \\
    &=\left\{
      \begin{aligned}
        &\int_{-\infty}^{\infty} dz\, \phi(z)\,, &&& n&=0\,,\\
        &0\,, &&& n&\ge 1\,, 
      \end{aligned}\right.
  \end{aligned}
\end{equation}
so $\rho_{1,\,\mathrm{sing}}(\lambda;m) \to a_0\delta(\lambda)$ near
$\lambda=0$ as $m\to 0$, but it gives no contribution to either
$\chi_\pi$ or $\Delta$. This explicitly shows how the appearance of a
Dirac delta at zero in $\rho_1(\lambda;0)$ is a necessary but not
sufficient condition for $\mathrm{U}(1)_A$ breaking. Notice also that
adding to this form of $\rho_{1,\,\mathrm{sing}}$ the one given in
Eq.~\eqref {eq:genU1abr12} one finds
$\lim_{\epsilon\to 0^+}\lim_{m\to
  0}\II_0(\epsilon;m)=\lim_{\epsilon\to 0^+}\lim_{m\to
  0}\II_1(\epsilon;m)\neq \lim_{m\to 0}\nn_1(\epsilon;m)$, showing
that limit commutativity does not require that
$2m^2\nn_1(\epsilon;m)\simeq \chit$.

\subsection{Remarks}
\label{sec:spdensrem}

The constraints Eqs.~\eqref{eq:firstorder_again_again} and
\eqref{eq:firstorder_again1_2_again} follow solely from the basic
requirement of chiral symmetry restoration at the level of scalar and
pseudoscalar susceptibilities [see Eq.~\exeqref{I-eq:susc_condition}].
The results in Sec.~\ref{sec:spdens_expl} provide simple functional
forms for the spectral density (even finite, including vanishing, at
$\lambda=0$) that comply with these constraints and lead to
$\Delta\neq 0$, breaking $\mathrm{U}(1)_A$ effectively in the chirally
symmetric phase.  The basic symmetry-restoration requirements are then
not sufficiently restrictive to exclude $\mathrm{U}(1)_A$ breaking,
under reasonably broad assumptions on the $\lambda$ dependence of
$\rho$.

However, if chiral symmetry remains manifest also in susceptibilities
involving non-local gauge operators, or in the presence of external
fermion fields [see Sec.~\exref{I-sec:gauge_ext} and
Appendix~\exref{I-sec:rhochirest}], then the ensuing
$m^2$-differentiability of the spectral density severely restricts the
possibilities. For a spectral density whose behavior near $\lambda=0$
is characterized by one or more power-law terms, as in
Eq.~\eqref{eq:spdens_gen1} (that generalizes the functional forms
considered in Refs.~\cite{Aoki:2012yj,Kanazawa:2015xna,
  Giordano:2024jnc}), the only way to obtain $\Delta\neq 0$ is by
means of a singular term
$\rho_{\mathrm{peak}}\simeq \f{\Delta}{2} m^2
\gamma(m^2)|\lambda|^{-1+\gamma(m^2)}$ with $m^2$-differentiable
$\gamma$, Eq.~\eqref{eq:rhom2diff_peak2}, tending to
$O(m^4)/|\lambda|$ in the chiral limit. A singular peak of this form
[as well as its generalization, Eq.~\eqref{eq:genU1abr12}] effectively
breaks $\mathrm{U}(1)_A$ in the chiral limit, while being compatible
with chiral symmetry restoration in the extended sense, and also not
in contradiction with commutativity of the thermodynamic and chiral
limits.

The functional forms of the spectral density parameterized by
Eq.~\eqref{eq:spdens_gen1} are of course far from exhausting all the
acceptable ones, and there is no reason to expect that
$\rho_{\mathrm{peak}}$ is the only mathematically allowed,
$\mathrm{U}(1)_A$-breaking functional form in the symmetric phase,
even assuming extended restoration --- it is indeed easy to produce
others [see Eqs.~\eqref{eq:genU1abr12} and
\eqref{eq:genU1abr19}]. Nonetheless, in order to obtain
$\Delta\neq 0$, under rather general, physically motivated conditions,
an $m^2$-differentiable $\rho$ must display some kind of singular
behavior near $\lambda=0$, that effectively leads to a term
$a_0 m^2 \delta(\lambda)$ in the chiral limit. Such a term may
originate from a singular near-zero peak,
$\lim_{\lambda\to 0}\rho(\lambda;m)=\infty$ for $m\neq 0$, but only if
the singularity tends to $1/|\lambda|$ in the chiral limit; or from a
spectral density finite at the origin, $\rho(0;m)< \infty$ for
$m\neq 0$, if this diverges in the chiral limit,
$\lim_{m\to 0}\rho(0;m)=\infty$. The appearance of such a term is,
however, a necessary but not sufficient condition that can be
satisfied in a variety of ways, that may or may not actually lead to
$\mathrm{U}(1)_A$ effective breaking, and may or may not be compatible
with commutativity of the thermodynamic and chiral limits, as the
examples discussed above in Sec.~\ref{sec:mdiffspdgen} show
explicitly.

The characterization of the functional form of the spectral density,
with its implications for the fate of $\mathrm{U}(1)_A$, should be
guided by physical rather than merely mathematical considerations, and
while the examples above are mathematically acceptable, whether or not
they are also physically plausible is a very different question.  A
near-zero peak has been observed in numerical simulations of lattice
QCD and pure $\mathrm{SU}(3)$ gauge
theory~\cite{Edwards:1999zm,HotQCD:2012vvd,Cossu:2013uua,Buchoff:2013nra,
  Dick:2015twa,Alexandru:2015fxa,Tomiya:2016jwr,Kovacs:2017uiz,
  Aoki:2020noz,Alexandru:2019gdm,Ding:2020xlj,Kaczmarek:2021ser,
  Vig:2021oyt,Kovacs:2021fwq,Meng:2023nxf,Kaczmarek:2023bxb,
  Alexandru:2024tel,JLQCD:2024xey,Fodor:2025yuj}, but its properties
and its fate in the chiral limit are still open questions. A clear
tendency towards a divergent peak in the thermodynamic limit is
observed when studying the spectrum of a chiral Dirac operator in the
background of gauge configurations obtained using non-chiral
fermions~\cite{Alexandru:2015fxa, Dick:2015twa,Alexandru:2019gdm,
  Kaczmarek:2021ser,Vig:2021oyt,Kovacs:2021fwq,Meng:2023nxf}. Of
course, the singular nature of the peak in this case could be an
artefact due to the use of a mixed action. Unfortunately, when using
instead a chiral Dirac operator also to produce the gauge
configurations~\cite{Cossu:2013uua,Buchoff:2013nra,Tomiya:2016jwr,
  Aoki:2020noz,JLQCD:2024xey,Fodor:2025yuj}, the limited precision due
to the inherent computational difficulties does not allow one yet to
distinguish a singular from a regular behavior at the origin. There
are, however, two theoretical proposals to explain the near-zero peak,
both suggesting a singular rather than a regular behavior.

References~\cite{Alexandru:2015fxa,Alexandru:2019gdm,Alexandru:2021pap,
  Alexandru:2021xoi,Alexandru:2023xho,Meng:2023nxf} proposed the
existence of a new phase of QCD, the ``IR phase'', appearing at some
temperature above the known chiral crossover
temperature~\cite{Aoki:2006we,Borsanyi:2010bp,Bazavov:2011nk,
  Bhattacharya:2014ara,Bazavov:2016uvm}, and being signalled by the
spectral density developing a singular behavior
$\rho\sim |\lambda|^\alpha$ with $\alpha<0$.\footnote{Note that
  partially quenched chiral perturbation theory predicts a logarithmic
  divergence in $\rho$ at $m\neq 0$ in the spontaneously broken
  phase~\cite{Osborn:1998qb,Damgaard:2008zs,Giusti:2008vb}.}  The
proposed IR phase is characterized by a partial restoration of
conformal symmetry, manifesting in particular in a behavior of the
spectral density close to $\rho\sim 1/|\lambda|$. According to
Refs.~\cite{Alexandru:2015fxa,Alexandru:2019gdm,Alexandru:2021pap,
  Alexandru:2021xoi,Alexandru:2023xho,Meng:2023nxf}, evidence for the
existence of the IR phase is provided by the near-zero spectral peak
observed in numerical simulations, and by the peculiar localization
properties of the near-zero modes~\cite{Alexandru:2021pap,
  Alexandru:2021xoi,Alexandru:2023xho,Meng:2023nxf}.  Notice, however,
that an exactly $1/|\lambda|$ behavior is strictly forbidden at any
$m$, as it gives a non-integrable singularity, and can only emerge
effectively in a limit where the ``height'' of the peak is at the same
time sufficiently suppressed, as is the case with
Eqs.~\eqref{eq:rhom2diff_peak2} and \eqref{eq:genU1abr12}.

References~\cite{Edwards:1999zm,HotQCD:2012vvd,Buchoff:2013nra,
  Dick:2015twa,Kovacs:2017uiz,Ding:2020xlj,Vig:2021oyt,
  Kaczmarek:2021ser,Kaczmarek:2023bxb,Kovacs:2023vzi,Fodor:2025yuj}
suggested instead that the spectral peak observed in numerical
simulations is of topological origin, emerging as the approximate zero
modes associated with local topological fluctuations mix and produce
nonzero modes that accumulate near
$\lambda=0$. Reference~\cite{Dick:2015twa} presented also direct
evidence of the relation between peak modes and topological objects in
the gauge-field configuration.  This mechanism was successfully tested
in simple matrix models based on a dilute gas of topological objects
in Ref.~\cite{Edwards:1999zm}, in the quenched case, and in
Ref.~\cite{Kovacs:2023vzi}, accounting for the effects of dynamical
fermions. These models produce a singular, power-law divergent
spectral peak, and in the unquenched case also important quantitative
features leading to effective $\mathrm{U}(1)_A$ breaking, such as the
scaling $\chit,\npeak \propto m^2$ of the topological susceptibility
and of the normalized mode number of the peak with the light-fermion
mass.  These features tie in nicely with the connection between
topology and a singular $\mathrm{U}(1)_A$-breaking peak implied by the
result
$\Delta=\lim_{m\to 0} \f{\npeak}{m^2} =\lim_{m\to 0}\f{\chit}{m^2}$,
see Eqs.~\eqref{eq:nasp9bis} and \eqref{eq:genU1abr_final}, and
support the possibility that the peak observed in numerical studies
eventually leads to the effective breaking of $\mathrm{U}(1)_A$ in the
symmetric phase. This is discussed in more detail in
Sec.~\ref{sec:u1breakingscenario}.

If the near-zero spectral peak observed in numerical simulations is
indeed produced by the topological mechanism outlined above, one
expects that it becomes power-law divergent [possibly up to
corrections as in Eq.~\eqref{eq:genU1abr12}] in the thermodynamic
limit. Moreover, the properties of the GW Dirac operator lead one to
similarly expect a (possibly singular) peak near $\lambda=2$, due to
the accumulation of doubler modes. The presence of this second peak
may discriminate between the instanton-gas origin and the
IR-phase/conformal symmetry origin of a singular near-zero peak, as
one would have no reason to expect the second peak in the latter
case.\footnote{The emergence of a near-zero peak of topological modes
  is contingent on the depletion of the low end of the spectrum due to
  the appearance of a mobility edge (see the discussion in
  Ref.~\cite{Bonanno:2025xuo}). The emergence of a peak near
  $\lambda=2$ would probably require the appearance of a mobility edge
  also in the ultraviolet region of the spectrum. Such a UV mobility
  edge has been observed in lower-dimensional pure gauge theories
  probed with staggered fermions~\cite{Baranka:2021san,
    Baranka:2022dib,Baranka:2024cuf}.}

A $\mathrm{U}(1)_A$-breaking spectral peak of topological origin would
become very hard to detect in numerical simulations as one approaches
the chiral limit. In order to have on average at least one mode in
the peak, one would need
\begin{equation}
  \label{eq:peak_numest1}
  1 \le \npeak \lvol \approx \Delta m^2 \lvol \approx \chit  \lvol\,,
\end{equation}
so a spatial volume $\svol$ at least of order
\begin{equation}
  \label{eq:peak_numest2}
  \svol \approx \f{\mathrm{T}}{\Delta m^2}  \approx \f{\mathrm{T}}{\chit}\,. 
\end{equation}
The sudden disappearance of the peak at a nonzero value of $m$
observed in
Refs.~\cite{Cossu:2013uua,Tomiya:2016jwr,Aoki:2020noz,JLQCD:2024xey}
could then be explained as a finite-volume effect: at the lowest
values of $m$ the peak is not yet fully formed on the available
lattice volumes. Such a peak would also be quickly suppressed and
would become hard to detect as the temperature increases, as $\chit$
is expected to approach zero with a high inverse power of $\mathrm{T}$
above $\mathrm{T}_c$.  For example, for gauge group $\mathrm{SU}(N_c)$
in the dilute instanton approximation one finds
$\chit \propto m^2 \mathrm{T}^{-\f{11 N_c}{3} +\f{10}{3} }$ for
$N_f=2$~\cite{Gross:1980br,Boccaletti:2020mxu}. Nonetheless, the
$\mathrm{U}(1)_A$-breaking effects of the peak, albeit small, would
persist to arbitrarily high temperature.

\section{Two-point function}
\label{sec:tpfunc}

I now discuss the consequences for the two-point function,
$\rho^{(2)}_c$, Eq.~\eqref{eq:spqrecap2}, of the constraints [see
Eqs.~\exeqref{I-eq:secondorder1_1} and \exeqref{I-eq:secondorder4}]
\begin{equation}
  \label{eq:tpf_constr1}
  \begin{aligned}
    -   4m^2\Id[\ff,\ff] &= \Delta'+ O(m^2)\,,    \\
    \Id[\ffh,\ffh] &=O(1)\,,
  \end{aligned}
\end{equation}
under additional technical assumptions. Here
\begin{equation}
  \label{eq:tpf_constr2}
  \begin{aligned}
    \Id[g_1,g_2]&= \int_0^2 d\lambda\int_0^2 d\lambda'\,  g_1(\lambda) g_2(\lambda')
                  \rho^{(2)}_{c}(\lambda,\lambda';m) \,, \\
    \Delta'&= \lim_{m\to 0}\lim_{\lvol\to\infty}
             \f{\la N_0\ra^2}{m^2 \lvol}\,,
  \end{aligned}
\end{equation}
and $\ffh(\lambda;m) = \ff(\lambda;m) - m^2\ff(\lambda;m)^2 $ [see
Eq.~\eqref{eq:firstorder_again_again2}]. Similarly to what was done
for the spectral density in Sec.~\ref{sec:detailed_spdens}, I will
assume that $\rho^{(2)}_c$ is an ordinary function, without any
delta-like singularity, for any $m\neq 0$. More precisely, in analogy
with Eq.~\eqref{eq:rho_distr_def} one defines $\rho_{c}^{(2)}$ as
\begin{equation}
  \label{eq:tpf_rig}
  \rho^{(2)}_{c}(\lambda,\lambda';m)
  = \de_\lambda \de_{\lambda'} \nnd(\lambda,\lambda';m)\,,
\end{equation}
where derivatives are in the sense of distributions, and where
\begin{equation}
  \label{eq:tpf_rig2}
  \begin{aligned}
    \nnd(\lambda,\lambda';m)
    &\equiv   \lim_{\lvol\to\infty}
      \f{\left\la\nn_U(\lambda)\nn_U(\lambda')\right\ra
      -\left\la\nn_U(\lambda)\right\ra \! \left\la\nn_U(\lambda')\right\ra }{\lvol} \\
    &\phantom{\equiv} -
      \mathrm{sgn}(\lambda \lambda') \nn(\min(|\lambda|,|\lambda'|);m)\,,
  \end{aligned}
\end{equation}
with $\nn_U(\lambda)$ and $\nn(\lambda;m)$ defined in
Eq.~\eqref{eq:rho_distr_def} [the mixed double derivative of the
second term produces the subtracted contact terms in
Eq.~\eqref{eq:spqrecap2}].  I assume that for any $m$,
$\nnd(\lambda,\lambda';m)$ is a differentiable function of both
$\lambda$ and $\lambda'$ (and so continuous and bounded) for
$\lambda,\lambda'\in [-2,2]$, with
$\de_{\lambda'} \nnd(\lambda,\lambda';m)$ continuous in $\lambda$ and
$\lambda'$, and differentiable in $\lambda$ everywhere except possibly
at $\lambda=\lambda'$, or when $\lambda=0$ or $\lambda'=0$ or
both. The existence of $\nnd(\lambda,\lambda';m)$ and
$\de_{\lambda'}\nnd(\lambda,\lambda';m)$ excludes the presence of
non-integrable singularities in $\rho^{(2)}_{c}$. Continuity of
$\de_{\lambda'} \nnd(\lambda,\lambda';m)$ in $\lambda$ prevents the
appearance of Dirac deltas in $\rho^{(2)}_{c}(\lambda,\lambda';m)$,
while integrable divergences are allowed at $\lambda=\lambda'$, or
when $\lambda$, $\lambda'$, or both vanish.  In particular, as $\nnd$
is an odd function of $\lambda$ and $\lambda'$, continuity requires
$\nnd(0,\lambda';m)= 0$ and
$\lim_{\lambda\to 0}\de_{\lambda'} \nnd(\lambda,\lambda';m)= 0$,
$\forall\lambda'\in[0,2]$. Since
$\nnd(\lambda',\lambda;m)= \nnd(\lambda,\lambda';m)$, the same applies
to $\de_{\lambda} \nnd(\lambda,\lambda';m)$ (with the roles of
$\lambda$ and $\lambda'$ interchanged).  Since
$\nnd(\lambda,\lambda';m)$ is bounded, $\forall m$, the contribution
to the integral $\Id[g_1,g_2]$ of the spectral region
$\delta\le \lambda,\lambda' \le 2$,
\begin{equation}
  \label{eq:bound_tpf_0}
  \begin{aligned}
    \Id_\delta[g_1,g_2]
    &   \equiv  \int_\delta^2 d\lambda\int_\delta^2 d\lambda'\,
      g_1(\lambda;m) g_2(\lambda';m)    \\
    &\hphantom{\equiv}
      \times  \rho^{(2)}_{c}(\lambda,\lambda';m)\,,
  \end{aligned}
\end{equation}
is finite in the chiral limit for any fixed $\delta> 0$, for any of
the relevant, $m$-dependent functions $g_{1,2}=\ff,\ffh$ (see Appendix
\ref{sec:twopointbound}). In particular, it can be ignored when
considering the chiral limit of $m^2\Id[g_1,g_2]$.  The same applies
also for a mass-dependent cutoff $\bar{\delta}(m)$ if
$\bar{\delta}(0)\neq 0$. Here and in the following argument, the
inequalities
\begin{equation}
  \label{eq:fineq}
  \begin{aligned}
    0\le \ff(\lambda;m) &\le \f{1}{\lambda^2 + m^2}\,, \\
    0\le \ffh(\lambda;m) &\le \ff(\lambda;m)\,,
  \end{aligned}
\end{equation}
are used.

\subsection{Finite two-point function}
\label{sec:tpfunc_finite}

The first consequence of the constraints Eq.~\eqref{eq:tpf_constr1} is
that if $\rho^{(2)}_c$ is finite at $\lambda=\lambda'=0$, then it must
vanish there in the chiral limit. Assume that
\begin{equation}
  \label{eq:ftpf1}
  \rho_c^{(2)}(\lambda,\lambda';m)=A(m)+B(\lambda,\lambda';m)\,, 
\end{equation}
with $B$ vanishing at the origin, and obeying the loose bound
\begin{equation}
  \label{eq:ftpf2}
  |B(\lambda,\lambda';m)|\le b (\lambda^2 +
  \lambda^{\prime\,2})^{\f{\beta}{2}} \,,
\end{equation}
for some $b>0$ and positive $\beta$, that can be taken without loss of
generality in the range $0<\beta< 1$.\footnote{If
  $|B(\lambda,\lambda';m)|\le \bar{b}\, (\lambda^2 +
  \lambda^{\prime\,2})^{\f{\bar{\beta}}{2}}$ with $\bar{b}>0$ and
  $\bar{\beta}\ge 1$, for any $0<\beta<1$ one has
  $\bar{\beta}-\beta> 0$, and so for $\lambda,\lambda'\in [0,2]$ one
  finds
  $ |B(\lambda,\lambda';m)|\le b \,(\lambda^2 +
  \lambda^{\prime\,2})^{\f{\beta}{2}}$ with
  $b=(2\sqrt{2})^{\bar{\beta}-\beta}\, \bar{b}$.}  One has
\begin{equation}
  \label{eq:ftpf3}
  \lim_{m\to 0} m^2 \Id[\ff,\ff]  = A(0)\left(\tf{\pi}{2}\right)^2 + \lim_{m\to 0} I_B(m)  \,,
\end{equation}
having used Eq.~\eqref{eq:ftpf3_bis}, and where
\begin{equation}
  \label{eq:IBinte}
  I_B(m)  \equiv m^2\int_0^2d\lambda \int_0^2 d\lambda'\,\ff(\lambda;m)\ff(\lambda';m)
  B(\lambda,\lambda';m)\,.
\end{equation}
Using the bound on $B$, the first inequality in Eq.~\eqref{eq:fineq},
and going over to polar coordinates, one finds
\begin{equation}
  \label{eq:ftpf4}
  \begin{aligned}
    &    | I_B(m)| 
      \le   bm^2 \int_0^2d\lambda \int_0^2 d\lambda'\,
      \f{\left(\lambda^2 + \lambda^{\prime\,2}\right)^{\f{\beta}{2}}}{\left(\lambda^2+m^2\right)\left(\lambda^{\prime\,2}+m^2\right)} \\ 
    &\le      bm^{\beta}\int_0^{\f{\pi}{2}}d\phi \int_0^\infty  dr\,
      \f{r^{1+\beta}}{ \left(\tf{1}{2}\sin\phi\right)^2r^4 +r^2+1} \\
    &    \le     bm^{\beta}\int_0^{\f{\pi}{2}}d\phi\left\{1 +
      \f{1}{\f{2}{\pi}\sin\f{\pi\beta}{2}}\left(\f{2}{\sin\phi}\right)^\beta \right\} \,,
  \end{aligned}
\end{equation}
where I made use of the following inequality,
\begin{equation}
  \label{eq:twop2_bis}
  \int_0^\infty dr\,\f{r^{1+\beta}}{x^2r^4+r^2+1} 
  \le 1 + x^{-\beta}\int_0^{\infty} dr\,\f{r^{1-\beta}}{r^2+1}\,,
\end{equation}
and of the results in Appendix~\ref{sec:inte} [see
Eqs.~\eqref{eq:inte1_again_app2} and \eqref{eq:nasp11_ter}]. The
integral over $\phi$ on the last line of Eq.~\eqref{eq:ftpf4} is
convergent, so $\lim_{m\to 0} I_B(m)=0$, and one finds
\begin{equation}
  \label{eq:ftpf5}
  \lim_{m\to 0} 4m^2 \Id[\ff,\ff] = \pi^2 A(0) = -\Delta'\,,
\end{equation}
having used the first constraint in Eq.~\eqref{eq:tpf_constr1} in the
last passage. [A result similar to Eq.~\eqref{eq:ftpf5} was reported
in Ref.~\cite{Kanazawa:2015xna}, Eq.~(4.29), where, however, the
contribution of zero modes was not fully taken into account.]  On the
other hand, using the finiteness of $ \Id[\ffh,\ffh]$ in the chiral
limit required by the second constraint in Eq.~\eqref{eq:tpf_constr1},
a similar calculation gives
\begin{equation}
  \label{eq:ftpf6}
  0=    \lim_{m\to 0} m^2 \Id[\ffh,\ffh]
  = A(0)\left(\tf{\pi}{4}\right)^2+ \lim_{m\to 0}\hat{I}_B(m) \,,
\end{equation}
having used Eq.~\eqref{eq:ftpf3_bis}, and where
\begin{equation}
  \label{eq:hatIBinte}
  \hat{I}_B(m) \equiv m^2\int_0^2d\lambda \int_0^2  d\lambda'\,
  \ffh(\lambda;m)\ffh(\lambda';m) B(\lambda,\lambda';m)\,.
\end{equation}
Using again the bound on $B$ and the inequalities in
Eq.~\eqref{eq:fineq}, one finds
\begin{equation}
  \label{eq:ftpf7}
  | \hat{I}_B(m)| \le
  bm^2 \int_0^2d\lambda \int_0^2  d\lambda'\,
  \f{\left(\lambda^2 + \lambda^{\prime\,2}\right)^{\f{\beta}{2}}}{\left(\lambda^2+m^2\right)\left(\lambda^{\prime\,2}+m^2\right)}\,,
\end{equation}
and so it follows from Eq.~\eqref{eq:ftpf4} that
$\lim_{m\to 0}\hat{I}_B(m)=0$, and therefore
\begin{equation}
  \label{eq:ftpf8}
  \lim_{m\to 0} m^2 \Id[\ffh,\ffh] = \f{\pi^2}{16}A(0) = 0\,.
\end{equation}
If in the symmetric phase the two-point function has a finite value
$A(m)$ at the origin for nonzero $m$, then this must vanish in the
chiral limit.

Combining Eqs.~\eqref{eq:ftpf5} and \eqref{eq:ftpf8} one finds that
the assumption of finiteness of $\rho^{(2)}_c$ at the origin implies
also that $\Delta'=0$. The quantity $\Delta'$ can be obtained from the
first moment of the probability distribution of a positive random
variable, i.e.,
\begin{equation}
  \label{eq:ftpf_extra2}
  \begin{aligned}
    \sqrt{\Delta'} &= \lim_{m\to 0}\lim_{\lvol\to\infty} \int_0^\infty dx\, P_{\lvol}(x;m) \,x \\
                   &  = \lim_{m\to 0} \int_0^\infty dx\, P_{\infty}(x;m) \,x \,,\\ 
    P_{\lvol}(x;m)&\equiv \left\la \delta\left(x-\f{N_0}{|m|\sqrt{\lvol}} \right)\right\ra\,.
  \end{aligned}
\end{equation}
A vanishing $\Delta'$ requires then that in the chiral limit the
measure $ P_{\infty}(x;m)$ vanishes almost everywhere in
$(0,\infty)$. This implies in turn that\footnote{More directly, in the
  large-volume limit the distribution of $Q$ is expected to be
  Gaussian, so $\Delta'=\f{2}{\pi}\Delta$, see
  footnote~\exref{I-foot:deltaprime}, and $\Delta$ vanishes if
  $\Delta'$ does.}
\begin{equation}
  \label{eq:ftpf_extra3}
  \Delta =  \lim_{m\to 0} \f{\chit}{m^2} = \lim_{m\to 0} \int_0^\infty dx\, P_{\infty}(x;m) \,x^2 = 0\,.
\end{equation}
In the symmetric phase, a two-point function regular at the origin for
$m\neq 0$ leads then to effective $\mathrm{U}(1)_A$ restoration in the
chiral limit, at least at the level of the simplest order parameter.

\subsection{Two-point function and localization}
\label{sec:tpfunc_loc}

Reversing the conclusion of the argument above, effective breaking of
$\mathrm{U}(1)_A$ by a nonzero $\Delta$ in the symmetric phase is
possible only if the two-point function is singular at the origin at
nonzero $m$. Such a behavior could in principle originate from a
strong repulsion between the lowest modes, or from a spectral density
divergent at the origin, or from both.

By itself, however, a divergent spectral density would not suffice to
yield $\Delta\neq 0$. Indeed, if the normalized two-point function
were bounded for all $\lambda,\lambda',m$, i.e.,
$ |\rho^{(2)}_c(\lambda,\lambda';m) |
/[\rho(\lambda;m)\rho(\lambda';m)] \le C $, one would find
$m^2\Id[\ff,\ff] \le C (m \chi_\pi / 4)^2$, that vanishes in the
chiral limit in the symmetric phase, independently of the behavior of
$\rho$.  By Eq.~\eqref{eq:tpf_constr1}, this would lead to $\Delta'=0$
and so, arguing as in Sec.~\ref{sec:tpfunc_finite}, to $\Delta =0$.
On the other hand, for systems of dense random matrices the two-point
function is known to have (integrable) singularities as
$\lambda\to\lambda'$ due to eigenvalue
repulsion~\cite{Ambjorn:1990ji,Brezin:1993qg,BEENAKKER1994515,
  Ambjorn:1996ga}, and so the bound above, while expected to hold for
well-separated $\lambda,\lambda'$, has no reason to hold in
general. This leaves open the possibility that $\Delta\neq 0$ due to
strong eigenvalue repulsion.

Strong repulsion between near-zero modes, however, seems at odds with
the numerical evidence accumulated in recent years, indicating that
low-lying Dirac modes are localized in the high-temperature phase of
QCD and other gauge theories (see Ref.~\cite{Giordano:2021qav} for a
review). I now argue that localization of the low modes closest to
zero is indeed incompatible with effective $\mathrm{U}(1)_A$ breaking
in the symmetric phase by $\Delta\neq 0$, independently of how the
spectral density behaves.

Localization is a well-known phenomenon in disordered condensed-matter
systems, where part of the spectrum of the Hamiltonian comprises
localized modes, whose size (averaged over disorder realizations) does
not grow with the system size~\cite{thouless1974electrons,
  lee1985disordered,kramer1993localization}. There is by now a large
amount of evidence showing that in gauge theories at high temperature
there is a critical point in the Dirac spectrum, the ``mobility edge''
$\lambda_c$, separating bulk modes delocalized over the whole system
from low modes localized on the scale of the inverse
temperature~\cite{Gockeler:2001hr,Gattringer:2001ia,
  GarciaGarcia:2005vj,GarciaGarcia:2006gr,Gavai:2008xe,Kovacs:2009zj,
  Kovacs:2010wx,Bruckmann:2011cc,
  Kovacs:2012zq,Giordano:2013taa,Nishigaki:2013uya, Ujfalusi:2015nha,
  Giordano:2015vla,Giordano:2016cjs,Giordano:2016vhx,
  Cossu:2016scb,Giordano:2016nuu,Kovacs:2017uiz,
  Holicki:2018sms,Giordano:2019pvc,Vig:2020pgq,Bonati:2020lal,
  Baranka:2021san,Kovacs:2021fwq,Cardinali:2021fpu,Baranka:2022dib,
  Kehr:2023wrs,Baranka:2023ani,Bonanno:2023mzj,Baranka:2024cuf,
  Bonanno:2025xuo}. The appearance of a mobility edge is ascribed to
the ordering of Polyakov loops at high
temperature~\cite{Bruckmann:2011cc,
  Giordano:2015vla,Giordano:2016cjs,Giordano:2016vhx,Giordano:2021qav,
  Baranka:2022dib,Bonanno:2025xuo}, and is therefore directly
connected to the confining properties of the theory. Indeed, in the
presence of a sharp deconfinement transition the mobility edge in the
Dirac spectrum appears precisely at the critical temperature in a
variety of models~\cite{Kovacs:2017uiz,Giordano:2019pvc,Vig:2020pgq,
  Bonati:2020lal,Baranka:2021san,Kovacs:2021fwq,Baranka:2022dib,
  Baranka:2024cuf,Bonanno:2025xuo}, including with dynamical
fermions~\cite{Giordano:2016nuu,Cardinali:2021fpu}.

Localization implies in turn that low modes fluctuate independently of
each other, and so the corresponding eigenvalues obey Poisson
statistics~\cite{altshuler1986repulsion}, with only weak repulsion
between them. Indeed, for a purely Poisson spectrum the two-point
function $\rho^{(2)}_{\mathrm{P}\, c}$ is known, and
reads~\cite{Guhr:1997ve,Kanazawa:2015xna}
\begin{equation}
  \label{eq:ltpf1}
  \rho_{\mathrm{P}\,c}^{(2)}(\lambda,\lambda') =
  -\f{1}{\mathrm{N}_{\mathrm{P}}}\rho_{\mathrm{P}}(\lambda)\rho_{\mathrm{P}}(\lambda')\,, 
\end{equation}
where $\mathrm{N}_{\mathrm{P}}\lvol$ is the total number of modes for
a system of size $\lvol$, and $\rho_{\mathrm{P}}$ is the spectral
density of the system. One is then led to expect that $\Delta=0$ if
localized modes extended all the way down to the origin also 
in the chiral limit.

To show this in detail, assume that modes are localized for
$0< \lambda < \lambda_c$. Based on Eq.~\eqref{eq:ltpf1}, one expects
that in the localized region of the spectrum the two-point function
obeys the bound
\begin{equation}
  \label{eq:ltpf2}
  |\rho_c^{(2)}(\lambda,\lambda';m)| \le C
  \rho(\lambda;m)\rho(\lambda';m)\,, 
\end{equation}
for some constant $C$, for $0<\lambda,\lambda'<\lambda_c$. The
correlation between localized modes and modes far beyond the mobility
edge is expected to obey a similar bound, but it is not clear
\textit{a priori} what happens as $\lambda$ and $\lambda'$ both
approach $\lambda_c$ from opposite sides. However, it was pointed out
in Ref.~\cite{Giordano:2024jnc} that since localized modes fluctuate
independently of each other, one expects that whatever correlation
they have with modes beyond the mobility edge, this will be
proportional to the spectral density of localized modes. One would
then expect that
\begin{equation}
  \label{eq:ltpf3}
  |\rho_c^{(2)}(\lambda,\lambda';m)| \le C' \rho(\lambda;m)\,,
\end{equation}
for some other constant $C'$, if $0<\lambda<\lambda_c<\lambda'$.
Taking into account more precisely how localized and delocalized modes
correlate with each other, one can actually argue for a sharper bound,
namely (see Appendix~\ref{sec:locdeloccorr})
\begin{equation}
  \label{eq:ltpf3_bis}
  |\rho_c^{(2)}(\lambda,\lambda';m)| \le C^{\prime\prime} \rho(\lambda;m)\rho(\lambda';m)\,,
\end{equation}
if $0<\lambda<\lambda_c<\lambda'$ or $0<\lambda'<\lambda_c<\lambda$,
i.e., a bound of the same form as Eq.~\eqref{eq:ltpf2}. This implies
Eq.~\eqref{eq:ltpf3} if the spectral density is bounded above
$\lambda_c$.  Since the bounds
Eqs.~\eqref{eq:ltpf2}--\eqref{eq:ltpf3_bis} originate simply in the
assumed localized nature of the eigenmodes below $\lambda_c$, one
expects that mass-independent, nonzero constants
$C,C', C^{\prime\prime}$ exist; this certainly applies if
$\lambda_c\not\to 0$.

The results of
Refs.~\cite{Baranka:2021san,Baranka:2022dib,Baranka:2024cuf} suggest
the presence of another mobility edge $\lambda_c'>\lambda_c$ in the
ultraviolet region of the spectrum, separating delocalized bulk modes
($\lambda_c<\lambda<\lambda_c'$) from localized high modes
($\lambda>\lambda_c'$).  Although this could be an artefact due to the
use of staggered fermions,\footnote{Notice, however, that since the
  mobility edge in units of the quark mass is a renormalization-group
  invariant~\cite{Kovacs:2012zq,Giordano:2022ghy}, the presence of a
  mobility edge should not depend on the type of discretization
  employed.} or specific to the models studied in those works, this
does not exclude that it is present also for GW fermions in realistic
theories. In this case, a bound of the form Eq.~\eqref{eq:ltpf2}
should apply for $\lambda,\lambda'>\lambda_c'$, as well as for
$0<\lambda<\lambda_c$, $\lambda'>\lambda_c'$ and $\lambda>\lambda_c'$,
$0<\lambda'<\lambda_c$, and a bound of the form
Eq.~\eqref{eq:ltpf3_bis} should apply for
$\lambda_c<\lambda<\lambda'_c$, $\lambda'>\lambda_c'$ and
$\lambda>\lambda_c'$, $\lambda_c<\lambda'<\lambda_c'$, possibly for
different values of the constants.

From the arguments above one concludes that if modes are localized
below $\lambda_c$, and if the mobility edge remains separated from the
origin in the chiral limit, $\lambda_c\not\to 0$, then one can
reasonably assume that
\begin{equation}
  \label{eq:ltpf2_gen}
  |\rho_c^{(2)}(\lambda,\lambda';m)| \le \bar{C}
  \rho(\lambda;m)\rho(\lambda';m)\,, 
\end{equation}
for a suitable mass-independent constant $\bar{C}$, as long as
$\lambda<\lambda_c$, or $\lambda'<\lambda_c$, or both. Using
Eqs.~\eqref{eq:ltpf2_gen} and \eqref{eq:bound_tpf_final} one finds
\begin{equation}
  \label{eq:ltpf4_gen}
    \begin{aligned}
      &      \lim_{m\to 0} \left| m^2 \Id[\ff,\ff]\right| \\
      &    \le   \lim_{m\to 0}m^2 \left\{\bar{C} \left[ \left( \Iu[\ff]\right)^2
        \!  - \! \left(\int_{\lambda_c}^2 d\lambda\,\ff(\lambda;m) \rho(\lambda;m)\right)^2\right] \right.\\
      & \left.  \hphantom{\le  \lim_{m\to 0}}
        \vphantom{\left( \int_{\lambda_c}^2 d\lambda\,\ff(\lambda;m) \rho(\lambda;m)\right)^2} +
        \left|\Id_{\lambda_c}[\ff,\ff] \right| \right\}  \\
      &   \le\bar{C}\lim_{m\to 0}  \left(m\Iu[\ff]  \right)^2
        = \bar{C}\lim_{m\to 0} \left( \f{m\chi_\pi}{4}\right)^2 =0\,,
  \end{aligned}
\end{equation}
since $\lim_{m\to 0} \chi_\pi<\infty$ in the symmetric
phase.\footnote{One reaches the same conclusion using the bound
  Eq.~\eqref{eq:ltpf3} for the correlation between localized modes in
  the region $0<\lambda<\lambda_c$ and modes above the mobility edge
  (independently of their localization properties), provided
  $\lambda_c$ does not vanish in the chiral
  limit~\cite{Giordano:2024jnc}.} One has then 
$\lim_{m\to 0} m^2 \Id[\ff,\ff]=0$, and so, by
Eq.~\eqref{eq:tpf_constr1}, that $\Delta'=0$, and so $\Delta=0$.
Under the same assumptions, using the second inequality in
Eq.~\eqref{eq:fineq}, one finds by means of a similar calculation
\begin{equation}
  \label{eq:ltpf6_gen}
    \lim_{m\to 0} \left|\Id[\ffh,\ffh]\right| 
  \le \lim_{m\to 0}  \left(\f{\bar{C}\chi_\pi^2}{16}
    +  \left|\Id_{\lambda_c}[\ffh,\ffh]  \right| \right) <\infty\,,
\end{equation}
so that the second constraint in Eq.~\eqref{eq:tpf_constr1} is
satisfied.\footnote{\label{foot:lcz}Strictly speaking, this proves
  finiteness of $\limsup_{m\to 0}\Id[\ffh,\ffh]$ and
  $\liminf_{m\to 0}\Id[\ffh,\ffh]$, but not their equality. Since
  $|\Id_{\lambda_c}[g,g] |\le \mathrm{const.}/\lambda_c^4$ for
  $g=\ff,\ffh$ [see Eq.~\eqref{eq:bound_tpf_6}],
  Eq.~\eqref{eq:ltpf4_gen} holds also if $\lambda_c\to 0$ in the
  chiral limit, as long as $m/\lambda_c^2 \to 0$ and the bound
  Eq.~\eqref{eq:ltpf2_gen} applies, while in this case
  Eq.~\eqref{eq:ltpf6_gen} could not be proved without making further
  assumptions. However, for the purposes of the present argument it is
  immaterial whether or not the bound Eq.~\eqref{eq:ltpf2_gen} implies
  that the second constraint in Eq.~\eqref{eq:tpf_constr1} is
  satisfied; in particular, $m/\lambda_c^2 \to 0$ and
  Eq.~\eqref{eq:ltpf2_gen} suffice to show that $\Delta=0$.}

One concludes that $\mathrm{U}(1)_A$ is effectively restored (at this
level) in the symmetric phase if the near-zero modes are localized and
remain so in the chiral limit. Effective $\mathrm{U}(1)_A$ breaking by
$\Delta\neq 0$ requires then that either $\lambda_c\to 0$ in the
chiral limit, or that another mobility edge, $\bar{\lambda}_c$, be
present near the origin (at least for small light-fermion mass),
separating localized low modes ($\bar{\lambda}_c<\lambda<\lambda_c$)
from delocalized near-zero modes
($0<\lambda<\bar{\lambda}_c$).\footnote{Another possibility would be
  that near-zero modes are ``critical'', i.e., their size grows with
  the system size but more slowly than $\svol$. An extended region of
  critical modes beyond the mobility edge has been observed in
  spatially two-dimensional disordered
  systems~\cite{xie1998kosterlitz,PhysRevB.60.5295,
    Wen-ShengLiu_1999,Giordano:2019pvc,Baranka:2024cuf}, but I do not
  know of any examples in three spatial dimensions.}  This conclusion
holds independently of the near-zero behavior of the spectral density
at $m\neq 0$. (Note that both a singular and a regular behavior at
$\lambda=0$ can lead to $\Delta\neq 0$ in the chiral limit, even if
$\rho$ is $m^2$-differentiable, see Secs.~\ref{sec:chipi} and
\ref{sec:mdiffspdgen}.)  The spectral statistical properties of
delocalized modes are of the same type observed in dense random
matrices~\cite{mehta2004random, Pastur2011,Verbaarschot:2000dy}, that
are largely
universal~\cite{Ambjorn:1990ji,Brezin:1993qg,BEENAKKER1994515,Ambjorn:1996ga,
  erdos2012,taovu2011}. One then expects in this case a normalized
two-point function that is singular for
$\lambda\to\lambda'$~\cite{Ambjorn:1990ji,Brezin:1993qg,BEENAKKER1994515},
thus making $\mathrm{U}(1)_A$ breaking possible.

\subsection{Remarks}
\label{sec:tpfrem}

The results above show that effective $\mathrm{U}(1)_A$ breaking in
the symmetric phase by $\Delta\neq 0$ is possible only if near-zero
modes strongly repel each other. This requires that they be not
localized (at least for small $m$); or if they are localized, that the
mobility edge approach zero as $m\to 0$ (sufficiently fast, see
footnote~\ref{foot:lcz}). The second possibility is disfavored if
$\Delta\neq 0$ due to a singular spectral peak at $\lambda=0$, as in
Eqs.~\eqref{eq:rhom2diff_peak2} or \eqref{eq:genU1abr12} [or a regular
peak whose height $\propto |m|/\varepsilon(m)$ diverges in the chiral
limit, Eq.~\eqref{eq:genU1abr19} for $\xi=0$ and
$\varepsilon(m)=o(|m|)$]: For spatial dimension higher than two, one
expects that all modes can be localized in a dense spectral region
only in the presence of strong
disorder~\cite{lee1985disordered,kramer1993localization}, which cannot
occur in a lattice gauge theory as the link variables take values in a
compact group. The expected delocalization of the peak modes is easy
to understand if these originate in the zero modes associated with
local topological fluctuations, as discussed in
Sec.~\ref{sec:spdensrem}. In this case the localized ``unperturbed'',
exact zero modes associated with isolated topological objects would
easily mix when accounting for the effects of the full gauge-field
configuration thanks to their degeneracy, even when the corresponding
objects are quite far from each other, leading to delocalized exact
eigenmodes near zero. Since a region of localized low modes is known
to exist in high-temperature gauge theories~\cite{Giordano:2021qav},
in this case one would expect to find another mobility edge,
$\bar{\lambda}_c$, near the origin. In the presence of a spectral peak
of topological origin effectively breaking $\mathrm{U}(1)_A$, one
expects then that modes are delocalized for
$0<\lambda<\bar{\lambda}_c$, localized for
$\bar{\lambda}_c<\lambda<\lambda_c$, and again delocalized above
$\lambda_c$ (possibly only up to another mobility edge,
$\lambda_c'$~\cite{Baranka:2021san,Baranka:2022dib,Baranka:2024cuf}).

The mechanisms leading to $\bar{\lambda}_c$ and $\lambda_c$ in the
scenario outlined above would be quite different from each other.  The
lower one, $\bar{\lambda}_c$, would presumably appear within the
near-zero peak, marking the point in the spectrum where it ceases to
be ``energetically'' convenient for a topological mode to hop on
topological objects all over the system, remaining essentially
confined to a subset of them. The higher one, $\lambda_c$, would
instead be the mobility edge generally expected in the
high-temperature phase of a gauge theory, already observed in a
variety of systems~\cite{Gockeler:2001hr,Gattringer:2001ia,
  GarciaGarcia:2005vj,GarciaGarcia:2006gr,Gavai:2008xe,Kovacs:2009zj,
  Kovacs:2010wx,Bruckmann:2011cc,
  Kovacs:2012zq,Giordano:2013taa,Nishigaki:2013uya, Ujfalusi:2015nha,
  Giordano:2015vla,Giordano:2016cjs,Giordano:2016vhx,
  Cossu:2016scb,Giordano:2016nuu,Kovacs:2017uiz,
  Holicki:2018sms,Giordano:2019pvc,Vig:2020pgq,Bonati:2020lal,
  Baranka:2021san,Kovacs:2021fwq,Cardinali:2021fpu,Baranka:2022dib,
  Kehr:2023wrs,Baranka:2023ani,Bonanno:2023mzj,Baranka:2024cuf,
  Bonanno:2025xuo,Giordano:2021qav}, and found far above the near-zero
peak when this is present~\cite{Alexandru:2021pap,
  Alexandru:2021xoi,Kovacs:2017uiz,Kovacs:2021fwq}.  This mobility
edge is driven by the ordering of the Polyakov loop and the resulting
depletion of the low-lying spectrum~\cite{Bruckmann:2011cc,
  Giordano:2015vla,Giordano:2016cjs,Giordano:2016vhx,
  Giordano:2021qav,Baranka:2022dib,Bonanno:2025xuo}. On the one hand,
this depletion makes it possible for the peak to emerge. On the other,
it allows Dirac modes to localize on suitable gauge-field fluctuations
in the intermediate, low-density region between the peak and the bulk
of the spectrum, as it typically happens at the edge of the spectrum
in disordered systems, where the spectral density is
low~\cite{thouless1974electrons,lee1985disordered,
  kramer1993localization}. As the Polyakov loop in the
high-temperature phase remains ordered also in the chiral
limit~\cite{Clarke:2020htu}, one expects that $\lambda_c$ remains
separated from zero even as $m\to 0$.

For a spectrum as just described, whether or not $\bar{\lambda}_c$
tends to zero in the chiral limit is not determined at this stage.
Assuming, as above, that Eq.~\eqref{eq:ltpf2_gen} holds if
$\bar{\lambda}_c<\lambda<\lambda_c$ or
$\bar{\lambda}_c<\lambda'<\lambda_c$ or both, and that
$\lambda_c\not\to 0$, and moreover that the normalized two-point
function is bounded if $\lambda$ and $\lambda'$ are well
separated,\footnote{It suffices to assume that
  $|\rho^{(2)}_c(\lambda,\lambda';m)|\le
  \hat{C}\rho(\lambda;m)\rho(\lambda';m)$, for some $m$-independent
  $\hat{C}$, for $\lambda<\bar{\lambda}_c< \lambda_c<\lambda'$ or
  $\lambda'<\bar{\lambda}_c< \lambda_c<\lambda$; or at least that
  $|\rho^{(2)}_c(\lambda,\lambda';m)|\le \hat{C}'\rho(\lambda;m)$, for
  some $m$-independent $\hat{C}'$, if
  $\lambda<\bar{\lambda}_c<\lambda_c<\lambda'$.}  one finds from the
first constraint in Eq.~\eqref{eq:tpf_constr1}
\begin{equation}
  \label{eq:barlcchiral2}
  \begin{aligned}
    - \Delta'
    & = \lim_{m\to 0} 4m^2 \int_0^{\bar{\lambda}_c}d\lambda\int_0^{\bar{\lambda}_c}d\lambda'\,\ff(\lambda;m) \ff(\lambda';m) \\
    &\hphantom{= \lim_{m\to 0}4m^2}\times  \rho^{(2)}_c(\lambda,\lambda';m)\,.
  \end{aligned}
\end{equation}
The right-hand side may or may not vanish both if $\bar{\lambda}_c$
remains nonzero or vanishes in the chiral limit, depending on the
specific form of $\rho^{(2)}_c$ and on how $\bar{\lambda}_c$ scales
with $m$. The second constraint in Eq.~\eqref{eq:tpf_constr1}
requires instead
\begin{equation}
  \label{eq:barlcchiral2_bis}
    \int_0^{\bar{\lambda}_c}d\lambda\int_0^{\bar{\lambda}_c}d\lambda'\,\ffh(\lambda;m)\ffh(\lambda';m)
    \rho^{(2)}_c(\lambda,\lambda';m) =  O(1)\,,
\end{equation}
with the other contributions guaranteed to be $O(1)$ under the stated
assumptions.

Notice that if the lower mobility edge were exactly at the origin,
$\bar{\lambda}_c\equiv 0$ (at least for small $m$), as proposed in
Refs.~\cite{Alexandru:2021pap,Alexandru:2021xoi,Alexandru:2023xho,
  Meng:2023nxf}, then one could ignore it entirely in this context, as
one can always exclude the zero-measure lines $\lambda=0$ and
$\lambda'=0$ from the integral defining $\Id[g_1,g_2]$ without any
effect (having assumed that $\rho^{(2)}_c$ is an ordinary
function). In this case Eq.~\eqref{eq:ltpf4_gen} would still apply,
and one would find $\Delta=0$ in the chiral limit (assuming
$\lambda_c\not\to 0$).
  
As already mentioned in Sec.~\ref{sec:spdensrem}, a near-zero spectral
peak has been observed in numerical simulations of lattice QCD and
pure $\mathrm{SU}(3)$ gauge theory in the high-temperature
phase~\cite{Edwards:1999zm,Cossu:2013uua,Buchoff:2013nra,Dick:2015twa,
  Alexandru:2015fxa,Tomiya:2016jwr,Kovacs:2017uiz,Aoki:2020noz,
  Alexandru:2021pap,Alexandru:2021xoi,Alexandru:2019gdm,Ding:2020xlj,
  Kaczmarek:2021ser,Vig:2021oyt,Kovacs:2021fwq,Meng:2023nxf,
  Kaczmarek:2023bxb,Alexandru:2024tel,Fodor:2025yuj}, with mixing of
the zero modes associated with localized topological objects as a
viable explanation for its appearance~\cite{Edwards:1999zm,
  HotQCD:2012vvd,Buchoff:2013nra,Dick:2015twa,Kovacs:2017uiz,
  Ding:2020xlj,Vig:2021oyt,Kaczmarek:2021ser,Kaczmarek:2023bxb,
  Kovacs:2023vzi,Fodor:2025yuj}. This leads one to expect the presence
of two mobility edges in the low-lying Dirac spectrum of these
theories if $\mathrm{U}(1)_A$ remains effectively broken in the chiral
limit.  There are numerical results indicating the existence of a
near-zero region of delocalized modes in QCD with physical and
lower-than-physical quark masses, supporting this scenario.
Non-Poissonian repulsion of the lowest modes toward the chiral limit
has been observed, although with staggered fermions, in
Ref.~\cite{Ding:2020xlj}. Direct evidence for the presence of two
mobility edges in the low-lying Dirac spectrum in QCD at physical
quark masses is provided by Ref.~\cite{Meng:2023nxf}, that finds at
$\mathrm{T}=187\,\mathrm{MeV}$ an ``infrared
dimension''~\cite{Horvath:2018aap,Alexandru:2021pap}
$d_{\mathrm{IR}}(0^+)\approx 3$ for a small but finite range of
near-zero modes of the overlap operator (computed on configurations
obtained using improved Wilson fermions).  This shows their full
spatial delocalization; since localized modes are found higher up in
the low-lying spectrum, it also indicates the presence of a mobility
edge near zero, and at a finite distance from
it.\footnote{Reference~\cite{Meng:2023nxf} reports also that
  $d_{\mathrm{IR}}(0^+)\approx 2$ at $\mathrm{T}=234\,\mathrm{MeV}$,
  which the authors find indicative of a transition to the IR
  phase~\cite{Alexandru:2015fxa,Alexandru:2019gdm,Alexandru:2021pap,
    Alexandru:2021xoi,Alexandru:2023xho,Meng:2023nxf} (see discussion
  in Sec.~\ref{sec:spdensrem}) at some critical temperature
  $187\,\mathrm{MeV}< \mathrm{T}_{\mathrm{IR}} < 234\,\mathrm{MeV}$.
  If the peak is a $\mathrm{U}(1)_A$-breaking singular peak of
  topological origin, this could be a finite-volume artefact due to
  its expected suppression with temperature, caused by the suppression
  of $\chit$ (see Sec.~\ref{sec:spdensrem}). In fact, the very large
  volumes required for the full development of the peak [see
  Eq.~\eqref{eq:peak_numest2}] are likely required also for the
  stabilization of the localization properties of the corresponding
  eigenvectors, with larger volumes required at higher temperatures.}

As a final comment, notice that the strong restrictions imposed on the
spectral density and on the two-point function if $\Delta\neq 0$,
derived in this section and in the previous one, depend majorly on the
assumption that $\rho$ and $\rho_{c}^{(2)}$ are ordinary functions, in
particular without Dirac deltas at the origin of the spectrum. In the
presence of such singularities, it is easy to obtain
$\mathrm{U}(1)_A$-breaking contributions yielding $\Delta\neq 0$
(e.g., $\rho_{\mathrm{sing}}$ mentioned in Sec.~\ref{sec:ctc}), and
one would not be able to conclude much in terms of restrictions.

\section{A scenario for
  \texorpdfstring{$\mathrm{U}(1)_A$}{U(1)\textunderscore A} breaking}
\label{sec:u1breakingscenario}

Summarizing the findings of the previous sections, effective breaking
of $\mathrm{U}(1)_A$ in the chiral limit by a nonzero $\Delta$ is
compatible with chiral symmetry restoration (in the extended sense),
but only if a rather demanding list of requirements is fulfilled.

As shown in Sec.~\exref{I-sec:top_iig} (and previously in
Ref.~\cite{Kanazawa:2014cua}), if $\mathrm{U}(1)_A$ is effectively
broken by $\Delta\neq 0$, chiral symmetry restoration requires that
the cumulants of the topological charge distribution be identical, to
leading order in $m$, to those found in an ideal gas of topological
objects of charge $\pm 1$, of equal densities $\chit/2$ with
$\chit=\Delta m^2 + O(m^4)$. These objects need not be the usual
instantons and anti-instantons (or, more precisely, their
finite-temperature analogs, i.e., calorons and
anti-calorons~\cite{Harrington:1978ve,Harrington:1978ua,Kraan:1998kp,
  Kraan:1998pm,Kraan:1998sn,Lee:1997vp,Lee:1998vu,Lee:1998bb,
  GarciaPerez:1999ux,Chernodub:1999wg,Diakonov:1995ea,Schafer:1996wv,
  Diakonov:2009jq}), but only effective topological degrees of freedom
fluctuating independently of each other.  Similarly, the required
instanton gas-like behavior need not be that of the usual
semiclassical dilute instanton gas~\cite{Gross:1980br,
  Boccaletti:2020mxu}.

Next, as shown in Sec.~\ref{sec:tpfunc}, $\Delta\neq 0$ is
incompatible with a two-point eigenvalue correlation function finite
at the origin, and with localized modes in the immediate vicinity of
$\lambda=0$.

Finally, as discussed in Sec.~\ref{sec:mdiffspdgen}, if chiral
symmetry is restored in the extended sense (or more generally if
$m^2$-differentiability applies), $\Delta\neq 0$ requires that the
spectral density effectively develops a term
$\propto m^2 \delta(\lambda)$ in the chiral limit, under rather
general assumptions. This can be achieved in a variety of ways, in
particular by a singular near-zero peak tending to $O(m^4)/|\lambda|$
in the chiral limit [see Eqs.~\eqref{eq:rhom2diff_peak2} and
\eqref{eq:genU1abr12}]. In this case the required divergent two-point
eigenvalue correlation function and delocalization of near-zero modes
are most likely due to the presence of another mobility edge near
$\lambda=0$, distinct from the well-established one already observed
in high-temperature gauge theories~\cite{Giordano:2021qav}.

Although the features listed above are mathematically consistent, one
would need a concrete physical mechanism implementing them to make
effective $\mathrm{U}(1)_A$ breaking a realistic possibility.
Expanding on previous comments in Secs.~\ref{sec:spdensrem} and
\ref{sec:tpfrem}, I now argue that such a mechanism is provided by the
mixing of the approximate zero modes associated with a dilute gas of
topological excitations. This was previously proposed in
Refs.~\cite{Edwards:1999zm,HotQCD:2012vvd,Buchoff:2013nra,
  Dick:2015twa,Kovacs:2017uiz,Ding:2020xlj,Vig:2021oyt,
  Kaczmarek:2021ser,Kaczmarek:2023bxb} as a qualitative explanation
for the near-zero spectral peak observed in numerical simulations, and
underlies the instanton-gas model developed in
Ref.~\cite{Kovacs:2023vzi}, on which the following discussion is
based, that describes it in a more quantitative fashion.

The model of Ref.~\cite{Kovacs:2023vzi} assumes that the zero modes
and the near-zero part of the Dirac spectrum can be described in terms
of the mixing of the exact chiral zero modes associated with isolated
topological objects of charge $Q=\pm 1$. In the basis of these zero
modes, the (continuum) Dirac operator $\slashed{D}$ has a block
diagonal structure, with nonzero matrix elements
$\slashed{D}_{\imath\bar{\imath}}$ and
$\slashed{D}_{\bar{\imath}\imath}= -
\slashed{D}_{\imath\bar{\imath}}^*$ only between modes associated with
objects $\imath$ and $\bar{\imath}$ of opposite charge, as dictated by
the chiral property $\{\gamma_5,\slashed{D}\}=0$. These matrix
elements are exponentially suppressed with the distance between the
oppositely charged objects, due to the expected localized nature of
the associated zero modes at finite temperature. The partition
function is then defined as
\begin{equation}
  \label{eq:model_Z}
  Z_{\text{{\protect\NoHyper  Ref.~\cite{Kovacs:2023vzi}}\protect\endNoHyper}}
  =\sum_{n,\bar{n}}p_np_{\bar{n}}
  \int d^{3n}x \int d^{3\bar{n}}\bar{x}\left[\det(\slashed{D}+m)\right]^2\,,
\end{equation}
where $n$ and $\bar{n}$ are the numbers of ``instantons'' and
``anti-instantons'' in a configuration, and the integral is over their
positions in a finite three-dimensional box.  (Here physical units are
used.) In the absence of interactions, that are mediated by the
determinant of the massive Dirac operator, the distributions $p_n$ and
$p_{\bar{n}}$ of topological objects are taken to be identical
independent Poisson distributions, motivated by the numerical results
obtained at high temperature in pure $\mathrm{SU}(3)$ gauge
theory~\cite{Bonati:2013tt}. Finally, the entry
$\slashed{D}_{\imath \bar{\imath}}$ corresponding to instanton
$\imath$ and anti-instanton $\bar{\imath}$ is taken of the form
\begin{equation}
  \label{eq:model_Z2}
  \slashed{D}_{\imath\bar{\imath}}= i ce^{-\pi \mathrm{T}| x_{\imath}-\bar{x}_{\bar{\imath}}|}\,,
\end{equation}
where $c\in\mathbb{R}$ sets the overall scale of the matrix element,
and $1/(\pi\mathrm{T})$ sets the localization scale of the zero modes.

The important features of this model, reported in
Ref.~\cite{Kovacs:2023vzi}, are the following.  (i.)\ After
interactions are switched on, the topological objects arrange into a
gas of instanton-anti-instanton molecules, plus a ``free-gas''
component of unpaired instantons and anti-instantons of density
$\nfree \propto m^2$, essentially non-interacting and
Poisson-distributed, that entirely determine the topological charge of
the configuration. (ii.)\ The topological susceptibility equals
$\f{\la (n-\bar{n})^2\ra}{\svol/\mathrm{T}}=\nfree$, and is
proportional to $m^2$ for small $m$. (iii.)\ Mixing of the zero modes
associated with the objects in the free-gas component leads to a
singular power-law near-zero peak in the spectral density of
$\slashed{D}$, with mass-dependent negative exponent $\alpha(m)$, and
with a number of modes per unit volume, $\npeak$ [corresponding to the
normalized mode number in Eq.~\eqref{eq:rhom2diff_peakdens}], matching
the density of the free-gas component, $\npeak\approx \nfree $.

The features (i.)--(iii.)\ fulfill almost completely the requirements
for $\mathrm{U}(1)_A$ breaking listed above. In
Ref.~\cite{Kovacs:2023vzi} it was not checked whether $\alpha\to -1$
in the chiral limit, and if near-zero modes are delocalized, but both
these crucial features are highly plausible. In fact, the model of
Ref.~\cite{Kovacs:2023vzi} is similar to a model of disordered
condensed-matter systems with chiral symmetry, discussed in
Refs.~\cite{S_N_Evangelou_2003,PhysRevB.74.113101}, where these
features have been demonstrated. This is a tight-binding model on a
bipartite cubic lattice with purely off-diagonal, uncorrelated,
nearest-neighbor hopping disorder, with positive hopping coefficients
$t_{ij}$ ranging in an exponentially wide interval, and with
$\ln t_{ij} \in \left[-\f{W}{2},\f{W}{2}\right]$ distributed
uniformly. The model of Ref.~\cite{Kovacs:2023vzi} is exactly of the
same type, with purely off-diagonal (although not only
nearest-neighbor, and not uncorrelated) disorder ranging over an
exponentially wide range.\footnote{I thank T.G.~Kov\'acs for pointing
  this out.} The analog of the disorder parameter $W$ is the mean free
path between topological objects, that sets the scale at which the
distribution of the distance between unpaired instantons and
anti-instantons is effectively cut off---it is very unlikely to find
objects whose nearest unpaired neighbor of opposite charge is much
farther than a few mean free paths. In turn, the mean free path is
inversely related to the density of objects, $\nfree\approx \chit$,
with much larger fluctuations allowed when the density decreases. The
amount of analog disorder in the system is then controlled by
$1/\chit \propto 1/m^2$.

In Ref.~\cite{S_N_Evangelou_2003} it was shown that the spectral
density of the tight-binding model displays a singular near-zero peak,
with exponent tending to $-1$ in the limit of large disorder,
$W\to \infty$. One then expects that $\alpha(m)\to -1$ as $m\to 0$ in
the model of Ref.~\cite{Kovacs:2023vzi}. In
Ref.~\cite{PhysRevB.74.113101} it was established that the
tight-binding model has a near-zero mobility edge, separating
delocalized near-zero modes from localized modes higher up in the
spectrum. This happens also when random phase factors are included in
the hopping terms, changing the symmetry class of the system from
chiral orthogonal to chiral
unitary~\cite{thouless1974electrons,lee1985disordered,
  kramer1993localization,Verbaarschot:2000dy}. Moreover, the mobility
edge gets closer to the origin as the disorder increases. One then
expects the presence of a similar mobility edge also in the model of
Ref.~\cite{Kovacs:2023vzi}.\footnote{A near-zero region of delocalized
  modes was found in Ref.~\cite{cain1999off} in the chiral orthogonal
  tight-binding model also for uniformly distributed nearest-neighbor
  hopping disorder, $t_{ij}\in [-\f{W}{2},\f{W}{2}]$.  A near-zero
  region of delocalized modes and a mobility edge getting closer to
  zero as the disorder increases were observed also in
  Ref.~\cite{Takaishi_2018}, in a tight-binding model in the chiral
  unitary class with correlated nearest-neighbor hopping disorder
  determined by the spin fluctuations in a separate spin model. The
  results of Refs.~\cite{PhysRevB.74.113101,Takaishi_2018} were
  reported incorrectly in Ref.~\cite{Giordano:2021qav}, \S 4.3.}

The results of Ref.~\cite{Kovacs:2023vzi}, together with those of
Refs.~\cite{S_N_Evangelou_2003,PhysRevB.74.113101}, provide strong
evidence that the model defined by Eq.~\eqref{eq:model_Z} fulfills all
the requirements for $\mathrm{U}(1)_A$ breaking listed above, thus
providing a concrete physical mechanism for effective
$\mathrm{U}(1)_A$ breaking in the symmetric phase.  Clearly, the
existence of such a mechanism is no guarantee that it is actually at
play in QCD or QCD-like gauge theories. Indications of a singular
spectral peak in QCD~\cite{Edwards:1999zm,HotQCD:2012vvd,
  Cossu:2013uua,Buchoff:2013nra,Dick:2015twa,Alexandru:2015fxa,
  Tomiya:2016jwr,Kovacs:2017uiz,Aoki:2020noz,
  Alexandru:2019gdm,Ding:2020xlj,Kaczmarek:2021ser,Vig:2021oyt,
  Kovacs:2021fwq,Meng:2023nxf,Kaczmarek:2023bxb,Alexandru:2024tel,
  JLQCD:2024xey,Fodor:2025yuj} have been mentioned repeatedly, and
while there is not yet a consensus about its persistence as one lowers
the fermion mass~\cite{Tomiya:2016jwr, Aoki:2020noz}, numerical
results are consistent with it. Even if confirmed, the persistence of
the peak is by itself not sufficient to break $\mathrm{U}(1)_A$, and
the specific features listed above (exponent tending to $-1$,
normalized mode number equal to $m^2\Delta$, delocalized near-zero
modes) have to be present. However, a sufficiently detailed
characterization of the properties of the peak, in particular of its
exponent and of its mass dependence, is still lacking.

Concerning topological aspects, there are indications of a gas-like
behavior of the topological charge in QCD in the symmetric phase, at
least for sufficiently high
temperatures~\cite{Bonati:2015vqz,Bonati:2018blm,Athenodorou:2022aay,
  Bonanno:2024zyn}.  Although the ideal-gas behavior must emerge only
in the chiral limit if $\mathrm{U}(1)_A$ is effectively broken by
$\Delta\neq 0$, the smallness of the $u$ and $d$ quark masses leads
one to expect that in this case a near-ideal behavior should manifest
not far above the pseudocritical temperature, $\mathrm{T}_c$.
Instead, a clear deviation from the ideal-gas behavior was observed
up to $2\mathrm{T}_c$ in Ref.~\cite{Bonati:2015vqz}. However, in the
light of the revised results obtained after algorithmic improvements
in Refs.~\cite{Bonati:2018blm,Athenodorou:2022aay,Bonanno:2024zyn},
this deviation could simply be a finite-volume or finite-spacing
artefact. On the other hand, it could be a genuine effect of
finite-$m$ corrections to the ideal-gas behavior [see
Eq.~\exeqref{I-eq:topconst10_quinquies}] being larger than expected
near $\mathrm{T}_c$. Of course, $\mathrm{U}(1)_A$ may as well be
effectively restored in the chiral limit (or effectively broken but
with $\Delta= 0$) and no ideal-gas behavior would be expected, except
at very high temperature, and for a different
reason~\cite{Gross:1980br,Boccaletti:2020mxu}.

Finally, as already mentioned in Sec.~\ref{sec:tpfrem}, the presence
of a near-zero mobility edge in high-temperature QCD (for physical
quark masses) is supported by the results of
Ref.~\cite{Meng:2023nxf}. This mobility edge, and in particular its
dependence on $m$, should be further studied in detail. If
$\mathrm{U}(1)_A$ is effectively broken in the chiral limit of
high-temperature QCD by the mechanism proposed in
Ref.~\cite{Kovacs:2023vzi}, the results of
Ref.~\cite{PhysRevB.74.113101} suggest that the near-zero mobility
edge should decrease as $m\to 0$, possibly tending to zero.

To summarize, a viable scenario for effective $\mathrm{U}(1)_A$
breaking in the symmetric phase of a (topologically nontrivial) gauge
theory is the formation of an ideal gas of topological objects in
typical gauge configurations, leading directly to an ideal-gas
behavior of the cumulants of the topological charge distribution, and
to a singular spectral peak and a near-zero mobility edge through the
mixing of the associated zero modes. The quantitative aspects of the
various requirements, namely $1/|\lambda|$ behavior of the peak in
the chiral limit, and a topological susceptibility proportional to
$m^2$ matching the normalized mode number of the peak, are expected to
be naturally satisfied. This appears at present the most natural
mechanism that could lead to effective $\mathrm{U}(1)_A$ breaking in
the chiral limit, and there are already partial indications that it
could actually be at play in QCD.

\section{Conclusions}
\label{sec:concl}

In this paper I have continued the investigation of the properties of
the Dirac spectrum in the symmetric phase of a gauge theory, started
in {\DSI} expanding on the results of
Refs.~\cite{Giordano:2024jnc,Giordano:2024awb} and of previous work by
others~\cite{Cohen:1997hz,Aoki:2012yj,Kanazawa:2015xna,
  Azcoiti:2023xvu,Kanazawa:2014cua} (see also
Ref.~\cite{Carabba:2021xmc}). In the first paper I worked on the
foundations of the approach, clarifying the assumptions of
Refs.~\cite{Cohen:1997hz,Aoki:2012yj,Kanazawa:2015xna,
  Azcoiti:2023xvu,Kanazawa:2014cua}, and providing a systematic way of
deriving constraints on the Dirac spectrum imposed by chiral symmetry
restoration.  Here I focussed on the consequences of these
constraints, in particular for the fate of $\mathrm{U}(1)_A$ symmetry
in the chiral limit, using additional technical assumptions on the
spectral density and the two-point correlation function of nonzero
eigenvalues. The main results are the following.

(1.)\ Assuming only chiral symmetry restoration at the level of scalar
and pseudoscalar susceptibilities, it is easy to find simple
functional forms of the spectral density that lead to effective
$\mathrm{U}(1)_A$ breaking in the chiral limit while complying with
the constraints imposed by chiral symmetry restoration
(Sec.~\ref{sec:chipi}). Assuming also symmetry restoration for
susceptibilities involving nonlocal gauge functionals (nonlocal
restoration), or for susceptibilities involving external fermion
fields in a partially quenched setting, one needs the spectral density
to be $m^2$-differentiable (i.e., a $C^\infty$ function of $m^2$ at
$m=0$), and strong restrictions apply on the possibility of
effectively breaking $\mathrm{U}(1)_A$ in the symmetric phase at the
level of the simplest order parameter,
$\Delta = \lim_{m\to 0} (\chi_\pi-\chi_\delta)/4$.

(1a.)\ Within a rather large class of functional forms of the spectral
density, with power-law behavior near $\lambda=0$ and
$m^2$-differentiable, the only allowed behavior compatible with
chiral symmetry restoration that breaks $\mathrm{U}(1)_A$ is a
singular near-zero peak
$\rho_{\mathrm{peak}}(\lambda;m) = \f{\Delta}{2}
m^2\gamma(m^2)|\lambda|^{-1+\gamma(m^2)}$, with $m^2$-differentiable
$\gamma>0$ and $\gamma=O(m^2)$ (Sec.~\ref{sec:mdiffspd}). In this case
the normalized mode number of the peak (i.e., number of modes in the
peak per unit four-volume) equals $\chit$ to leading order in $m$,
showing a close relation between the singular peak and topology
(Sec.~\ref{sec:spdenstop}).  Surprisingly, and contrary to what was
stated in Refs.~\cite{Giordano:2024jnc,Giordano:2024awb}, this
behavior is also compatible with commutativity of the thermodynamic
and chiral limits (Sec.~\ref{sec:ctc}).

(1b.)\ Under more general assumptions on the spectral density, if chiral
symmetry is restored in its extended form (requiring
$m^2$-differentiability) then a singular near-zero behavior of some
sort is required to obtain $\Delta \neq 0$. In fact, a necessary (but
not sufficient) condition for it is that the spectral density
effectively develops a term $\propto m^2\delta(\lambda)$ in the chiral
limit. This can be achieved in a variety of ways, including the
singular peak $\rho_{\mathrm{peak}}$ or generalizations thereof, that
may or may not be compatible with commutativity of the thermodynamic
and chiral limits, and may or may not have the same sharp relation
between mode number and topological susceptibility
(Sec.~\ref{sec:mdiffspdgen}).

(2.)\ If $\mathrm{U}(1)_A$ is effectively broken in the chiral limit
by a nonzero $\Delta$, the two-point eigenvalue correlation function
must be singular at the origin, indicating strong eigenvalue repulsion
(Sec.~\ref{sec:tpfunc_finite}). The required singularity cannot be
obtained if near-zero modes are localized, not even if the spectral
density diverges at zero. A nonzero $\Delta$ implies then that
near-zero modes cannot be localized in the chiral limit, requiring a
mobility edge in the near-zero region (Sec.~\ref{sec:tpfunc_loc}).
This result is obtained making use only of general bounds on the
two-point function (including a new one on the correlation function of
localized and delocalized modes, see Appendix~\ref{sec:locdeloccorr}),
well motivated by the study of random matrix systems.  In the presence
of a divergent near-zero spectral peak, the near-zero mobility edge is
most likely a new mobility edge, distinct from the well-known one in
the bulk of the spectrum~\cite{Giordano:2021qav}.

(3.)\ The results above, together with numerical results indicating
the presence of a near-zero spectral peak~\cite{Edwards:1999zm,
  HotQCD:2012vvd,Cossu:2013uua,Buchoff:2013nra,Dick:2015twa,
  Alexandru:2015fxa,Tomiya:2016jwr,Kovacs:2017uiz,Aoki:2020noz,
  Alexandru:2019gdm,Ding:2020xlj,Kaczmarek:2021ser,Vig:2021oyt,
  Kovacs:2021fwq,Meng:2023nxf,Kaczmarek:2023bxb,Alexandru:2024tel,
  JLQCD:2024xey,Fodor:2025yuj}, and of a near-zero mobility
edge~\cite{Meng:2023nxf} lead to a plausible but highly constrained
scenario for effective $\mathrm{U}(1)_A$ breaking in the symmetric
phase, requiring very specific spectral features, namely: a singular
peak tending to $O(m^4)/|\lambda|$ in the chiral limit; topological
susceptibility proportional to $m^2$ matching the normalized mode
number of the peak; and a near-zero mobility edge. These features emerge
naturally in the QCD-inspired model of weakly interacting instantons
and anti-instantons of Ref.~\cite{Kovacs:2023vzi}, that provides an
explicit mechanism realizing the proposed $\mathrm{U}(1)_A$-breaking
scenario, showing that it is physically viable
(Sec.~\ref{sec:u1breakingscenario}).

The specific functional form of the singular spectral peak
$\rho_{\mathrm{peak}}$ is certainly not the most general, but is
physically motivated by the available numerical results for the Dirac
spectrum in the symmetric phase of QCD and in high-temperature pure
$\mathrm{SU}(3)$ gauge
theory~\cite{Edwards:1999zm,HotQCD:2012vvd,Cossu:2013uua,Buchoff:2013nra,Dick:2015twa,
  Alexandru:2015fxa,Tomiya:2016jwr,Kovacs:2017uiz,Aoki:2020noz,
  Alexandru:2019gdm,Ding:2020xlj,Kaczmarek:2021ser,Vig:2021oyt,
  Kovacs:2021fwq,Meng:2023nxf,Kaczmarek:2023bxb,Alexandru:2024tel,
  JLQCD:2024xey,Fodor:2025yuj}. The fact that in this case the
normalized mode number of the peak equals $\chit$ is in agreement with
the expected topological origin of the peak~\cite{Edwards:1999zm,
  HotQCD:2012vvd,Buchoff:2013nra,Dick:2015twa,Kovacs:2017uiz,Ding:2020xlj,Vig:2021oyt,
  Kaczmarek:2021ser,Kaczmarek:2023bxb,Kovacs:2023vzi,Fodor:2025yuj}.
This holds true also for its most straightforward generalization [see
Eq.~\eqref{eq:genU1abr12}].

The stated compatibility of $\rho_{\mathrm{peak}}$ (and of some of its
generalizations) with commutativity of the thermodynamic and chiral
limits is in contradiction with the conclusions of
Ref.~\cite{Azcoiti:2023xvu} taken at face
value. Reference~\cite{Azcoiti:2023xvu} identifies the behavior
$\rho_{\mathrm{sing}}= \Delta m^2\delta(\lambda)$ as the only one
leading to $\mathrm{U}(1)_A$ breaking under the assumptions of
$m^2$-differentiability of the free energy density and commutativity of
limits. The contradiction, however, is only apparent. In fact,
Ref.~\cite{Azcoiti:2023xvu} proved first a condition on the spectral
density implied by the assumptions above [Eq.~\eqref{eq:commlim17_0},
rederived here in Appendix~\ref{sec:app_commlim}], and then singled
out $\rho_{\mathrm{sing}}$ as the only acceptable functional form
among those satisfying this condition. This second step, however, is
unjustified, as producing perfectly acceptable examples shows
explicitly.

For more general functional forms of the spectral density, compatible
with (extended) chiral symmetry restoration and with $\Delta\neq 0$,
discussed in Sec.~\ref{sec:mdiffspdgen}, the same close relation of
the near-zero modes with topology found for $\rho_{\mathrm{peak}}$,
and the compatibility with limit commutativity, can be achieved but
are not guaranteed.

The physical viability of the singular-peak scenario discussed in
(3.)\ is supported by the instanton-based mechanism proposed in
Ref.~\cite{Kovacs:2023vzi}, that provides also a very concrete and
natural way for it to be realized in practice. Conversely, fulfilling
the constraints from chiral symmetry restoration makes the model of
Ref.~\cite{Kovacs:2023vzi} a mathematically acceptable description of
the near-zero Dirac spectrum in the chiral limit of QCD in the
$\mathrm{U}(1)_A$-breaking case. It should be noted that although they
are most likely present, certain detailed features required by chiral
symmetry restoration still need to be explicitly confirmed in the
model, which calls for further studies. It would also be interesting
to go beyond Refs.~\cite{Edwards:1999zm,HotQCD:2012vvd,
  Buchoff:2013nra,Dick:2015twa,Kovacs:2017uiz,Ding:2020xlj,
  Vig:2021oyt,Kaczmarek:2021ser,Kaczmarek:2023bxb} and check directly
whether the mechanism proposed in Ref.~\cite{Kovacs:2023vzi} is
actually at play in QCD and other realistic gauge theories with
fermions; and to possibly connect it more tightly with the
first-principles results obtained here. Independently of what
mechanism actually drives it, the singular-peak scenario leads to a
highly constrained set of detailed predictions for the behavior of the
Dirac spectrum, that should be carefully tested in numerical lattice
calculations.
  
From the theoretical point of view, it would be interesting to
characterize scalar and pseudoscalar susceptibilities in the symmetric
phase, and the topological features of gauge field configurations, for
an arbitrary number of flavors $N_f\ge 2$. This would in turn provide
constraints on the Dirac spectrum, which would again help in getting
insight into the issue of effective $\mathrm{U}(1)_A$ breaking in the
chiral limit (this time by studying the global $\mathrm{U}(1)_A$
condensates of Ref.~\cite{Carabba:2021xmc}). Further insight could
also be obtained by studying sectors other than the scalar and
pseudoscalar one.

In conclusion, this series of works shows how studying the Dirac
spectrum and its interplay with the topological features of
gauge-field configurations can lead to considerable progress in
understanding the relation between the
$\mathrm{SU}(2)_L\times \mathrm{SU}(2)_R$ and $\mathrm{U}(1)_A$
symmetries in the chirally symmetric phase. This paves the way to
finally settling the outstanding issue of the fate of
$\mathrm{U}(1)_A$ in the symmetric phase of QCD and other gauge
theories in the chiral limit.

\begin{acknowledgments}
  I thank V.~Azcoiti, C.~Bonanno, G.~Endr{\H o}di, I.~Horv\'ath,
  S.~D.~Katz, D.~N\'ogr\'adi, A.~Patella, A.~P\'asztor, Zs.~Sz\'ep,
  and especially T.~G.~Kov{\'a}cs for discussions. This work was
  partially supported by the NKFIH grants K-147396, NKKP Excellence
  151482, and TKP2021-NKTA-64.
\end{acknowledgments}

\appendix

\section{Constraints on the spectral density: details}
\label{sec:app_spdens_det}

\subsection{Finiteness of
  \texorpdfstring{$\chi_\pi$}{chi\textunderscore pi} and contributions
  to \texorpdfstring{$\Delta$}{Delta} in the chiral limit}
\label{sec:coeff_contr}

After splitting the integrals $\Iu[\ff]$ and $m^2\Iu[\ff^2]$ as
follows,
\begin{equation}
  \label{eq:int_new20}
  \begin{aligned}
    \Iu[\ff]
    &= \int_0^\delta d\lambda\, \rho(\lambda;m)\ff(\lambda,m)\\
    &\phantom{=} + \int_\delta^2 d\lambda\, \rho(\lambda;m)\ff(\lambda,m)\,,\\
    m^2\Iu[\ff^2]
    &= m^2    \int_0^\delta d\lambda\, \rho(\lambda;m)\ff(\lambda,m)^2\\
    &\phantom{=}+  m^2 \int_\delta^2 d\lambda\, \rho(\lambda;m)\ff(\lambda,m)^2\,,
  \end{aligned}
\end{equation}
where $0<\delta\le 2$ is an arbitrary $m$-independent cutoff, and
further
\begin{equation}
  \label{eq:commlim_gna0}
  \begin{aligned}
    \int_0^\delta d\lambda\, \rho(\lambda;m)\ff(\lambda,m)
    &= \II_0(\delta;m)  -\f{1}{4} R_1(\delta;m)\,,\\
    m^2    \int_0^\delta d\lambda\, \rho(\lambda;m)\ff(\lambda,m)^2
    &= \II_1(\delta;m) 
      - \f{m^2}{2} R_2(\delta;m)\\
    &\phantom{=}   + \f{m^2}{16} R_3(\delta;m)\,,
  \end{aligned}
\end{equation}
where [see Eq.~\eqref{eq:int_new20_def}]
\begin{equation}
  \label{eq:commlim_gna1}
  \begin{aligned}
    \II_n(\delta;m)&\equiv    \int_0^\delta d\lambda\, \f{m^{2n}\rho(\lambda;m)}{(\lambda^2 + m^2)^{n+1}}\,,\\
    R_1(\delta;m)&\equiv     \int_0^\delta d\lambda\, 
                   \f{\lambda^4 \rho(\lambda;m)}{(\lambda^2+ m^2)(\lambda^2+ m^2 h(\lambda))} \,,\\
    R_2(\delta;m)&\equiv   \int_0^\delta d\lambda\, 
                   \f{\lambda^4\rho(\lambda;m)}{(\lambda^2+ m^2)^2(\lambda^2+ m^2 h(\lambda))}\,, \\
    R_3(\delta;m)&\equiv  \int_0^\delta d\lambda\, 
                   \f{\lambda^8\rho(\lambda;m)}{(\lambda^2+ m^2)^2(\lambda^2+ m^2 h(\lambda))^2} \,,
  \end{aligned}
\end{equation}
one finds
\begin{equation}
  \label{eq:int_new21}
  \begin{aligned}
    R_1(\delta;m)
    &\le \int_0^\delta d\lambda\, \rho(\lambda;m) < \infty\,, \\
    \int_\delta^2 d\lambda\, \rho(\lambda;m)\ff(\lambda;m)
    & \le \f{1}{\delta^2} \int_\delta^2 d\lambda\, \rho(\lambda;m)<\infty  \,,
  \end{aligned}
\end{equation}
so $\Iu[\ff]$, and therefore $\chi_\pi$, is finite in the chiral limit
if and only if $\II_0$ is finite. Moreover,
\begin{equation}
  \label{eq:commlim_gna2_0}
  \begin{aligned}
    R_3(\delta;m)
    &\le \int_0^\delta d\lambda\, \rho(\lambda;m)<\infty\,,\\
    \int_\delta^2 d\lambda\, \rho(\lambda;m)\ff(\lambda;m)^2 
    &\le \f{1}{\delta^4}\int_\delta^2 d\lambda\, \rho(\lambda;m) <\infty\,,
  \end{aligned}
\end{equation}
and imposing finiteness of $\II_0$ one finds also
\begin{equation}
  \label{eq:commlim_gna2}
  R_2(\delta;m) \le \int_0^\delta d\lambda\, \f{\rho(\lambda;m)}{\lambda^2 + m^2}=
  \II_0(\delta;m) <\infty\,,
\end{equation}
so the chiral limit of $m^2\Iu[\ff^2]$, i.e., $\f{\Delta}{2}$, equals
the chiral limit of $\II_1(\delta;m)$ in the symmetric phase.  From
Eqs.~\eqref{eq:int_new20} and \eqref{eq:commlim_gna0} follows then
$m^2\Iu[\ff^2]-\II_1(\delta;m)=O(m^2)$.

For $m^2$-differentiable
$\rho(\lambda;m)=\rho(\lambda;0)+m^2\rho_1(\lambda;m)$ as in
Sec.~\ref{sec:mdiffspdgen}, and choosing $\delta<\lambda_0$ [see after
Eq.~\eqref{eq:rho1_integral}], one proves as follows that the
existence of $\lim_{m\to 0}\II_0(\delta;m)$ is equivalent to that of
$\lim_{m\to 0}\chi_\pi$. One has
\begin{equation}
  \label{eq:lim_mdiff1}
\begin{aligned}
  R_1(\delta;m) &= \int_0^\delta d\lambda\,\f{\lambda^4 \rho(\lambda;0)}{(\lambda^2+ m^2)(\lambda^2+ m^2 h(\lambda))} \\
                &\phantom{=}+ m^2\int_0^\delta d\lambda\,\f{\lambda^4 \rho_1(\lambda;m)}{(\lambda^2+ m^2)(\lambda^2+ m^2 h(\lambda))}\,.
\end{aligned}
\end{equation}
The second term is $O(m^2)$ since the integral is bounded from above
by $\nn_1(\delta;m)\ge 0$; and since
\begin{equation}
  \label{eq:lim_mdiff2}
  \int_0^\epsilon d\lambda\, \f{\lambda^4 \rho(\lambda;0)}{(\lambda^2+ m^2)(\lambda^2+ m^2 h(\lambda))}\le
  \int_0^\epsilon d\lambda\,  \rho(\lambda;0)\,,
\end{equation}
that is independent of $m$ and has a vanishing $\epsilon\to 0^+$
limit, one finds that $\lim_{m\to 0} R_1(\delta;m)$ exists and is
finite,
\begin{equation}
  \label{eq:lim_mdiff3}
\begin{aligned}
  &   \lim_{m\to 0}  R_1(\delta;m)\\
  &= \lim_{\epsilon\to 0^+} \lim_{m\to 0}\int_\epsilon^\delta d\lambda\,\f{\lambda^4 \rho(\lambda;0)}{(\lambda^2+ m^2)(\lambda^2+ m^2 h(\lambda))}\\
  &= \lim_{\epsilon\to 0^+} \int_\epsilon^\delta d\lambda\, \rho(\lambda;0)  = n(\delta;0)\,.
\end{aligned}
\end{equation}
Similarly, one finds that
\begin{equation}
  \label{eq:lim_mdiff4}
  \lim_{m\to 0}      \int_\delta^2 d\lambda\, \rho(\lambda;m)\ff(\lambda;m) =
  \int_\delta^2 d\lambda\, \rho(\lambda;0)\f{h(\lambda)}{\lambda^2} 
\end{equation}
exists and is finite. Existence and finiteness of
$\lim_{m\to 0}\chi_\pi$ requires then the existence and finiteness of
$\lim_{m\to 0}\II_0(\delta;m)$ (and vice versa).

\subsection{Integrals}
\label{sec:inte}

The quantities in Eq.~\eqref{eq:inte1_again} are obtained from the
integrals
\begin{equation}
  \label{eq:inte1_again_app}
  \mathcal{I}_{\gamma}(\delta;m)\equiv \int_0^\delta d\lambda\, \f{\lambda^{\gamma}}{\lambda^2+ m^2 }\,,
\end{equation}
with $\delta>0$ and $\gamma>-1$, by setting $\gamma=\alpha(m)$, where
$\alpha(m)$ is continuous, $\alpha(m)>-1$ for $m\neq 0$, and
$\alpha(0)\ge - 1$.  $\mathcal{I}_{\gamma}(\delta;m)$ is a continuous
function of $\gamma$ for $\gamma>-1$ and $m\neq 0$.  Notice the
recursion relation
\begin{equation}
  \label{eq:inteI_add}
  \mathcal{I}_{\gamma+2}(\delta;m) 
  = \f{\delta^{\gamma+1}}{\alpha +1} 
  -m^2\mathcal{I}_{\gamma}(\delta;m)\,.
\end{equation}
After changing variables to $\lambda=|m|z$ one finds
\begin{equation}
  \label{eq:inte1_again_app2}
  \mathcal{I}_{\gamma}(\delta;m)= |m|^{\gamma-1} \int_0^{\f{\delta}{|m|}} dz\, \f{z^{\gamma}}{z^2+ 1}
  = |m|^{\gamma-1}  \bar{\mathcal{I}}_{\gamma}\left(\tf{\delta}{|m|}\right) \,.
\end{equation}
The integral $\bar{\mathcal{I}}_\gamma(\Lambda)$ can be evaluated
using the residue theorem on a half-circular contour of radius
$\Lambda$ centered at the origin (excluding a half-circle of radius
$\epsilon$ around zero, whose contribution vanishes in the limit
$\epsilon\to 0$). For $\Lambda > 1$ (corresponding to $|m|<\delta$)
one finds
\begin{equation}
  \label{eq:nasp11}
  \begin{aligned}
    & \f{2}{\pi}\cos\left(\f{\pi}{2}\gamma\right)
      \bar{\mathcal{I}}_\gamma(\Lambda) = 1- \f{\bar{\mathcal{R}}_\gamma(\Lambda)}{\Lambda^{1-\gamma}} \,, \\
    & \bar{\mathcal{R}}_\gamma(\Lambda) \equiv 
      \f{1}{2\pi}\int_{-\pi}^{\pi} d\theta \, \f{e^{i\theta\f{1-\gamma}{2}}}{1- \f{1}{\Lambda^2}e^{i\theta}}\,.
  \end{aligned}
\end{equation}
To evaluate the integral $\bar{\mathcal{R}}_\gamma(\Lambda)$ one
expands the integrand in powers of $\Lambda^{-2}$, and since the
expansion converges uniformly in $\theta$ one can exchange integration
and summation to get
\begin{equation}
  \label{eq:nasp11_bis}
    \bar{\mathcal{R}}_\gamma(\Lambda)
    = \left\{
      \begin{aligned}
        &    \f{2}{\pi }\cos\left(\f{\pi}{2}\gamma\right)
          \sum_{n=0}^\infty \f{(-1)^n}{\Lambda^{2n}} \f{1}{2n+1-\gamma} \,, \\
        &\text{if } \gamma\neq 2n_0+1\,, \forall n_0\in\mathbb{N}_0\,, \\
        &    \f{1}{\Lambda^{2n_0} }    \,, \\
        &\text{if } \gamma= 2n_0+1\,, n_0\in\mathbb{N}_0 \,.
      \end{aligned}\right.
\end{equation}
The series on the first line in Eq.~\eqref{eq:nasp11_bis} is
convergent if $\gamma\neq 2n_0+1\,, \forall n_0\in\mathbb{N}_0$, and
diverges like
$ \f{1}{\Lambda^{2n_0} } \left[
  \f{2}{\pi}\cos\left(\f{\pi}{2}\gamma\right)\right]^{-1}$ when
$\gamma\to 2n_0+1 $, $n_0\in\mathbb{N}_0$. Substituting
Eq.~\eqref{eq:nasp11_bis} into Eq.~\eqref{eq:nasp11} one finds
\begin{equation}
  \label{eq:nasp11_ter}
  \bar{\mathcal{I}}_\gamma(\Lambda) = \f{1}{\f{2}{\pi}\cos\left(\f{\pi}{2}\gamma\right)} -
  \f{1}{\Lambda^{1-\gamma}}  \sum_{n=0}^\infty \f{(-1)^n}{\Lambda^{2n}} \f{1}{2n+1-\gamma}\,,
\end{equation}
if $\gamma\neq 2n_0+1\,, \forall n_0\in\mathbb{N}_0$; the case
$\gamma= 2n_0+1 $, $n_0\in\mathbb{N}_0$, can be obtained by
continuity. Plugging Eq.~\eqref{eq:nasp11_ter} into
Eq.~\eqref{eq:inte1_again_app2} one finds after setting
$\gamma=\alpha(m)$
\begin{equation}
  \label{eq:nasp12_new}
  \begin{aligned}
    \mathcal{I}_{\alpha(m)}(\delta;m)
    &=  \f{|m|^{\alpha(m)-1}}{\f{2}{\pi}\cos\left(\f{\pi}{2}\alpha(m)\right)} \\
    &\phantom{=}-\delta^{\alpha(m)-1}
      \sum_{n=0}^\infty   \left(\f{m}{\delta}\right)^{2n} \f{(-1)^n}{2n+1-\alpha(m)}\,.
  \end{aligned}
\end{equation}
To leading order in $m$, if $\alpha(0)>1$ is not a positive odd
integer then
\begin{equation}
  \label{eq:nasp12_new_1}
  \mathcal{I}_{\alpha(m)}(\delta;m) \sim \f{\delta^{\alpha(0)-1}}{\alpha(0)-1}\,,
\end{equation}
as one could obtain directly by taking $m\to 0$ in
Eq.~\eqref{eq:inte1_again_app}. Corrections are $O(m^2)$, or
$O(|m|^{\alpha(m)-1})$ if $\alpha(0)<3$. If $\alpha(0)=2n_0+1$ is a
positive odd integer then setting $2n_0+1-\alpha(m)=\epsilon(m)$ one
has
\begin{equation}
  \label{eq:nasp12_new_2}
  \begin{aligned}
    \mathcal{I}_{\alpha(m)}(\delta;m)
    &= (-1)^{n_0}m^{2n_0}\left\{ \f{|m|^{-\epsilon(m)}}{\f{2}{\pi
      }\sin\left(\f{\pi}{2}\epsilon(m)\right)}
      - \f{\delta^{-\epsilon(m)}}{\epsilon(m)}\right\} \\
    &\phantom{=}     -\delta^{2n_0-\epsilon(m)}\!
      \sum_{\substack{n=0, \\n\neq n_0}}^\infty  \! \left(\f{m}{\delta}\right)^{2n}
    \!\f{(-1)^n}{2(n-n_0)+\epsilon(m)}\,.
  \end{aligned}
\end{equation}
In the chiral limit the quantity in braces behaves as
\begin{equation}
  \label{eq:nasp12_new_3}
  \begin{aligned}
    &\f{|m|^{-\epsilon(m)}}{\f{2}{\pi}\sin\left(\f{\pi}{2}\epsilon(m)\right)}
      - \f{\delta^{-\epsilon(m)}}{\epsilon(m)}\\
    &= \f{e^{l(m)}-1}{l(m)}\ln\f{1}{|m|} + 
      \ln\delta +O(|m|^{-\epsilon(m)}\epsilon(m),\epsilon(m))\,,
  \end{aligned}
\end{equation}
where $l(m)=\epsilon(m)\ln\f{1}{|m|}$. If $\lim_{m\to 0} l(m) = c$ is
finite (possibly zero) this quantity diverges logarithmically in $|m|$
[this includes the case of constant $\epsilon(m)= 0$, where one finds
by continuity the leading behavior $\ln\f{1}{|m|}$]; if
$\lim_{m\to 0} l(m) = +\infty$ it diverges faster than a logarithm,
but more slowly than any inverse power times a logarithm; if
$\lim_{m\to 0} l(m) = -\infty$ it diverges but more slowly than a
logarithm. Unless $n_0=0$ then
\begin{equation}
  \label{eq:nasp12_new_4}
  \mathcal{I}_{\alpha(m)}(\delta;m) \sim
  \f{\delta^{2n_0} }{2n_0}=  \f{\delta^{\alpha(0)-1} }{\alpha(0)-1}\,,
\end{equation}
again as one could obtain directly from
Eq.~\eqref{eq:inte1_again_app}. Corrections are $O(m^2)$, or
$O(m^2\f{e^{l(m)}-1}{l(m)}\ln\f{1}{|m|})$, i.e.,
$O(m^2\f{m^{-\epsilon(m)}-1}{\epsilon(m)})$, if $\alpha(0)=3$
($n_0=1$). If $\alpha(0)=1$ ($n_0=0$) then
\begin{equation}
  \label{eq:nasp12_new_5}
  \mathcal{I}_{\alpha(m)}(\delta;m) \sim  \f{e^{l(m)}-1}{l(m)}\ln\f{1}{|m|}
  = \f{m^{\alpha(m)-1}-1}{1-\alpha(m)}
\end{equation}
diverges, with $O(1)$ corrections. If $-1\le \alpha(0)<1$ then
$\mathcal{I}_{\alpha(m)}(\delta;m) $ diverges in the chiral limit,
with
\begin{equation}
  \label{eq:nasp12_new_6}
  \mathcal{I}_{\alpha(m)}(\delta;m) \sim 
  \f{|m|^{\alpha(m)-1}}{\f{2}{\pi
    }\cos\left(\f{\pi}{2}\alpha(0)\right)} \,,
\end{equation}
if $ \alpha(0)\neq -1$,  and
\begin{equation}
  \label{eq:nasp12_new_7}
  \mathcal{I}_{\alpha(m)}(\delta;m) \sim   \f{|m|^{\alpha(m)-1}}{1+\alpha(m)} \,,
\end{equation}
if $ \alpha(0)= -1$, in both cases with corrections of order 
$O(|m|^{\alpha(m)-1}(\alpha(m)-\alpha(0)))$, or $O(1)$ if $\alpha(m)$
is constant (at least for small $|m|$).

Setting $M=m/\sqrt{1-m^2/4}$, and using the results above and the
relation
$\f{\lambda^2}{(\lambda^2+M^2)^2} =
-\f{\lambda}{2}\de_\lambda\f{1}{\lambda^2+M^2}$, one finds
\begin{equation}
  \label{eq:ftpf3_bis}
  \begin{aligned}
    &    \lim_{m\to 0}  m \int_0^2d\lambda \,\ff(\lambda;m)
      =\lim_{M\to 0}  M \int_0^2d\lambda \,\f{h(\lambda)}{\lambda^2+M^2} \\
    &=             \lim_{M\to 0}      M\mathcal{I}_{0}(2;M)  =
      \mathrm{sgn}(m)\f{\pi}{2}\,, \\
    &\lim_{m\to 0}  m \int_0^2d\lambda \,\ffh(\lambda;m) =
      \lim_{M\to 0}  M \int_0^2d\lambda \,\f{\lambda^2h(\lambda)}{\left(\lambda^2+M^2\right)^2} \\
    &= \f{1}{2}            \lim_{M\to 0}  M\mathcal{I}_{0}(2;M) =
      \mathrm{sgn}(m)\f{\pi}{4}\,.
  \end{aligned}
\end{equation}
To obtain Eq.~\eqref{eq:int_new20_bis_delta2bis_0} one needs the
integrals [see Eq.~\eqref{eq:u1abreak}]
\begin{equation}
  \label{eq:inteJ_0}
  \mathcal{J}_{\gamma}(\delta;m)\equiv\int_0^\delta d\lambda\,\f{m^2\lambda^\gamma}{(\lambda^2+m^2)^2}\,,
\end{equation}
for $\gamma=\alpha_i(m)$. The following relation holds,
\begin{equation}
  \label{eq:inteJ_1}
  \mathcal{J}_{\gamma}(\delta;m)   = \f{1}{2}\f{\delta^{\gamma+1}}{\delta^2+m^2} 
  + \f{1-\gamma}{2}\mathcal{I}_{\gamma}(\delta;m)\,,
\end{equation}
that corresponds to Eq.~\eqref{eq:spdens_gen7} for $n=0$. Combined
with Eq.~\eqref{eq:inteI_add} this gives
\begin{equation}
  \label{eq:inteJ_2}
  \mathcal{J}_{\gamma+2}(\delta;m) 
  = \f{1+\gamma}{2}m^2\mathcal{I}_{\gamma}(\delta;m)
  -\f{m^2}{2}\f{\delta^{\gamma+1}}{\delta^2+m^2} 
  \,.
\end{equation}
Notice the following results for $\mathcal{I}_{\alpha(m)}(\delta;m)$
and $\mathcal{J}_{\alpha(m)}(\delta;m)$ for constant integer
$\alpha(m)=n$,
\begin{equation}
  \label{eq:inteI_integer}
  \begin{aligned}
    \mathcal{I}_{0}(\delta;m) &= \f{1}{|m|}\arctan\f{\delta}{|m|} = \f{\pi}{2|m|}+ O(1)\,,\\
    \mathcal{I}_{1}(\delta;m) &= \f{1}{2}\ln\left(1+\f{\delta^2}{m^2}\right) = \ln\f{1}{|m|}+ O(1)\,,
  \end{aligned}
\end{equation}
and $\mathcal{I}_{n}(\delta;m) = O(1)$ for $ n\ge 2$, and moreover
\begin{equation}
  \label{eq:inteJ_integer}
  \begin{aligned}
    \mathcal{J}_{0}(\delta;m) &=\f{\pi}{4|m|}+ O(1)\,,\\
    \mathcal{J}_{1}(\delta;m) &=\f{1}{2}+ O(m^2)\,,\\
    \mathcal{J}_{2}(\delta;m) &=\f{\pi |m|}{4}+ O(1)\,,\\
    \mathcal{J}_{3}(\delta;m) &=m^2\ln\f{1}{|m|}+ O(m^2)\,,
  \end{aligned}
\end{equation}
obtained using Eqs.~\eqref{eq:inteJ_1} and \eqref{eq:inteJ_2}.

\subsection{Finiteness of
  \texorpdfstring{$C_i \mathcal{I}_{\alpha_i}$}{C\textunderscore i
    I\textunderscore alpha\textunderscore i}}
\label{sec:finX0}

For a spectral density of the form Eq.~\eqref{eq:spdens_gen1}, one
finds
\begin{equation}
  \label{eq:spdens_gen2}
  \II_n(\delta;m) = \sum_{i=1}^s C_i(m) X_i^{(n)}(\delta;m) +\bar{\II}_n(\delta;m)\,,
\end{equation}
where $\II_n$ is defined in Eq.~\eqref{eq:commlim_gna1}, with
\begin{equation}
  \label{eq:spdens_gen5}
  \begin{aligned}
    X_i^{(n)}(\delta;m) &\equiv \int_0^\delta d\lambda\,  \f{m^{2n}\lambda^{\alpha_i(m)}}{(\lambda^2+m^2)^{n+1}}  \,, \\
    \bar{\II}_n(\delta;m) &\equiv \int_0^\delta d\lambda\, \f{m^{2n}\bar{\rho}(\lambda;m)}{(\lambda^2+m^2)^{n+1}}\,.
  \end{aligned}
\end{equation}
Finiteness of $\chi_\pi$ requires finiteness of $\II_{0}$ in the
chiral limit [see Eq.~\eqref{eq:int_new20_bis}].  Of course,
$X_i^{(0)}(\delta;m)=\mathcal{I}_{\alpha_i(m)}(\delta;m)$
[Eqs.~\eqref{eq:int_new15} and \eqref{eq:nasp12_new}], and
$X_i^{(1)}(\delta;m)=\mathcal{J}_{\alpha_i(m)}(\delta;m)$
[Eq.~\eqref{eq:inteJ_0}]. In the following the arguments $\delta$ and
$m$ will be mostly dropped to avoid clutter.

I show now that in the symmetric phase $C_i \mathcal{I}_{\alpha_i}$
must be separately finite in the chiral limit. Crucially, since
$\rho\ge 0$, one has $0\le \II_n\le \II_0$, and since chiral symmetry
restoration requires that $\II_0$ be finite as $m\to 0$, all $\II_n$
will be finite in this limit as well.  Using the bound on
$\bar{\rho}$, Eq.~\eqref{eq:barrhobound}, one finds that
$\bar{\II}_{0}$ is finite in the chiral limit, and for $n\ge 1$ [see
Eqs.~\eqref{eq:nasp12_new_1}--\eqref{eq:nasp12_new_7}]
\begin{equation}
  \label{eq:spdens_gen4}
  |\bar{\II}_{n}(\delta;m)  | \le \int_0^\delta d\lambda\,  \f{A m^2\lambda^{\zeta-1}}{\lambda^2+m^2} 
  = Am^2\mathcal{I}_{\zeta-1}(\delta;m)   =o(1)  \,,
\end{equation}
and more precisely $O(|m|^\zeta)$ if $0<\zeta<1$,
$O(m^2\ln \f{1}{|m|})$ if $\zeta=1$, and $O(m^2)$ if $\zeta>1$, so
$\bar{I}_{n}$ vanishes in the chiral limit for $n\ge 1$. Finiteness of
$\II_{0}$ in the chiral limit requires then
\begin{equation}
  \label{eq:X0fin}
  \sum_{i=1}^s C_i X_i^{(n)} = O(1)\,, \quad \forall n\ge 0.
\end{equation}
A direct calculation shows that $X_i^{(n)}$, $n\ge 0$, obeys the
following recursion relation,
\begin{equation}
  \label{eq:spdens_gen7}
  X_i^{(n+1)} = \f{m^{2n}\delta^{2(1-\ah_i)}}{2(n+1)(\delta^2+m^2)^{n+1}}
  + \f{n + \ah_i}{n+1}X_i^{(n)}\,,
\end{equation}
where $\ah_i\equiv \f{1-\alpha_i}{2}$.
Equation~\eqref{eq:inteJ_1} corresponds to $n=0$.
The first term is $o(1)$ for
$n\ge 1$ and $O(1)$ for $n=0$, so iterating over $n$ one finds for
$n\ge 1$
\begin{equation}
  \label{eq:spdens_gen12_0}
  \begin{aligned}
    X_i^{(n)}   = O(1) + \f{1}{n!}P_n(\ah_i) X_i^{(0)}\,, 
  \end{aligned}
\end{equation}
where $P_n(x)\equiv \prod_{j=0}^{n-1}(j+x)$ is a polynomial of order
$n$. Setting also $P_0(x)\equiv 1$, one has for $n\ge 0$ the
recursion relation $P_{n+1}(x) = (n+x)P_n(x)$.

For $n\ge 1$, the second term in Eq.~\eqref{eq:spdens_gen12_0} is of
order $O(\ah_i X_i^{(0)})$, and so it diverges in the chiral limit for
$i\neq s$. Depending on the behavior of the $i=s$ term, one
distinguishes two cases.  (1.)\ If $\alpha_s(0)\neq 1$, or if
$\alpha_s(0)=1$ and $\lim_{m\to 0}|m|^{\alpha_s(m)-1}=\infty $,
$P_n(\ah_s) X_s^{(0)}$ diverges for $n\ge 1$.  (2.)\ If
$\alpha_s(0)=1$ with $\lim_{m\to 0}|m|^{\alpha_s(m)-1}< \infty$,
including zero, one has that $\ah_s X_s^{(0)}$ is at most $O(1)$ in
the chiral limit, and so $P_n(\ah_s) X_s^{(0)} = O(1)$ for $n\ge
1$. Setting $Y_i^{(0)}\equiv C_iX_i^{(0)}$, the finiteness requirement
Eq.~\eqref{eq:X0fin} reduces in case (1.)\ to
\begin{equation}
  \label{eq:spdens_gen13}
  \sum_{i=1}^s P_n(\ah_i)   Y_i^{(0)} = O(1)\,,\qquad\forall n\ge 0\,,
\end{equation}
and in case (2.)\ to
\begin{equation}
  \label{eq:spdens_gen13_bis}
  \begin{aligned}
    \sum_{i=1}^s   Y_i^{(0)}
    &= O(1)\,,\\
    \sum_{i=1}^{s-1} P_{n}(\ah_i)   Y_i^{(0)}
    &   = O(1)\,,\qquad\forall n\ge 1\,.
  \end{aligned}
\end{equation}
In both cases, combining the finiteness conditions for $n$ and $n-1$
and using the recursion relation for $P_n$, one finds for $n\ge 1$
\begin{equation}
  \label{eq:spdens_gen14}
  \begin{aligned}
    O(1)&=  \sum_{i=1}^{\hat{s}} P_n(\ah_i) Y_i^{(0)}- (n-1)\sum_{i=1}^{\hat{s}} P_{n-1}(\ah_i) Y_i^{(0)} \\
        &= \sum_{i=1}^{\hat{s}} P_{n-1}(\ah_i) \ah_i Y_i^{(0)}\,, \quad \forall n\ge 1\,,
  \end{aligned}
\end{equation}
where $\hat{s}=s$ in case (1.)\ and $\hat{s}=s-1$ in case (2.). [For
$n=1$ the relation is correct also in case (2.)\ since the second term
on the first line vanishes identically.] Iterating the procedure one
finds
\begin{equation}
  \label{eq:spdens_gen14_bis}
  O(1)=   \sum_{i=1}^{\hat{s}} P_{0}(\ah_i) \ah_i^n Y_i^{(0)}=   \sum_{i=1}^{\hat{s}} \ah_i^n Y_i^{(0)}\,,
  \quad \forall n\ge 1\,.
\end{equation}
Including also the request of finiteness of
$\sum_{i=1}^s C_i X_i^{(0)}=\sum_{i=1}^s Y_i^{(0)}$ this leads in case
(1.)\ to
\begin{equation}
  \label{eq:spdens_gen15}
  \sum_{i=1}^s \ah_i^n Y_i^{(0)} = O(1)\,, \quad \forall n\ge 0\,,
\end{equation}
and in case (2.)\ to
\begin{equation}
  \label{eq:spdens_gen15_bis}
  \begin{aligned}
    \sum_{i=1}^s  Y_i^{(0)} &= O(1)\,, \\
    \sum_{i=1}^{s-1} \ah_i^n Y_i^{(0)} &= O(1)\,, \quad \forall n\ge 1\,.
  \end{aligned} 
\end{equation}
In case (1.), one takes the first $s$ relations in
Eq.~\eqref{eq:spdens_gen15}, with $n=0,1,\ldots,s-1$, and writes them
in matrix form,
\begin{equation}
  \label{eq:spdens_gen16}
  V Y = O(1)\,,
\end{equation}
where $Y$ collects $Y^{(0)}_i$, $i=1,\ldots,s$, in a vector, and $V$
is the Vandermonde matrix $V_{ij}=(\ah_j)^{i-1}$, $i,j=1,\ldots,s$.
Since
\begin{equation}
  \label{eq:spdens_gen17}
  \det V = \prod_{1\le i<j\le s}(\ah_j - \ah_i) = \prod_{1\le i<j\le s}\f{\alpha_i-\alpha_j}{2}  \,,
\end{equation}
$V$ is invertible as long as the exponents are all different.  Since
by assumption $\alpha_i(0)\neq \alpha_j(0)$ $\forall i\neq j$,
$\det V$ is nonzero, at least for small $m$ and in the limit $m\to 0$.
Equation~\eqref{eq:spdens_gen16} implies then $Y=O(1)$, i.e.,
\begin{equation}
  \label{eq:spdens_gen18}
  Y_i^{(0)} =  C_i X_i^{(0)} =O(1)\,,
\end{equation}
separately for $1\le i\le s$. In case (2.), one takes instead the
$s-1$ relations with $1\le n\le s-1$ and write them in matrix form,
\begin{equation}
  \label{eq:spdens_gen16_bis0}
  \tilde{V} \tilde{Y} = O(1)\,,
\end{equation}
where now $\tilde{Y}$ collects $Y_i^{(0)}$ for $i=1,\ldots, s-1$, and
$\tilde{V}_{ij}=(\ah_j)^{i}$, $i,j=1,\ldots,s-1$.  The determinant of
$\tilde{V}$ is
\begin{equation}
  \label{eq:spdens_gen17_bis}
  \begin{aligned}
    \det\tilde{V}
    &=\left(\prod_{i=1}^{s-1}\ah_i\right) \prod_{1\le i<j\le s-1}(\ah_j - \ah_i)\\
    & = \left(\prod_{i=1}^{s-1}\f{1-\alpha_i}{2}\right)\prod_{1\le i<j\le s-1}\f{\alpha_i-\alpha_j}{2}  \,, 
  \end{aligned}
\end{equation}
so $\det\tilde{V}\neq 0$ in the chiral limit, since
$\alpha_i(0)\neq \alpha_j(0)$ $\forall i\neq j$ and
$1-\alpha_i(0)\neq 0$ for $1\le i\le s-1$, and therefore $\tilde{V}$
is invertible in the chiral limit, implying $\tilde{Y}=O(1)$, i.e.,
\begin{equation}
  \label{eq:spdens_gen18_bis}
  Y_i^{(0)} =  C_i X_i^{(0)} =O(1)\,,
\end{equation}
separately for $1\le i\le s-1$. Using now the first equation in
Eq.~\eqref{eq:spdens_gen13_bis} one concludes that
$C_s X_s^{(0)} =O(1)$ as well.

For a spectral density of the form Eq.~\eqref{eq:spdens_gen1} one has
then that finiteness of $\chi_\pi$ in the chiral limit requires
$C_i X_i^{(0)} =O(1)$, $1\le i\le s$. Since
$ X_i^{(0)}=\mathcal{I}_{\alpha_i}$ diverges if
$-1\le \alpha_i(0)\le 1$, these conditions require that
\begin{equation}
  \label{eq:X0fin1}
  C_i = O(1/\mathcal{I}_{\alpha_i})=o(1)\,, \quad 1\le i\le s\,. 
\end{equation}
Using now Eqs.~\eqref{eq:int_new20_bis_delta}, \eqref{eq:inteJ_0},
\eqref{eq:spdens_gen4}, and \eqref{eq:spdens_gen7} for $n=0$ [i.e.,
Eq.~\eqref{eq:inteJ_1}], one finds
\begin{equation}
  \label{eq:u1abreak}
  \begin{aligned}
    \f{\Delta}{2}
    &= \lim_{m\to 0}\II_1(\delta;m)
      =       \lim_{m\to 0}\sum_{i=1}^s C_i(m) \mathcal{J}_{\alpha_i(m)}(\delta;m) \\
    &= \lim_{m\to 0}\sum_{i=1}^s C_i(m) X^{(1)}_i(\delta;m)\\
    &= \lim_{m\to 0}\sum_{i=1}^s C_i(m)\left(O(1)+\ah_i(m) 
      X^{(0)}_i(\delta;m) \right)\\
    &= \sum_{i=1}^s   \f{1-\alpha_i(0)}{2}
      \lim_{m\to 0}\left(C_i(m) \mathcal{I}_{\alpha_i(m)}(\delta;m) \right) \,.
  \end{aligned}
\end{equation}
Since $C_i X^{(0)}_i=O(1)$ by the argument above, if $\alpha_s(0)=1$
the corresponding term does not contribute, and $\Delta$ receives
contributions only from terms with $\alpha_i(0)< 1$.

The same reasoning used above allows one to prove that $\Delta=0$ if
and only if $\lim_{m\to 0}C_i X^{(0)}_i=0$ for $i=1,\ldots s$, if
$\alpha_s(0)\neq 1$, and for $i=1,\ldots s-1$, if $\alpha_s(0)= 1$.
In fact, since $0\le I_{n+1}\le I_n$, if $\Delta = \lim_{m\to 0}I_1=0$
then $\lim_{m\to 0}I_n=0$, $\forall n\ge 1$. Setting
$\hat{Y}_i^{(0)} = \lim_{m\to 0} C_iX_i^{(0)}$, one has for $n\ge 1$
\begin{equation}
  \label{eq:zeroDel1}
  \begin{aligned}
    0&=\lim_{m\to 0} \sum_{i=1}^s C_i X_i^{(n)} =
       \lim_{m\to 0} \sum_{i=1}^s C_i \left(O(1) + \f{P_n(\ah_i)}{n!}X_i^{(0)}\right) \\
     &=\f{1}{n!} \sum_{i=1}^s \left(\lim_{m\to 0} C_iX_i^{(0)}\right)  P_n(\ah_i) =
       \f{1}{n!} \sum_{i=1}^s \hat{Y}_i^{(0)}  P_n(\ah_i) \,,
  \end{aligned}
\end{equation}
so 
\begin{equation}
  \label{eq:zeroDel2}
  \begin{aligned}
    0 &= \sum_{i=1}^s \hat{Y}_i^{(0)}  P_n(\ah_i) - (n-1)\sum_{i=1}^s \hat{Y}_i^{(0)}  P_{n-1}(\ah_i) \\
      &= \sum_{i=1}^s \hat{Y}_i^{(0)}\ah_i  P_{n-1}(\ah_i) \,, 
  \end{aligned}
\end{equation}
and iterating
\begin{equation}
  \label{eq:zeroDel3}
  0 = \sum_{i=1}^s P_{0}(\ah_i)\ah_i^{n}\hat{Y}_i^{(0)}  = \sum_{i=1}^s \ah_i^n\hat{Y}_i^{(0)}  \,,
  \qquad \forall n\ge 1\,.
\end{equation}
If $\alpha_s(0)\neq 1$, so that $\ah_s(0)\neq 0$, one writes the first
$s$ relations in Eq.~\eqref{eq:zeroDel3}, corresponding to
$n=1,\ldots,s$, as $\hat{V}^{[s]} \hat{Y}^{[s]}=0$, where
$\hat{Y}^{[s]}$ collects $\hat{Y}_i^{(0)}$ for $i=1,\ldots s$ in a
vector, and $\hat{V}^{[s]}_{ij} = (\ah_j)^i$, $i,j=1,\ldots s$. Since
$\hat{V}^{[s]}$ is invertible, $\hat{Y}^{[s]}=0$. If $\alpha_s(0)= 1$,
so that $\ah_s(0)= 0$, the corresponding term does not contribute to
$\Delta$ and can be ignored from the outset.
One then repeats the same argument using $ \hat{Y}^{[s-1]}$ and
$\hat{V}^{[s-1]}$, and since $\hat{V}^{[s-1]}$ is again invertible,
one finds $\hat{Y}^{[s-1]}=0$. This completes the proof.

Finally, the very same argument can be used to single out
$\tilde{\rho}_{\mathrm{peak}}$, Eq.~\eqref{eq:rhom2diff_peak2_commut},
as the only $\mathrm{U}(1)_A$-breaking behavior compatible with
commutativity of the thermodynamic and chiral limits. The proof is
based on the fact that limit commutativity requires
$\lim_{\epsilon\to 0^+}\lim_{m\to
  0}\left(\II_n(\epsilon;m)-\II_{n+1}(\epsilon;m)\right)=0$, $n\ge 0$
[see Eq.~\eqref{eq:commlim6_0_gen} and \eqref{eq:commlim6_gen}]. For
the functional form Eq.~\eqref{eq:spdens_gen1} this amounts to require
\begin{equation}
  \label{eq:peak_cl_unique}
\begin{aligned}
  0&=\lim_{\epsilon\to 0^+}\lim_{m\to 0}\sum_{i=1}^s  C_i \left( X_i^{(n)} -X_i^{(n+1)}\right)  \\
   &=\lim_{\epsilon\to 0^+}\lim_{m\to 0}\sum_{i=1}^s  C_i \left[O(1)+
     \f{1-\ah_i}{(n+1)!}       P_n(\ah_i) X_i^{(0)}\right]\\
   &=\f{1}{(n+1)!}\sum_{i=1}^s  \left(\lim_{\epsilon\to 0^+}\lim_{m\to 0}C_i
     \f{1+\alpha_i}{2}X_i^{(0)}\right)P_n(\ah_i) \,,
\end{aligned}
\end{equation}
for $n\ge 0$, having used Eq.~\eqref{eq:spdens_gen12_0} and the
properties of $P_n$. Setting now
$\hat{Y}_i^{(0)}=\lim_{\epsilon\to 0^+}\lim_{m\to 0} C_i
\f{1+\alpha_i}{2}X_i^{(0)}$ and proceeding as above [starting from
Eq.~\eqref{eq:zeroDel2}], one concludes that
$\lim_{\epsilon\to 0^+}\lim_{m\to 0} C_i
\f{1+\alpha_i}{2}X_i^{(0)}=0$, $i=1,\ldots, s$, and so
$\lim_{\epsilon\to 0^+}\lim_{m\to 0} C_i X_i^{(0)}$, $i=1,\ldots, s$,
except possibly for $i=1$ if $\alpha_1(0)=-1$.

\subsection{\texorpdfstring{$m^2$}{m\textasciicircum 2}-differentiable
  spectral density: power-law or power series behavior}
\label{sec:mdiff}

For $\rho$ of the form Eq.~\eqref{eq:spdens_gen1} to be
$m^2$-differentiable, one needs that
\begin{equation}
  \label{eq:rhom2diff1}
  \begin{aligned}
    \de_{m^2}^k\rho(\lambda;m) 
    & = \sum_{i=1}^s\sum_{l=0}^k
      \begin{pmatrix}
        k\\ l
      \end{pmatrix}
      \left(\de_{m^2}^lC_i(m)\right)\left(\de_{m^2}^{k-l} |\lambda|^{\alpha_i(m)}
      \right)\\
    & \phantom{=} + \de_{m^2}^k\bar{\rho}(\lambda;m)
  \end{aligned}
\end{equation}
remains finite in the chiral limit for all $k$. Using the Fa\`a di
Bruno formula one finds
\begin{equation}
  \label{eq:rhom2diff2}
  \begin{aligned}
    & |\lambda|^{-\alpha(m)}  \de_{m^2}^n |\lambda|^{\alpha(m)} \\
    &=\sum_{\substack{\{n_j\}_{j=1,\ldots,n}\,,\\ n_j\ge 0,\\ \sum_{j=1}^n j n_j=n}}
    \f{(\ln |\lambda|)^{\sum_{j=1}^n n_j} n!}{\prod_{j=1}^nn_j!}
    \prod_{j=1}^n \left(\f{\alpha^{(j)}(m)}{j!}\right)^{n_j}\,,
  \end{aligned}
\end{equation}
where $\alpha^{(j)}\equiv \de_{m^2}^j\alpha $. Then
\begin{equation}
  \label{eq:rhom2diff3}
  \begin{aligned}
    & \de_{m^2}^k\rho(\lambda;m)\\
    &    = \sum_{i=1}^s|\lambda|^{\alpha_i(m)} \left\{C_i^{(k)}(m) + \sum_{l=0}^{k-1}
      \begin{pmatrix}
        k\\ l
      \end{pmatrix}
      C_i^{(l)}(m)\right.\\
    &\phantom{=}\left.   \times \left[\alpha^{(k-l)}(m)\ln |\lambda|
      + O\left((\ln |\lambda|)^2\right)\right]\vphantom{ \sum_{l=0}^{k-1}}\right\}
      +  \de_{m^2}^k\bar{\rho}(\lambda;m)\,,
  \end{aligned}
\end{equation}
where $C_i^{(j)}\equiv \de_{m^2}^jC_i $. The omitted terms in square
brackets involve products of $\alpha^{(j)}$ with $1\le j< k-l$. For a
given $k$ the two sets of quantities $C_i^{(k)}(m)$ and
$\alpha_i^{(k)}(m)$ are singled out, as they do not appear already in
lower-order derivatives. Assuming that $C_i^{(l)}(m)$,
$\alpha_i^{(l)}(m)$, and $\de_{m^2}^l\bar{\rho}$ are finite in the
chiral limit for all $l<k$, if $C_i^{(k)}(m)$,
$C_i(m)\alpha_i^{(k)}(m)$, or $\de_{m^2}^k\bar{\rho}$ diverged then
the corresponding divergences could not cancel each other, as they
would pertain to terms with different $\lambda$-dependence; and could
not be canceled by the remaining, finite contributions. Since
$C_i^{(0)}(m)=C_i(m)$, $\alpha_i^{(0)}(m)=\alpha_i(m)$, and
$\bar{\rho}$ are finite in the chiral limit (see
Sec.~\ref{sec:spdens_expl}), $C_i^{(k)}(m)$, $\alpha_i^{(k)}(m)$, and
$\de_{m^2}^k\bar{\rho}$ are then shown to be finite for all $k$ by
induction.

This applies in particular if $\rho = \rho_{\mathrm{series}}$,
Eq.~\eqref{eq:app_sppow1} [plus possibly non-integer power laws
$\sim |\lambda|^{\alpha(m)}$ with $-1\le \alpha(0)<1$ and
$\alpha(0)\neq 0$], proving that $\rho_0(m)$, $\rho_1(m)$, and
$\tilde{\rho}_2(\lambda;m)$, are $m^2$-differentiable (and that the
exponents and coefficients of non-integer power-law terms are
$m^2$-differentiable), where
\begin{equation}
  \label{eq:mdiff_rhoseries_0}
  |\lambda|^N\tilde{\rho}_N(\lambda;m)\equiv\rho_{\mathrm{series}}(\lambda;m)
  -\sum_{n=0}^{N-1}\rho_n(m)|\lambda|^n\,,
\end{equation}
with $\tilde{\rho}_N=O(|\lambda|^0$). Note that
$ \tilde{\rho}_N(\lambda;m) = \rho_N(m) +
|\lambda|\tilde{\rho}_{N+1}(\lambda;m)$. One shows by induction that
also $\rho_n$, $n\ge 2$ are $m^2$-differentiable. Denote
$\tilde{\rho}_N^{(k)}\equiv\de_{m^2}^k\tilde{\rho}_N$ and
$\rho_N^{(k)}(m)\equiv \de_{m^2}^k\rho_N(m)$, and assume
$\tilde{\rho}_{n}(\lambda;m)$ is $m^2$-differentiable
$\forall n\le N$, which is certainly true for $N=2$.  If
$\rho_N^{(k)}(m)$ were divergent in the chiral limit, then labeling
divergent parts of the same type by the same index $j$, one would have
for the corresponding coefficients $\rho_N^{(k)\mathrm{div},j}$ and
$\tilde{\rho}^{(k)\mathrm{div},j}_{N+1}(\lambda)=O(|\lambda|^0)$
\begin{equation}
  \label{eq:mdiff_rhoseries}
  0 = \tilde{\rho}^{(k)\mathrm{div},j}_{N}(\lambda) = \rho_N^{(k)\mathrm{div},j}
  + |\lambda|\tilde{\rho}^{(k)\mathrm{div},j}_{N+1}(\lambda)\,,
\end{equation}
$\forall k,j$, for $\lambda\in[
0,\delta_\rho
)$, and so in particular
$0= \tilde{\rho}^{(k)\mathrm{div},j}_{N}(0) =
\rho_N^{(k)\mathrm{div},j}$, $\forall k,j$, and therefore
$\tilde{\rho}^{(k)\mathrm{div},j}_{N+1}(\lambda)=0$, including at
$\lambda=0$ by continuity.  Then $\rho_N(m)$ and
$\tilde{\rho}_{N+1}(\lambda;m)$ are $m^2$-differentiable, and by
induction all $\rho_n(m)$ are $m^2$-differentiable.

\subsection{\texorpdfstring{$m^2$}{m\textasciicircum 2}-differentiable
  spectral density: general case -- divergent
  \texorpdfstring{$\nn_1(0^+;0)$}{n\textunderscore 1(0\textasciicircum
    +;0)}}
\label{sec:genm2rho_1}

A divergence at $\lambda=0$ in
$\de_{m^2}\nn(\lambda;m)|_{m=0}=\nn_1(\lambda;0)$ can lead to a
divergent $\chi_\pi$, although this depends on additional details. If
$\nn_1(\lambda;m)>0$ for $|\lambda|<\lambda_0$, $|m|<m_0$, and
$|\lambda|^\gamma \nn_1(\lambda;m)\ge C>0$, $0<\gamma<1$, for
$|m|<m_1\le m_0$ and $0\le a(m) \le |\lambda| \le b(m)$ with $a(0)=0$
and $b(0)\neq 0$, since
\begin{equation}
  \label{eq:nn1div}
\begin{aligned}
  &\lim_{m\to 0}  \int_0^\delta d\lambda \, \f{\rho(\lambda;m)}{\lambda^2+m^2} 
    \ge  \lim_{m\to 0} \int_0^{\min(\lambda_0,\delta)} d\lambda \,
    \f{2m^2 \lambda \nn_1(\lambda;m)}{(\lambda^2+m^2)^2}\\
  &\ge 2C|m|^{-\gamma} \int_{\f{a(m)}{|m|}}^{\f{b(m)}{|m|}}dz \, \f{z^{1-\gamma}}{(z^2+1)^2}\,,
\end{aligned}
\end{equation}
one finds a divergent contribution to $\chi_\pi$ if $a(m)/|m|$ has a
finite limit (including zero) as $m\to 0$. If $a(m)/|m|$ diverges, one
finds a divergent contribution if also
$|m|^{-\gamma}\left[\,|m|/a(m)\,\right]^{2+\gamma}=
m^2/a(m)^{2+\gamma}$ diverges, otherwise the lower bound derived above
remains finite, and a $\nn_1(\lambda;0)$ divergent at the origin may
be compatible with chiral symmetry restoration.

\subsection{\texorpdfstring{$m^2$}{m\textasciicircum 2}-differentiable
  spectral density: general case -- relaxing the monotonicity
  assumption}
\label{sec:genm2rho_2}

The monotonicity assumption $\de_\lambda\de_{m^2}\nn(\lambda;m)\ge 0$
for small $\lambda$ and $m$ can be replaced by the assumption that the
positive and negative components of $\rho_1$,
$\rho_1= \rho_{1,+}-\rho_{1,-}$ with
$\rho_{1,\pm}(\lambda;m)\equiv \f{1}{2}\left(
  |\rho_{1}(\lambda;m)|\pm\rho_{1}(\lambda;m)\right)$, have separately
well-defined chiral limits. This implies that both $\rho_1$ and
$|\rho_1|$ have well-defined chiral limits. This requirement excludes
a $\rho_1$ oscillating ever more wildly in the chiral limit as a
function of $\lambda$, which has no physical reason to be expected.
The precise request is that
$n_{1,\pm}(\delta;m)\equiv \int_0^\delta
d\lambda\,\rho_{1,\pm}(\lambda;m)$ separately have finite chiral
limits $n_{1,\pm}(\delta;0)$, differentiable in $\delta$,
$\forall \delta\neq 0$, and with finite limits as $\delta\to 0^+$.
This leads to
$\rho_{1,\pm}(\lambda;0)= a_\pm \delta(\lambda) + b_\pm (\lambda)$,
with $a_\pm,b_\pm(\lambda)\ge 0$, and $b_\pm(\lambda)$ integrable.

Under this assumption, finiteness of
$ \lim_{m\to 0} \II_0^{(0)}(\delta;m)$ is shown by modifying
Eq.~\eqref{eq:lim_mdiff_text} to
\begin{equation}
  \label{eq:lim_mdiff_text_app}
\begin{aligned}
  & \lim_{m\to 0}\II_0(\delta;m) \\ 
  &\ge \lim_{m\to 0} \II_0^{(0)}(\delta;m)
    + \liminf_{m\to 0}\left(\II_0(\delta;m)-\II_0^{(0)}(\delta;m)\right)\\
  &\ge \lim_{m\to 0} \II_0^{(0)}(\delta;m)
    - \lim_{m\to 0}\int_0^\delta d\lambda\,\left|\rho_1(\lambda;m)\right|\\
  &=\lim_{m\to 0} \II_0^{(0)}(\delta;m)   - n_{1,+}(\delta;0)-n_{1,-}(\delta;0)\,,
  \end{aligned}
\end{equation}
and so integrability of $\rho(\lambda;0)/\lambda^2$ still
follows. Finally, since positivity of the spectral density requires
that $m^2\rho_{1,-}(\lambda;m)\le \rho(\lambda;0)$, the result above  implies that
\begin{equation}
  \label{eq:ua1b_gen7_app}
  \begin{aligned}
    & \lim_{\epsilon\to 0^+}\lim_{m\to 0}\int_0^\epsilon d\lambda\,
      \f{m^4\rho_{1,-}(\lambda;m)}{(\lambda^2+m^2)^2} \\
    &  \le\lim_{\epsilon\to 0^+} \int_0^\epsilon d\lambda\,  \f{\rho(\lambda;0)}{\lambda^2} =0\,,
  \end{aligned}
\end{equation}
so $\rho_{1,-}(\lambda;m)$ plays no role in the fate of
$\mathrm{U}(1)_A$, even if it develops a term $b_-\delta(\lambda)$ in
the chiral limit, and Eq.~\eqref{eq:ua1b_gen11} becomes
\begin{equation}
  \label{eq:ua1b_gen10_app}
  \begin{aligned}
    \f{\Delta}{2}
    &= \lim_{\epsilon\to 0^+}\lim_{m\to 0}\int_0^\epsilon d\lambda\,
      \f{m^4\rho_{1,+}(\lambda;m)}{(\lambda^2+m^2)^2} \\
    & \le  \lim_{\epsilon\to 0^+}\lim_{m\to 0}\int_0^\epsilon d\lambda\,  \rho_{1,+}(\lambda;m) \\
    &= \lim_{\epsilon\to 0^+}n_{1,+}(\epsilon;0) = \f{a_+}{2}\,.
  \end{aligned}
\end{equation}

\subsection{\texorpdfstring{$m^2$}{m\textasciicircum 2}-differentiable
  spectral density: general case -- examples}
\label{sec:genm2rho_3}

I report here the detailed calculations related to the examples of
$m^2$-differentiable spectral density,
$\rho(\lambda;m) = \rho(\lambda;0) + m^2 \rho_1(\lambda;m)$, with
$\rho_1(\lambda;m)=
\rho_{1,\,\mathrm{sing}}(\lambda;m)+\rho_{1,\,\mathrm{reg}}(\lambda;m)$,
discussed in Sec.~\ref{sec:mdiffspdgen}.  For
$\rho_{1,\,\mathrm{sing}}(\lambda;m) =
\f{\gamma(m^2)}{|\lambda|}\phi\left(\gamma(m^2)\ln\tf{2}{|\lambda|}\right)$,
Eq.~\eqref{eq:genU1abr12}, with $\phi(x)$ positive and $C^\infty$,
integrable in $[0,\infty)$, and with $m^2$-differentiable
$\gamma=O(m^2)$, one finds after the change of variables
$\lambda=2e^{-w}$
\begin{equation}
  \label{eq:genU1abr13}
  \begin{aligned}
    & \int_0^\epsilon d\lambda \,\left(\f{m^2}{\lambda^2+m^2}\right)^n
      \rho_{1,\,\mathrm{sing}}(\lambda;m) \\
    & = \int_{\ln\f{2}{\epsilon}}^\infty dw 
      \left(\f{m^2}{4e^{-2w}+m^2}\right)^n \gamma(m^2)\phi\left(\gamma(m^2)w\right)\,.
  \end{aligned}
\end{equation}
Since $\phi$ is bounded, 
\begin{equation}
  \label{eq:genU1abr14}
\begin{aligned}
  &\lim_{m\to 0}\left|\int_0^{\ln\f{2}{\epsilon}} dw
    \left(\f{m^2}{4e^{-2w}+m^2}\right)^n \gamma(m^2)\phi\left(\gamma(m^2)w\right)\right| \\
  &\le \ln\tf{2}{\epsilon} \max_{x\ge 0}|\phi(x)|\lim_{m\to 0}\gamma(m^2)=0\,,
\end{aligned}
\end{equation}
so one can replace the lower limit of integration in
Eq.~\eqref{eq:genU1abr13} with $0$, and write
\begin{equation}
  \label{eq:genU1abr15}
  \begin{aligned}
    &  \lim_{m\to 0}\int_0^\epsilon d\lambda \left(\f{m^2}{\lambda^2+m^2}\right)^n
      \rho_{1,\,\mathrm{sing}}(\lambda;m) \\
    &  = \lim_{m\to 0} \int_{0}^\infty dw \,\left(1 +\f{4}{m^2}e^{-\f{2}{\gamma(m^2)}w}\right)^{-n} \phi(w) \,.
  \end{aligned}
\end{equation}
Splitting the integral as $\int_0^\infty=\int_0^\eta+\int_\eta^\infty$
and taking $\eta\to 0^+$ (after $m\to 0$), since for arbitrary
$\eta>0$
\begin{equation}
  \label{eq:genU1abr16}
  \left|\int_{0}^\eta dw \left(1 +\f{4}{m^2}e^{-\f{2}{\gamma(m^2)}w}\right)^{-n} \phi(w)\right|
  \le \eta  \max_{x\ge 0}\left|\phi(x)\right|\,,
\end{equation}
one concludes
\begin{equation}
  \label{eq:genU1abr17}
  \begin{aligned}
    &  \lim_{m\to 0}\int_0^\epsilon d\lambda \left(\f{m^2}{\lambda^2+m^2}\right)^n
      \rho_{1,\,\mathrm{sing}}(\lambda;m) \\
    &= \lim_{\eta\to 0^+}\lim_{m\to 0} \int_{\eta}^\infty dw
      \left(1 +\f{4}{m^2}e^{-\f{2}{\gamma(m^2)}w}\right)^{-n} \phi(w)   \\
    & = \lim_{\eta\to 0^+} \int_{\eta}^\infty dw \,\phi(w)
      =  \int_{0}^\infty dw \,\phi(w)\,.
  \end{aligned}
\end{equation}
For
$\rho_{1,\,\mathrm{sing}}(\lambda;m) = \f{1}{|m|\varepsilon(m)}
\phi\left(\f{|\lambda|- |m|\xi}{|m|\varepsilon(m)}\right)$,
Eq.~\eqref{eq:genU1abr19}, with positive and $m^2$-differentiable
$|m|\varepsilon(m)=O(m^2)$, and with $\phi(x)$ positive, $C^\infty$,
and integrable in $(-\infty,\infty)$, one finds after changing
variables to $z=\f{\lambda- |m|\xi}{|m|\varepsilon(m)}$
\begin{equation}
  \label{eq:genU1abr20}
  \begin{aligned}
    & \int_0^\epsilon d\lambda\,\left(\f{m^2}{\lambda^2+m^2}\right)^n
      \rho_{1,\,\mathrm{sing}}(\lambda;m)  \\
    & =\int_{-\f{\xi}{\varepsilon(m)}}^{\f{\f{\epsilon}{|m|}-\xi}{\varepsilon(m)}} dz\,
      \left[(\varepsilon(m)z+\xi)^2+1\right]^{-n} \phi(z)\,.
  \end{aligned}
\end{equation}
Since $\phi(z)$ is integrable, one has
\begin{equation}
  \label{eq:genU1abr20_1}
  \begin{aligned}
    \lim_{m\to 0} \int_{\f{\f{\epsilon}{|m|}-\xi}{\varepsilon(m)}}^\infty dz\,\phi(z)&=0\,,
    &&& \xi &\ge 0\,,\\
    \lim_{m\to 0}  \int_{-\infty}^{-\f{\xi}{\varepsilon(m)}}dz\, \phi(z)&= 0\,,
    &&& \xi &> 0\,,\\
    \lim_{m\to 0} \int_{-\f{\xi}{\varepsilon(m)}}^{\f{\f{\epsilon}{|m|}-\xi}{\varepsilon(m)}} dz\, \phi(z)&=0\,,
    &&&\xi &<0\,,
  \end{aligned}
\end{equation}
and so
\begin{equation}
  \label{eq:genU1abr20_2}
  \begin{aligned}
    &\lim_{m\to 0} \int_0^\epsilon d\lambda\,\left(\f{m^2}{\lambda^2+m^2}\right)^n
      \rho_{1,\,\mathrm{sing}}(\lambda;m)    \\
    &= \left\{
      \begin{aligned}
        &   \left(\xi^2+1\right)^{-n}  \int_{-\infty}^{\infty} dz\,\phi(z)\,, &&& \xi &>0\,,\\
        &       \int_{0}^{\infty} dz\,\phi(z)\,, &&& \xi &=0\,,\\
        &       0\,, &&&   \xi &<0\,.
      \end{aligned}\right.
  \end{aligned}
\end{equation}
If $\xi(m)>0$ depends on $m$, with $\lim_{m\to 0}\xi(m) =\infty$ but
$\lim_{m\to 0}m\xi(m) =0$, one finds instead
\begin{equation}
  \label{eq:genU1abr20_bis_again_0}
  \begin{aligned}
    & \lim_{m\to 0} \int_0^\epsilon d\lambda\,\left(\f{m^2}{\lambda^2+m^2}\right)^n
      \rho_{1,\,\mathrm{sing}}(\lambda;m)\\
    &=\lim_{m\to 0} \int_{-\f{\xi(m)}{\varepsilon(m)}}^{\f{\f{\epsilon}{|m|}-\xi(m)}{\varepsilon(m)}} dz\,
      \left[(\varepsilon(m)z+\xi(m))^2+1\right]^{-n} \phi(z) \\
    &= \lim_{\Lambda\to \infty}   \lim_{m\to 0}     \int_{-\Lambda}^{\Lambda} dz\,
      \left[(\varepsilon(m)z+\xi(m))^2+1\right]^{-n} \phi(z) \\
    &\le \lim_{\Lambda\to \infty}   \lim_{m\to 0}  \left( \f{\int_0^{\Lambda} dz\, \phi(z)}{\xi(m)^{2n}}  
      + \f{ \int_{-\Lambda}^{0} dz\, \phi(z)}{\left[\xi(m) - \varepsilon(m)\Lambda\right]^{2n}} \right)\,,
  \end{aligned}
\end{equation}
having used integrability of $\phi(z)$ on the third line, and so
\begin{equation}
  \label{eq:genU1abr20_bis_again}
  \begin{aligned}
    & \lim_{m\to 0} \int_0^\epsilon d\lambda\,\left(\f{m^2}{\lambda^2+m^2}\right)^n
      \rho_{1,\,\mathrm{sing}}(\lambda;m) \\
    & =  \left\{
      \begin{aligned}
        &\int_{-\infty}^{\infty} dz\, \phi(z)\,, &&& n&=0\,,\\
        &0\,, &&& n&\ge 1\,.
      \end{aligned}\right.
  \end{aligned}
\end{equation}

\section{Commutativity of the thermodynamic and chiral limits:
  derivations}
\label{sec:app_commlim}

In this Appendix I rederive the results of Ref.~\cite{Azcoiti:2023xvu}
discussed in Sec.~\ref{sec:ctc} using the formalism of the present
paper. To avoid notational ambiguities, let $\rho_{\lvol}(\lambda;m)$
denote the (normalized) spectral density in a finite four-volume
$\lvol=\svol/\mathrm{T}$,
\begin{equation}
  \label{eq:therm_rho_distr_def1}
  \begin{aligned}
    \rho_{\lvol}(\lambda;m) &\equiv \f{1}{\lvol}\la \rho_U(\lambda)\ra
                              =\de_\lambda \nn_{\lvol}(\lambda;m)\,,  \\
    \nn_{\lvol}(\lambda;m) &\equiv \int_0^\lambda d\lambda'\, \rho_{\lvol}(\lambda';m)\,.
  \end{aligned}
\end{equation}
The spectral density in infinite volume is denoted
$\rho(\lambda;m)=\lim_{\lvol\to\infty}\rho_{\lvol}(\lambda;m) $, with
the thermodynamic limit defined in the distributional sense through
Eq.~\eqref{eq:rho_distr_def}, i.e.,
$ \rho(\lambda;m)= \de_\lambda
\lim_{\lvol\to\infty}\nn_{\lvol}(\lambda;m)$.  One similarly defines
the spectral density in the chiral limit in a finite volume,
$\rho_{\lvol}(\lambda;0) \equiv \de_\lambda \lim_{m\to
  0}\nn_{\lvol}(\lambda;m)$ (although this limit should be
unproblematic), and its thermodynamic limit,
$ \rho_0(\lambda)\equiv\lim_{\lvol\to\infty}\rho_{\lvol}(\lambda;0)$,
i.e.,
\begin{equation}
  \label{eq:therm_rho_distr_def2}
  \rho_0(\lambda)\equiv \de_\lambda \lim_{\lvol\to \infty}\lim_{m\to 0}\nn_{\lvol}(\lambda;m)
  = \de_\lambda \lim_{\lvol\to\infty}\nn_{\lvol}(\lambda;0)\,,
\end{equation}
having used the subscript $0$ to indicate the use of the wrong order
of limits.

\subsection{Derivation of Eq.~\eqref{eq:commlim6_0}}
\label{sec:app_commlim_der1}

Assuming commutativity of the thermodynamic and chiral limits for
$\chi_\pi$ and $\chi_\delta$ amounts to stating that
\begin{equation}
  \label{eq:commlim1}
\begin{aligned}
  \lim_{m\to 0}\chi_{\pi}
  &= \lim_{m\to 0} \lim_{\lvol\to\infty} \f{\left\la (iP_a)^2\right\ra}{\lvol}
    \!\!  &&  \stackrel{!}{=} \lim_{\lvol\to\infty} \lim_{m\to 0} \f{\left\la (iP_a)^2\right\ra}{\lvol}\,,  \\
  \lim_{m\to 0}\chi_{\delta}
  &=  \lim_{m\to 0} \lim_{\lvol\to\infty} \f{\left\la (S_a)^2\right\ra}{\lvol}
    \! \!  && \stackrel{!}{=} \lim_{\lvol\to\infty}\lim_{m\to 0}  \f{\left\la (S_a)^2\right\ra}{\lvol}\,.
\end{aligned}
\end{equation}
Both order of limits can be obtained from
Eqs.~\exeqref{I-eq:coeff_rel1} and \exeqref{I-eq:first_order_expl},
and for the correct order of limits they are reported in
Eq.~\exeqref{I-eq:firstorder_again0_0}.  The corresponding expressions
taking limits in the wrong order are obtained by noticing that the
partition function, Eq.~\exeqref{I-eq:general_pf2}, is of the form
$Z=Z_0 + Z_1 + Z_{-1} + O(m^4)$, where $Z_Q$ is the partition function
restricted to the topological sector of charge $Q$. Moreover,
$Z_0 = O(m^0)$ and $Z_{1} = Z_{-1} = O(m^2)$, having used
Eq.~\exeqref{I-eq:fdet_final_det0}, $CP$ invariance, and that
$N_0\ge |Q|$ [and $N_0=|Q|$ if the index theorem is realized in a
minimal fashion, i.e., if $N_+ N_-=0$ almost everywhere in
configuration space]. The contribution of the exact zero modes does
not vanish in this case, and from Eq.~\eqref{eq:commlim1} one finds
\begin{equation}
  \label{eq:commlim2_0}
  \begin{aligned}
    &   \lim_{m\to 0}  \f{\chi_{\pi}}{2}
      = \lim_{m\to 0} 2\int_0^2 d\lambda\, \rho(\lambda;m)\ff(\lambda;m)  \\
    & \stackrel{!}{=} \lim_{\lvol\to\infty}\lim_{m\to 0} \left(\f{\la N_0\ra}{m^2 \lvol}
      + 2\int_0^2 d\lambda\, \rho_{\lvol}(\lambda;m)\ff(\lambda;m)\right)  \\
    &  = \lim_{\lvol\to\infty}\lim_{m\to 0} \left(\f{1}{m^2 \lvol}\f{2Z_{1}}{ Z_{0}}\right)
      + 2\int_0^2 d\lambda\, \f{\rho_0(\lambda)}{\lambda^2}  \,,
  \end{aligned}
\end{equation}
and
\begin{equation}
  \label{eq:commlim2_1}
  \begin{aligned}
    &\Delta = \lim_{m\to 0}  2m^2\int_0^2 d\lambda\, \rho(\lambda;m) \ff(\lambda;m)^2    \\
    & \stackrel{!}{=} \lim_{\lvol\to\infty}\lim_{m\to 0} \left(\f{\la N_0\ra}{m^2 \lvol}
      + 2m^2\int_0^2 d\lambda\, \rho_{\lvol}(\lambda;m)\ff(\lambda;m)^2\right)\\
    & = \lim_{\lvol\to\infty}\lim_{m\to 0} \left(\f{1}{m^2 \lvol}\f{2Z_{1}}{ Z_{0}}\right)\,.
  \end{aligned}
\end{equation}
In Eq.~\eqref{eq:commlim2_0}, finiteness of $\chi_\pi$ in the chiral
limit requires that both contributions on the second line be
separately finite since they are positive. This justifies the exchange
of the chiral limit with integration over the spectrum in the second
term made in the last passage. Indeed, $\rho_{\lvol}(\lambda;m)$ is a
$C^\infty$ function of $m^2$ in a finite volume, $\forall \lambda$,
including $\lambda=0$ where $\rho_{\lvol}$ vanishes, and one can write
$\rho_{\lvol}(\lambda;m)=\rho_{\lvol}(\lambda;0)
+m^2\rho_{1\,\lvol}(\lambda;m)$ with $\rho_{\lvol}(\lambda;0)$ and
$\rho_{1\,\lvol}(\lambda;m)$ bounded functions, $\forall
\lambda,m$. The contribution of $m^2\rho_{1\,\lvol}(\lambda;m)$ to
$\chi_\pi$ then vanishes as $m\to 0$, so
$\int_0^2 d\lambda\, \rho_{\lvol}(\lambda;0)\ff(\lambda;m)$ must have
a finite chiral limit, and by the same argument as in
Eq.~\eqref{eq:ua1b_gen7} one finds that
$\rho_{\lvol}(\lambda;0)/\lambda^2$ is integrable and, after taking
the thermodynamic limit (see footnote~\ref{foot:limex}), that
$\rho_0(\lambda)/\lambda^2$ is integrable. This implies also that
when taking limits in the wrong order, the contribution of non-zero
modes to $\Delta$ vanishes, since
\begin{equation}
  \label{eq:commlim2_delta}
  \begin{aligned}
    &\lim_{\lvol\to\infty}\lim_{m\to 0}  m^2\int_0^2 d\lambda\, \rho_{\lvol}(\lambda;m)\ff(\lambda;m)^2\\
    &= \lim_{\epsilon\to 0^+}\lim_{\lvol\to\infty}\lim_{m\to 0} m^2\int_0^\epsilon d\lambda\,
      \rho_{\lvol}(\lambda;m)\ff(\lambda;m)^2 \\
    &\le      \lim_{\epsilon\to 0^+}      \lim_{\lvol\to\infty}\lim_{m\to 0}  \int_0^\epsilon d\lambda\,
      \f{\rho_{\lvol}(\lambda;m)}{\lambda^2 }\\
    &=\lim_{\epsilon\to 0^+}      \int_0^\epsilon d\lambda\, \f{\rho_0(\lambda)}{\lambda^2 }=0\,,
  \end{aligned}
\end{equation}
having used the first inequality in Eq.~\eqref{eq:fineq}. Assuming
that commutativity of limits holds also for the normalized mode number
and so for the spectral density, i.e.,
$\rho_{0}(\lambda) \stackrel{!}{=}\rho(\lambda;0)\equiv \lim_{m\to
  0}\rho(\lambda;m)$, and combining Eqs.~\eqref{eq:commlim2_0} and
Eq.~\eqref{eq:commlim2_1} one ends up with the relation
\begin{equation}
  \label{eq:commlim3}
  \begin{aligned}
    &  \lim_{m\to 0} \int_0^2 d\lambda\, \rho(\lambda;m)\ff(\lambda;m) \\
    &=\lim_{m\to 0} m^2\int_0^2 d\lambda\,
      \rho(\lambda;m)\ff(\lambda;m)^2 + \int_0^2 d\lambda\,
      \f{\rho(\lambda;0)}{\lambda^2} \,.
  \end{aligned}
\end{equation}
Following Ref.~\cite{Azcoiti:2023xvu}, this can be further simplified
by splitting the integrals at an arbitrary point $\epsilon\in (0,2)$
and taking the limit $\epsilon\to 0^+$, resulting in
\begin{equation}
  \label{eq:commlim4}
  \begin{aligned}
    & \lim_{m\to 0} \int_0^2 d\lambda\, \rho(\lambda;m)\ff(\lambda;m) \\
    &=\lim_{\epsilon\to 0^+}\lim_{m\to 0} \left[\int_0^\epsilon d\lambda\,
      \rho(\lambda;m)\ff(\lambda,m) \right. \\
    &\phantom{=\lim_{\epsilon\to 0^+}\lim_{m\to 0}}
      \left.+ \int_\epsilon^2 d\lambda\, \rho(\lambda;m)\ff(\lambda,m) \right] \\
    &=\lim_{\epsilon\to 0^+}\lim_{m\to 0} \left[\int_0^\epsilon d\lambda\, \f{\rho(\lambda;m)}{\lambda^2+m^2}\right]
      + \lim_{\epsilon\to 0^+}\int_\epsilon^2 d\lambda\, \f{\rho(\lambda;0)}{\lambda^2}\\
    &=\lim_{\epsilon\to 0^+}\lim_{m\to 0} \left[\int_0^\epsilon d\lambda\, \f{\rho(\lambda;m)}{\lambda^2+m^2}\right]
      + \int_0^2 d\lambda\, \f{\rho(\lambda;0)}{\lambda^2}\,,
  \end{aligned}
\end{equation}
for the left-hand side, and
\begin{equation}
  \label{eq:commlim5}
  \begin{aligned}
    & \lim_{m\to 0} m^2\int_0^2 d\lambda\, \rho(\lambda;m)\ff(\lambda;m)^2   \\
    &= \lim_{\epsilon\to 0^+}  \lim_{m\to 0}  \int_0^\epsilon d\lambda\, \f{m^2\rho(\lambda;m)}{(\lambda^2+m^2)^2}
      \,,
  \end{aligned}
\end{equation}
for the first term on the right-hand side. Here I have made use of the
fact that replacing $\ff(\lambda;m)$ with $1/(\lambda^2+m^2)$ in
Eqs.~\eqref{eq:commlim4} and \eqref{eq:commlim5} does not change the
result, see Eqs.~\eqref{eq:commlim_gna0}--\eqref{eq:commlim_gna2}.
One concludes that
\begin{equation}
  \label{eq:commlim6}
  \begin{aligned}
    &  \lim_{\epsilon\to 0^+}\lim_{m\to 0} \int_0^\epsilon d\lambda\,\f{\rho(\lambda;m)}{\lambda^2+m^2} \\
    &= \lim_{\epsilon\to 0^+} \lim_{m\to 0} \int_0^\epsilon d\lambda\,\f{m^2\rho(\lambda;m)}{(\lambda^2+m^2)^2}
      = \f{\Delta}{2}\,,
  \end{aligned}
\end{equation}
which is Eq.~\eqref{eq:commlim6_0}, and the same as Eq.~(49) of
Ref.~\cite{Azcoiti:2023xvu}, up to terms that vanish in the chiral
limit. Since
\begin{equation}
  \label{eq:commlim6_gen}
  \begin{aligned}
    0&\le   \II_n(\epsilon;m)-\II_{n+1}(\epsilon;m)
       = \int_0^\epsilon d\lambda\f{m^{2n}\lambda^2\rho(\lambda;m)}{(\lambda^2+m^2)^{n+1}}\\
     &  \le   \II_0(\epsilon;m)-\II_{1}(\epsilon;m)\,,
  \end{aligned}
\end{equation}
if Eq.~\eqref{eq:commlim6_0} holds then
$\lim_{\epsilon\to 0^+}\lim_{m\to 0}\II_{n}(\epsilon;m)$ is
independent of $n$, and Eq.~\eqref{eq:commlim6_0_gen} follows.

\subsection{Derivation of Eq.~\eqref{eq:commlim17_0}} 
\label{sec:app_commlim_der2}

Following again Ref.~\cite{Azcoiti:2023xvu}, one recasts
Eq.~\eqref{eq:commlim6} in the equivalent form
\begin{equation}
  \label{eq:commlim11}
  \begin{aligned}
    \f{\Delta}{2}
    &=\lim_{\epsilon\to 0^+}\lim_{m\to 0} \int_0^\epsilon d\lambda\,\f{\rho(\lambda;m)}{\lambda^2+m^2} \\
    &=\lim_{\epsilon\to 0^+}\lim_{m\to 0}\left[ \int_0^{|m|} d\lambda\,\f{\rho(\lambda;m)}{\lambda^2+m^2}
    + \int_{|m|}^\epsilon d\lambda\, \f{\rho(\lambda;m)}{\lambda^2+m^2}\right]\,,\\
    0&=\lim_{\epsilon\to 0^+} \lim_{m\to 0} \int_0^\epsilon d\lambda\, \f{\lambda^2\rho(\lambda;m)}{(\lambda^2+m^2)^2} \\
    &=\lim_{\epsilon\to 0^+} \lim_{m\to 0}\left[ \int_0^{|m|} d\lambda\, \f{\lambda^2\rho(\lambda;m)}{(\lambda^2+m^2)^2}\right.\\
    &\phantom{=\lim_{\epsilon\to 0^+} \lim_{m\to 0}}\left.+ \int_{|m|}^\epsilon d\lambda\,\f{\lambda^2\rho(\lambda;m)}{(\lambda^2+m^2)^2}\right]\,.
\end{aligned}
\end{equation}
In the second equation both terms must vanish in the relevant limit
due to their positivity. Using this fact, one finds
\begin{equation}
  \label{eq:commlim12}
\begin{aligned}
  & 0\le \lim_{\epsilon\to 0^+}\lim_{m\to 0} \int_{|m|}^\epsilon d\lambda\, \f{\rho(\lambda;m)}{\lambda^2+m^2}  \\
  &=  \lim_{\epsilon\to 0^+}\lim_{m\to 0} \int_{|m|}^\epsilon d\lambda\, \f{\rho(\lambda;m)}{\lambda^2+m^2}\f{m^2}{\lambda^2+m^2} \\
  & \le \f{1}{2}\lim_{\epsilon\to 0^+}\lim_{m\to 0} \int_{|m|}^\epsilon d\lambda\, \f{\rho(\lambda;m)}{\lambda^2+m^2}\,,
\end{aligned}
\end{equation}
which is possible only if
\begin{equation}
  \label{eq:commlim13}
  \lim_{\epsilon\to 0^+}\lim_{m\to 0} \int_{|m|}^\epsilon d\lambda\, \f{\rho(\lambda;m)}{\lambda^2+m^2}=0\,.
\end{equation}
Equation~\eqref{eq:commlim11} requires then
\begin{equation}
  \label{eq:commlim14}
  \begin{aligned}
    \f{\Delta}{2}
    &=  \lim_{\epsilon\to 0^+}\lim_{m\to 0} \int_0^{|m|} d\lambda\, \f{\rho(\lambda;m)}{\lambda^2+m^2}    \\
    & =  \lim_{m\to 0} \int_0^1 dz\,\f{1}{z^2+1}   \f{\rho(|m| z;m)}{|m|} \,,\\
    0
    &=  \lim_{\epsilon\to 0^+} \lim_{m\to 0} \int_0^{|m|} d\lambda\,  \f{\lambda^2\rho(\lambda;m)}{(\lambda^2+m^2)^2}    \\
    & = \lim_{m\to 0} \int_0^1 dz\,\f{z^2}{(z^2+1)^2} \f{\rho(|m| z;m)}{|m|}\,.
  \end{aligned}
\end{equation}
Since the integrand is nonnegative, the second equation requires that
in the chiral limit it must vanish almost everywhere in $[0,1]$, so
$ \lim_{m\to 0}\f{\rho(|m| z;m)}{|m|}$ vanishes almost everywhere in
$ (0,1]$, and one finds
\begin{equation}
  \label{eq:commlim15}
  0 = \lim_{m\to 0} \int_{0}^1 dz\,\f{z^\kappa}{(z^2+1)^2} \f{\rho(|m|z;m)}{|m|}\,,
\end{equation}
for any $\kappa>0$. From this and from the symmetry of $\rho$ one
concludes that
\begin{equation}
  \label{eq:commlim16}
  \lim_{m\to 0}\f{1}{z^2+1}\f{\rho(|m| z;m)}{|m|} = \Delta\delta(z)\,,
\end{equation}
for smooth functions in the interval $[-1,1]$, and for this class of
functions one has equivalently
\begin{equation}
  \label{eq:commlim17}
  \lim_{m\to 0}\f{\rho(|m| z;m)}{|m|} = \Delta\delta(z)\,.
\end{equation}
The factor $\f{1}{2}$ in Eq.~\eqref{eq:commlim14} is recovered by
symmetrizing the integral,
$\int_0^1dz\, s(z) = \f{1}{2}\int_{-1}^1dz\,s(z)$ if $s(-z)=s(z)$. 

Equation~\eqref{eq:commlim17} requires that
$\delta_{m}(z)\equiv\f{\rho(|m| z;m)}{\Delta |m|}$ be a nascent delta
function if $\Delta\neq 0$. One can show that the singular peaks in
Eqs.~\eqref{eq:rhom2diff_peak2} and \eqref{eq:rhom2diff_peak2_commut}
indeed obey Eq.~\eqref{eq:commlim14} and so
Eq.~\eqref{eq:commlim17}. One has for
$\delta_{m\,\mathrm{peak}}(z)=\f{\rho_{\mathrm{peak}}(|m|z;m)}{\Delta|m|}$
that
\begin{equation}
  \label{eq:commlim18}
  \delta_{m\,\mathrm{peak}}(z) 
  = \left(\f{1}{2}+o(1)\right) \gamma(m^2) |z|^{-1+\gamma(m^2)}\,,
\end{equation}
having used $\lim_{m\to 0}|m|^{\gamma(m^2)}=1$, since $\gamma>0$ for
small $m$; similarly, for
$\tilde{\delta}_{m\,\mathrm{peak}}(z)=\f{\tilde{\rho}_{\mathrm{peak}}(|m|z;m)}{\Delta|m|}$
one has
\begin{equation}
  \label{eq:commlim_nonm2_1}
  \tilde{\delta}_{m\,\mathrm{peak}}(z)
  = \left(\f{1}{2}+o(1)\right) \tilde{\gamma}(m) |z|^{-1+\tilde{\gamma}(m)}\,.
\end{equation}
One finds then
\begin{equation}
  \label{eq:commlim19}
  \begin{aligned}
    0&\le \lim_{m\to 0} 2\int_{0}^1 dz\,\f{z^\kappa \delta_{m\,\mathrm{peak}}(z)}{(z^2+1)^2} \\
     &  = \lim_{m\to 0} \gamma(m^2)\int_0^1 dz\,\f{ z^{\kappa-1+\gamma(m^2)}}{(z^2+1)^2}  \\
     & \le \int_0^1 dz\,\f{ z^{\kappa-1}}{(z^2+1)^2} \,  \lim_{m\to 0} \gamma(m^2)=0\,,
\end{aligned}
\end{equation}
for any $\kappa>0$, and
\begin{equation}
  \label{eq:commlim20}
  \begin{aligned}
    &\lim_{m\to 0} 2\int_{0}^1 dz\,\f{\delta_{m\,\mathrm{peak}}(z)}{z^2+1} \\
    &= \lim_{\eta\to 0^+}\lim_{m\to 0}\left[  \int_\eta^\infty dw\,\f{ e^{-w}}{e^{-\f{2}{\gamma(m^2)}w}+1}\right.\\
    &\phantom{=\lim_{\eta\to 0^+}\lim_{m\to 0}}\left. + \int_0^\eta dw\,\f{ e^{-w}}{e^{-\f{2}{\gamma(m^2)}w}+1}\right]  \\
    & =  \lim_{\eta\to 0^+}  \int_\eta^\infty dw\,e^{-w}  = 1\,. 
  \end{aligned}
\end{equation}
The proof for $\tilde{\delta}_{m\,\mathrm{peak}}(z)$ requires only
replacing $\gamma$ with $\tilde{\gamma}$ in Eqs.~\eqref{eq:commlim19}
and \eqref{eq:commlim20}.

\section{Constraints on the two-point function: details}
\label{sec:tpf_app}

\subsection{Finiteness of  \texorpdfstring{$\Id_\delta[g]$}{I\textunderscore delta[g]}}
\label{sec:twopointbound}

The contribution $ \Id_\delta[g_1,g_2]$, Eq.~\eqref{eq:bound_tpf_0},
of the region $\delta\le \lambda,\lambda'\le 2$ to $\Id[g_1,g_2]$,
Eq.~\eqref{eq:tpf_constr2}, is
\begin{equation}
  \label{eq:bound_tpf_1}
  \begin{aligned}
    \Id_\delta[g_1,g_2]
    &= \int_\delta^2 d\lambda\int_\delta^2 d\lambda'\, g_1(\lambda;m) g_2(\lambda';m)\\
    &\hphantom{\equiv}\times 
      \de_\lambda \de_{\lambda'} \nnd(\lambda,\lambda';m)\,,   
  \end{aligned}
\end{equation}
see Eqs.~\eqref{eq:tpf_rig} and \eqref{eq:tpf_rig2}. Integrating by parts, one finds
\begin{widetext}
  \begin{equation}
    \label{eq:bound_tpf_3}
    \begin{aligned}
      &      \Id_\delta[g_1,g_2]  \\
      &=  g_1(2;m)g_2(2;m) \nnd(2,2;m)
        - \left[ g_1(2;m) g_2(\delta;m) + g_1(\delta;m)g_2(2;m) \right]
        \nnd(2,\delta;m)  +   g_1(\delta;m)    g_2(\delta;m)
        \nnd(\delta,\delta;m)  \\
      &\phantom{=}- \int_\delta^2 d\lambda\,   \left[ \dot{g}_1(\lambda;m)\left(g_2(2;m)
        \nnd(\lambda,2;m) -g_2(\delta;m)
        \nnd(\lambda,\delta;m)\right)\right.\\
      &\hphantom{-=} \left. +  \dot{g}_2(\lambda;m)\left(g_1(2;m)
        \nnd(\lambda,2;m) -g_1(\delta;m)
        \nnd(\lambda,\delta;m)\right)\right] +\int_\delta^2 d\lambda\int_\delta^2 d\lambda'\,
        \dot{g}_1(\lambda;m) \dot{g}_2(\lambda';m)  
        \nnd(\lambda,\lambda';m)\,,
    \end{aligned}
  \end{equation}
\end{widetext}
where $\dot{g}_{1,2}(\lambda;m)\equiv \de_\lambda g_{1,2}(\lambda;m)$,
and having used the obvious symmetry
$\nnd(\lambda',\lambda;m)= \nnd(\lambda,\lambda';m)$.  The functions
of interest, $g_{1,2}=\ff,\ffh$, are positive, and their derivatives,
\begin{equation}
  \label{eq:bound_tpf_4}
  \begin{aligned}
    \de_\lambda \ff(\lambda;m) &= -\f{2\lambda}{\left[\lambda^2+m^2h(\lambda)\right]^2}\,,\\
    \de_\lambda \ffh(\lambda;m) &= \left(1-2m^2\ff(\lambda;m)\right)\de_\lambda \ff(\lambda;m)\,,
  \end{aligned}
\end{equation}
are of fixed negative sign in any integration range $[\delta,2]$ for
sufficiently small $m$ (and actually $\de_\lambda \ff<0 $
$\forall m$), and moreover
\begin{equation}
  \label{eq:bound_tpf_5}
  \ffh  \le \ff \le \f{1}{\delta^2}\,. 
\end{equation}
This remains true also for an integration range $[\bar{\delta}(m),2]$
with mass-dependent $\bar{\delta}(m)$ if $\bar{\delta}(0)\neq 0$, of
course replacing $\delta$ with $\bar{\delta}(m)$ in
Eq.~\eqref{eq:bound_tpf_5}. Setting
$\nndmax(m) \equiv \max_{\bar{\delta}(m) \le \lambda,\lambda'\le
  2}\left| \nnd(\lambda,\lambda';m)\right|$, one finds for
$g_{1,2}=\ff,\ffh$
\begin{equation}
  \label{eq:bound_tpf_6}
  \begin{aligned}
    &    \lim_{m\to 0}  \left|  \Id_{\bar{\delta}(m)}[g_1,g_2]\right| \\
    &\le 4\lim_{m\to 0}  \nndmax(m) g_1(\delta;m)g_2(\delta;m)
      \le  \f{4 \nndmax(0)}{\bar{\delta}(0)^4}  \,. 
  \end{aligned}
\end{equation}
Note that $\lim_{m\to 0} \nndmax(m)$ does not depend on the sign of
$m$, see footnote~\ref{foot:rhomz}. For more general $g_{1,2}$,
bounded with bounded derivatives for $\lambda\in [\bar{\delta}(m),2]$
and for all sufficiently small $m$, $|g_{1,2}(\lambda;m)|\le a_{1,2}$,
$|\dot{g}_{1,2}(\lambda;m)|\le b_{1,2}$, one finds
\begin{equation}
  \label{eq:bound_tpf_6_bis}
  \lim_{m\to 0}  \left|  \Id_{\bar{\delta}(m)}[g_1,g_2]\right|\le  C
  \nndmax(0)\,,
\end{equation}
with
\begin{equation}
  \label{eq:bound_tpf_6_ter}
\begin{aligned}
  C  &  =   4a_{1}a_{2} + 2(2-\bar{\delta}(0))  (b_{1}a_{2}+a_{1}b_{2})  +(2-\bar{\delta}(0))^2  b_{1}b_{2}\\
     &  \le   4(a_{1} + b_{1})(a_{2}+b_{2})\,.
\end{aligned}
\end{equation}
In conclusion, for the relevant functions
\begin{equation}
  \label{eq:bound_tpf_final_0}
  \lim_{m\to 0}  \left|\Id_{\bar{\delta}(m)}[g_1,g_2]  \right|<\infty\,,
\end{equation}
as long as $\bar{\delta}(m)$ does not vanish in the chiral limit. In
particular, in the presence of a mobility edge, $\lambda_c$, that does
not vanish in the chiral limit, one has
\begin{equation}
  \label{eq:bound_tpf_final}
  \lim_{m\to 0}  m^2\Id_{\lambda_c}[\ff,\ff]  =0\,,  \qquad
  \left|  \lim_{m\to 0}  \Id_{\lambda_c}[\ffh,\ffh]\right|  <\infty\,.  
\end{equation}

\subsection{Correlation of localized and delocalized modes}
\label{sec:locdeloccorr}

For a spectrum comprising both localized and delocalized modes,
integrating the two-point function $\rho^{(2)}_c$,
Eq.~\eqref{eq:spqrecap2}, over intervals $\Delta_l$ and $\Delta_d$ in
the localized and delocalized regions of the spectrum one finds
\begin{equation}
  \label{eq:corrld1}
  \begin{aligned}
    C(\Delta_l,\Delta_d) &\equiv    \int_{\Delta_l}d\lambda \int_{\Delta_d}d\lambda'\,  \rho^{(2)}_c(\lambda,\lambda';m) \\
                         &=  \lim_{\lvol\to\infty}\f{1}{\lvol}  \la \delta N(\Delta_l)\delta N(\Delta_d)\ra\,,
  \end{aligned}
\end{equation}
where $\delta N(\Delta)\equiv N(\Delta) - \la N(\Delta)\ra$ with
\begin{equation}
  \label{eq:corrld2}
  N(\Delta)\equiv  \int_{\Delta}d\lambda\,   \rho_U(\lambda)
\end{equation}
the number of modes in the spectral interval $\Delta$ in a given
configuration, $U$. The Dirac-delta term in $\rho^{(2)}_c$ does not
contribute, even for adjacent spectral intervals, for in that case
$\lambda=\lambda'$ only at a single point, i.e., the mobility edge
between them. For the total number of localized and delocalized modes
(including zero and doubler modes), $N_l$ and $N_d$, one obviously has
that the fluctuations $\delta N_{l,d} = N_{l,d} - \la N_{l,d}\ra$ obey
$ \delta N_l + \delta N_d =0$, as any fluctuation in $N_l$ on a given
configuration is compensated by an opposite fluctuation in $N_d$.

An increase in the number of localized modes as the result of a
perturbation in the disorder (i.e., a perturbation of the gauge
configuration, in the present case) requires that the (unperturbed)
delocalized modes interfere destructively all over the spectrum, all
contributing comparably to the magnitude of the newly formed localized
modes. The change in the number of modes in a delocalized spectral
region is then expected to be proportional to the number of modes
there, with a (mostly negative) proportionality constant of order
$O(\lvol^0)$. Suppressing factors of order 1, one then expects
$\delta N(\Delta_d) \sim \f{N(\Delta_d)}{N_d}\delta N_d =
-\f{N(\Delta_d)}{N_d}\delta N_l$ for the mode number
fluctuation. Furthermore, localized modes fluctuate independently (up
to finite-size effects), so $\delta N(\Delta_l)$ is independent of the
analogous fluctuations in other intervals in the localized regime of
the spectrum. One finds then [ignoring $O(1)$ factors and retaining
only contributions leading in volume]
\begin{equation}
  \label{eq:corrld4}
  \begin{aligned}
    & -\la \delta N(\Delta_l) \delta N(\Delta_d)\ra
      \sim   \left\la \delta N(\Delta_l) \delta N_l\f{N(\Delta_d)}{N_d}\right\ra\\
    & \sim  \left\la \delta N(\Delta_l)^2 \f{N(\Delta_d)}{N_d}\right\ra
      \sim   \f{\la N(\Delta_d)\ra}{\la N_d\ra}\left\la\delta N(\Delta_l)^2\right\ra\\
    &   \sim   \f{\la N(\Delta_d)\ra}{\la N_d\ra}\left\la N(\Delta_l)\right\ra\,,
\end{aligned}
\end{equation}
where in the last passage I have used the fact that localized modes
are Poisson distributed,
$\left\la\delta N(\Delta_l)^2\right\ra=\left\la N(\Delta_l)\right\ra$
(again up to finite-size effects). One has then
\begin{equation}
  \label{eq:corrld5}
  \begin{aligned}
    C(\Delta_l,\Delta_d) &\sim  -  \lim_{\lvol\to\infty}\f{\left\la N(\Delta_l)\right\ra \la N(\Delta_d)\ra}{\lvol^2} \f{\lvol}{\la N_d\ra} \\
                         & = -  \f{1}{\nu_d(m)}
                           \int_{\Delta_l}d\lambda \,\rho(\lambda;m) \int_{\Delta_d}d\lambda' \,\rho(\lambda';m)   \,,
  \end{aligned}
\end{equation}
where $\nu_d \equiv \lim_{\lvol\to\infty} \la N_d\ra/\lvol$, up to
$\Delta_{l,d}$-dependent factors of order $O(\lvol^0)$.  Making the
integration intervals infinitesimal one concludes that for $\lambda$
and $\lambda'$ in the localized and delocalized regions of the
spectrum, respectively, one has
\begin{equation}
  \label{eq:corrld6}
  \rho^{(2)}_c(\lambda,\lambda';m) =
  -C^{(2)}(\lambda,\lambda';m)\rho(\lambda;m)\rho(\lambda';m)\,,
\end{equation}
for some bounded and (mostly) positive function $C^{(2)}$. Since the
same procedure applies, of course, if the roles of $\lambda$ and
$\lambda'$ are interchanged, $C^{(2)}$ is symmetric under exchange of
its arguments. Finally, Eq.~\eqref{eq:ltpf3_bis} follows from
boundedness of $C^{(2)}$.  This property is expected to hold also in
the chiral limit, at least if the localized region does not disappear.

\mbox{}

\mbox{}

\bibliographystyle{../apsrev4-2_mod}
\bibliography{../references_chi_PRD}

\end{document}